\documentclass[twocolumn,times]{aastex631}

\usepackage{newtxtext,newtxmath}
\usepackage{pifont}

\newcommand{\nut}{{\sc nut}}
\newcommand{\MBdiez}{\textcolor{MB10}{\textbf{MB10}}}
\newcommand{\MBonce}{\textcolor{MB11}{\textbf{MB11}}}
\newcommand{\MBdoce}{\textcolor{MB12}{\textbf{MB12}}}
\newcommand{\MBinj}{\textcolor{MBinj}{\textbf{MBinj}}}
\newcommand{\MBveinte}{\textcolor{MB20}{\textbf{MB20}}}
\newcommand{\K}{\mathrm{K}}
\newcommand{\erg}{\mathrm{erg}}
\newcommand{\s}{\mathrm{s}}
\newcommand{\g}{\mathrm{g}}
\newcommand{\cm}{\mathrm{cm}}
\newcommand{\cc}{\mathrm{cm}^{-3}}
\newcommand{\pc}{\mathrm{pc}}
\newcommand{\kpc}{\mathrm{kpc}}
\newcommand{\muG}{\mu\mathrm{G}}
\newcommand{\Msun}{\mathrm{M}_\odot}
\newcommand{\redshift}{z_\text{redshift}}
\newcommand{\zcoord}{z}
\newcommand{\hgal}{h_\text{CR,e}}
\newcommand{\Rgal}{R_\text{CR,e}}
\newcommand{\pFIR}{p_{0, \text{FIR}}}
\newcommand{\pCR}{p_\text{CR}}

\newcommand{\hIFIR}{h_\text{I\,\text{FIR}}}
\newcommand{\hPIFIR}{h_\text{PI\,\text{FIR}}}
\newcommand{\hIradio}{h_\text{I\,\text{radio}}}
\newcommand{\hPIradio}{h_\text{PI\,\text{radio}}}
\newcommand{\hSFR}{h_{\Sigma\,\text{SFR10}}}
\newcommand{\hSFRlong}{h_{\Sigma\,\text{SFR100}}}
\newcommand{\hCNM}{h_{\Sigma\,\text{CNM}}}

\newcommand{\hCNMWNM}{h_{\Sigma\,\text{CNM+WNM}}}

\newcommand{\dx}{\text{dx}}%
\newcommand{\dxmin}{\Delta\text{x}_\text{min}}%
\newcommand{\ndust}{n_\text{dust}}%
\newcommand{\ramses}{{\sc ramses}}
\newcommand{\xmark}{\ding{55}}%

\shorttitle{a tomographic view of fir and radio in galaxy simulations}
\shortauthors{martin-alvarez et al.}

\begin{document}

\title{Extragalactic Magnetism with SOFIA (SALSA Legacy Program). VII. A Tomographic View of Far-infrared and Radio Polarimetric Observations through MHD Simulations of Galaxies}

\author[0000-0002-4059-9850]{Sergio Martin-Alvarez}
\affiliation{Kavli Institute for Particle Astrophysics \& Cosmology (KIPAC), Stanford University, Stanford, CA 94305, USA}

\author[0000-0001-5357-6538]{Enrique Lopez-Rodriguez}
\affiliation{Kavli Institute for Particle Astrophysics \& Cosmology (KIPAC), Stanford University, Stanford, CA 94305, USA}

\author[0000-0002-4746-2128]{Tara Dacunha}
\affiliation{Kavli Institute for Particle Astrophysics \& Cosmology (KIPAC), Stanford University, Stanford, CA 94305, USA}

\author[0000-0002-7633-3376]{Susan E. Clark} 
\affiliation{Kavli Institute for Particle Astrophysics \& Cosmology (KIPAC), Stanford University, Stanford, CA 94305, USA}
\affiliation{Department of Physics, Stanford University, Stanford, CA 94305, USA}

\author[0000-0003-3249-4431]{Alejandro S. Borlaff}
\affiliation{NASA Ames Research Center, Moffett Field, CA 94035, USA}
\affiliation{Bay Area Environmental Research Institute, Moffett Field, CA 94035, USA}

\author[0009-0001-8154-3562]{Rainer Beck}
\affiliation{Max-Planck-Institut für Radioastronomie, Auf dem Hügel 69, D-53121, Bonn, Germany}

\author[0000-0001-6535-1766]{Francisco Rodríguez Montero} \affiliation{Sub-department of Astrophysics, University of Oxford, Keble Road, Oxford, OX1 3RH, UK}

\author[0000-0001-5512-3735]{S. Lyla Jung}
\affiliation{Research School of Astronomy \& Astrophysics, The Australian National University, Canberra ACT 2611, Australia}
\affiliation{Sub-department of Astrophysics, University of Oxford, Keble Road, Oxford, OX1 3RH, UK}

\author[0000-0002-8140-0422]{Julien Devriendt}
\affiliation{Sub-department of Astrophysics, University of Oxford, Keble Road, Oxford, OX1 3RH, UK}

\author{Adrianne Slyz}
\affiliation{Sub-department of Astrophysics, University of Oxford, Keble Road, Oxford, OX1 3RH, UK}

\author[0000-0001-6326-7069]{Julia Roman-Duval}
\affiliation{Space Telescope Science Institute, 3700 San Martin Drive, Baltimore, MD 21218, USA}

\author[0000-0002-4324-0034]{Evangelia Ntormousi}
\affiliation{Scuola Normale Superiore di Pisa, Piazza dei Cavalieri 7, 56126 Pisa, Italy}

\author[0000-0001-8749-1436]{Mehrnoosh Tahani}
\affiliation{Kavli Institute for Particle Astrophysics \& Cosmology (KIPAC), Stanford University, Stanford, CA 94305, USA}

\author[0000-0001-7039-9078]{Kandaswamy Subramanian} 
\affiliation{IUCAA, Post Bag 4, Ganeshkhind, Pune 411007, India}
\affiliation{Department of Physics, Ashoka University, Rajiv Gandhi Education City, Rai, Sonipat 131029, Haryana, India}

\author[0000-0002-5782-9093]{Daniel~A.~Dale}
\affiliation{Department of Physics and Astronomy, University of Wyoming, Laramie, WY 82071, USA}

\author{Pamela M. Marcum}
\affiliation{NASA Ames Research Center, Moffett Field, CA 94035, USA}

\author[0000-0002-8831-2038]{Konstantinos Tassis}
\affiliation{Department of Physics, and Institute for Theoretical and Computational Physics, University of Crete, Voutes University campus, GR-70013 Heraklion, Greece}
\affiliation{Institute of Astrophysics, Foundation for Research and Technology-Hellas, N. Plastira 100, Vassilika Vouton, GR-71110 Heraklion, Greece}

\author[0000-0001-8931-1152]{Ignacio del Moral-Castro}
\affiliation{Kapteyn Astronomical Institute, University of Groningen, PO Box 800, 9700 AV Groningen, The Netherlands}
\affiliation{Facultad de F\'isica, Universidad de La Laguna, Avda. Astrof\'isico Fco. S\'anchez s/n, 38200, La Laguna, Tenerife, Spain}
\affiliation{Instituto de Astrofísica, Facultad de Física, Pontificia Universidad Católica de Chile, Campus San Joaquín, Av. Vicuña Mackenna 4860, Macul, 7820436, Santiago, Chile}

\author[0000-0002-6488-8227]{Le Ngoc Tram}
\affiliation{Max-Planck-Institut für Radioastronomie, Auf dem Hügel 69, D-53121, Bonn, Germany}

\author[0000-0001-7039-9078]{Matt J. Jarvis}
\affiliation{Subdepartment of Astrophysics, University of Oxford, Keble Road, Oxford, OX1 3RH, UK}
\affiliation{Department of Physics and Astronomy, University of the Western Cape, Robert Sobukwe Road, 7535 Bellville, Cape Town, South Africa}



\begin{abstract}
The structure of magnetic fields in galaxies remains poorly constrained, despite the importance of magnetism in the evolution of galaxies. Radio synchrotron and far-infrared (FIR) polarization and polarimetric observations are the best methods to measure galactic scale properties of magnetic fields in galaxies beyond the Milky Way. We use synthetic polarimetric observations of a simulated galaxy to identify and quantify the regions, scales, and interstellar medium (ISM) phases probed at FIR and radio wavelengths. Our studied suite of magnetohydrodynamical cosmological zoom-in simulations features high-resolutions (10~pc full-cell size) and multiple magnetization models. Our synthetic observations have a striking resemblance to those of observed galaxies. We find that the total and polarized radio emission extends to approximately double the altitude above the galactic disk (half-intensity disk thickness of $h_\text{I\,\text{radio}} \sim h_\text{PI\,\text{radio}} = 0.23 \pm 0.03$ kpc) relative to the total FIR and polarized emission that are concentrated in the disk midplane ($h_\text{I\,\text{FIR}} \sim h_\text{PI\,\text{FIR}} = 0.11 \pm 0.01$ kpc). Radio emission traces magnetic fields at scales of $\gtrsim 300$ pc, whereas FIR emission probes magnetic fields at the smallest scales of our simulations. These scales are comparable to our spatial resolution and well below the spatial resolution ($<300$ pc) of existing FIR polarimetric measurements. Finally, we confirm that synchrotron emission traces a combination of the warm neutral and cold neutral gas phases, whereas FIR emission follows the densest gas in the cold neutral phase in the simulation. These results are independent of the ISM magnetic field strength. The complementarity we measure between radio and FIR wavelengths motivates future multiwavelength polarimetric observations to advance our knowledge of extragalactic magnetism.
\end{abstract}


\section{Introduction} 
\label{s:Introduction}

\definecolor{MB10}{rgb}{0.4550, 0.271 , 0.424}
\definecolor{MB11}{rgb}{0.6150, 0.375 , 0.572}
\definecolor{MB12}{rgb}{0.7550, 0.532 , 0.714}
\definecolor{MBinj}{rgb}{0.667, 0.000 , 0.000}
\definecolor{MB20}{rgb}{0.5000, 0.500 , 1.000}

Magnetic fields are a fundamental constituent of the interstellar medium (ISM) for every galaxy in our Universe. Theoretical and observational studies predict that the magnetic energy budget is comparable to that of the thermal and turbulent components in the ISM \citep{Beck2007, Bernet2008, Mao2017, Pakmor2017, Martin-Alvarez2020,Lopez-Rodriguez2021,Geach2023Polarized2.6, Lopez-Rodriguez2023}, and dominates over the thermal component in cold molecular clouds \citep{Crutcher1999}. Magnetic fields impact the ISM, due to their significant strengths, with typical values in the $\sim3-7~\mu$G range for the large-scale ordered component and are typically on the order of $\sim17~\mu$G for the total field strength \citep{Fletcher2010,Beck2019b}. Some of their effects are the co-regulation of star formation \citep{McKeeOstriker2007}, modifying the gas distribution across ISM phases \citep{Iffrig2017,Hennebelle2019}, the fragmentation of gas \citep{Inoue2019}, and influencing the formation of molecular clouds \citep{Inoue2018, Tahani2022a, Tahani2022bb}. Their effects expand beyond small galactic scales, as magnetic fields are likely to modify gas mixing in halos and outflows \citep{Cottle2020,Lopez-Rodriguez2021,vandeVoort2021,Buie2022,Lopez-Rodriguez2023}, general properties of galaxies \citep{Pillepich2018a,Martin-Alvarez2020}, and even be non-negligible during galaxy hierarchical growth \citep{Whittingham2021}. One of the few galaxy parameters that appear to not be directly affected by the predicted magnetic fields of $\sim\muG$ are galaxy stellar masses \citep{Su2017,Martin-Alvarez2020}, except for dwarf galaxies, where the effects of magnetic fields are still not fully understood \citep{Martin-Alvarez2023, Whitworth2023, Sanati2024}.

Observations of magnetic fields are required to characterize and constrain their properties and their relationship to ISM structure. Mapping magnetic fields in galaxies beyond the Milky Way is mostly limited to polarimetric observations in the FIR and at radio wavelengths. 
Radio emission is generated by a combination of thermal and synchrotron emission. Below 10 GHz, the synchrotron dominates the total observed and polarized emission. This synchrotron emission is generated by energetic particles in the form of cosmic rays (CRs) spiraling along magnetic field lines. The radiation emitted by these particles traces the orientation of the field perpendicular to the line of sight (LOS). The synchrotron emission is frequently observed to be widespread over the surface of galaxy disks and extends to high altitudes above their midplanes (e.g., \citealt{BeckLive}\footnote{The latest revision (currently 2023 September) is available at: \href{https://arxiv.org/abs/1302.5663}{https://arxiv.org/abs/1302.5663}}; \citealt{Krause2020}). Consequently, it has the potential to probe the distribution of magnetic fields at larger scales than the FIR emission, as well as the number density of CR electrons responsible for the majority of the observed radio emission \citep{Sun2008}.

Polarized thermal emission at FIR wavelengths arises from magnetically aligned dust grains \citep[e.g.,][]{Purcell1979, HL2016} and traces dust content in a density-weighted manner along the LOS \citep{Seifried2019}. FIR ($53-214~\mu$m) polarimetric observations were possible using the imaging polarimetric mode with the High-resolution Airborne Wideband Camera-Plus (HAWC+) on board the Stratospheric Observatory For Infrared Astronomy (SOFIA). Using HAWC+/SOFIA and archival radio polarimetric observations, the Survey of extragALactic magnetiSm with SOFIA \citep[SALSA SOFIA Legacy Program; PI: Lopez-Rodriguez, E. \& Mao, S. A.;][]{SALSAIV} has enabled for the first time a combined analysis of extragalactic magnetic fields in the multiphase ISM of nearby galaxies. These observations have shown that the FIR polarized emission can be associated with dense, N$_{\text{HI}+\text{H}_{2}} \in [10^{20},10^{23}]\, \rm{cm}^{-2})$, and cold, $T_\text{d} \in [20,50]\, \K$, regions of the ISM \citep{SALSAIV}.
Furthermore, the angular variations of the plane-of-the-sky magnetic field orientations inferred from FIR observations are larger than those measured by the radio magnetic fields \citep{SALSAV}. \citet{Surgent2023} showed that the FIR magnetic field orientations are $7-25$\% more disordered than those from the magnetic fields inferred from the radio emission. Note that these variations are at the scales of the angular resolution of the observations, typically $300-500$ pc. These results indicate that the large-scale ordered FIR magnetic fields are sensitive to physical structures (e.g., star-forming regions, molecular clouds) below the angular resolution of the observations.
These works illustrate the remarkable potential for the combination of radio and FIR observations. However, a precise understanding of their complementarity requires a quantitative framework associating specific observables with the underlying physical quantities. For example, estimating the scale depth probed by each type of emission, unambiguously determining depolarization mechanisms at play, and segregating each wavelength range by ISM phase, if possible. Such quantitative relations are best studied through theoretical methods capable of predicting specific observables from intrinsic physical quantities. Taking a first step into establishing such a framework is precisely the objective of this work.

In this context, magnetohydrodynamical (MHD) simulations of galaxy formation emerge as a powerful method to unravel the complexity of magnetic fields and their connection with galaxy properties and observables. During the last decade, multiple MHD high-resolution studies have been able to simulate realistic galaxies with magnetic properties in broad agreement with those inferred by global properties of observed galaxies \citep[e.g.,][]{Dubois2010, Pakmor2017, Hopkins2020, Martin-Alvarez2021, Martin-Alvarez2023}. These simulations have been able to capture both the turbulent and large-scale dynamo amplification processes. The turbulent dynamo is believed to bridge the gap separating the $\muG$ observed in galaxies from the extremely weak magnetic fields at the beginning of the universe \citep[e.g.,][]{Su2017,Martin-Alvarez2021,Gent2023,Kriel2023,Pakmor2023}. The large-scale dynamo reorganizes the magnetic field on galactic scales and may be responsible for the kiloparsec-scale ordered spiral magnetic fields observed at radio and FIR wavelengths \citep{BeckLive,SALSAIV}. Some detailed reviews regarding the origin and amplification of magnetic fields in galaxies can be found in \citet{Widrow2002}, \citet{Brandenburg2005} and \citet{Shukurov2021}. In addition to exploring the role of magnetization on the evolution of galaxies \citep{Steinwandel2019, Buck2020, Martin-Alvarez2020}, MHD simulations can be employed to gain a more detailed understanding of observables (i.e., orientation and fraction of the measured polarization at different wavelength regimes) related to magnetic fields structure and strength \citep[e.g.,][]{Pakmor2018, Reissl2019, Werhahn2021c, Werhahn2021a, Werhahn2021b, Pfrommer2022, Ponnada2022, Rodriguez-Montero2022, Jung2023, Martin-Alvarez2023, Ponnada2023a}. By generating synthetic observations that mimic those performed by telescopes, MHD simulations are a promising avenue to establish direct and quantitative relations between the observables and the underlying physical properties of magnetic fields across various wavelength regimes. This work can guide the interpretation of current and future observational data. 

In this work we bring together MHD simulations of galaxy formation and polarimetric observations of galaxies by generating synthetic polarimetric observations at FIR and radio wavelengths. We measure the galactic scales associated with each type of emission, determine the interrelation of magnetic fields with the different phases in the ISM, determine the regions responsible for the observed polarized intensities, and quantify the galactic depths probed by each frequency. 

This study is structured as follows: the high-resolution MHD simulations are presented in Section~\ref{s:Methods}. The methods employed to generate our synthetic observations are described in Section~\ref{s:Mocks}. Section~\ref{ss:FIRMethod} describes our synthetic FIR observations and Section~\ref{ss:RadioMethod} focuses on the synthetic radio observations. We analyze our results in Section~\ref{s:Results} and summarize our work in Section~\ref{s:Conclusions}.

\section{Numerical Methods}
\label{s:Methods}
\subsection{The {\sc ramses} Code}
The numerical simulations explored in this work were presented in \citet{Martin-Alvarez2020} and \citet{Martin-Alvarez2021}, and further extended down to redshift $\redshift = 0.95$ in this work. The MHD simulations have been generated with our own modified version of the public code \ramses~\citep{Teyssier2002}. To solve the evolution of the baryonic and dark matter components, \ramses~couples an Eulerian treatment of the gas with the use of stellar and dark matter particles. All these components are coupled through the {\ramses} gravity solver. The MHD evolution of the gas fluid is solved on a discretized grid using an adaptive mesh refinement (AMR) octree. Crucial to this work, the MHD solver in \ramses~employs a constrained transport (CT) treatment of the magnetic field \citep{Fromang2006, Teyssier2006}. The CT method ensures that the solenoidal constraint is satisfied ($\vec{\nabla} \cdot \vec{B} = 0$) down to numerical precision \citep[e.g., for cosmological galaxy formation simulations,][]{Martin-Alvarez2018, Martin-Alvarez2020}. The fulfillment of this constraint is required to avoid spurious artifacts in the MHD evolution as well as for the preservation of conserved quantities, which is not guaranteed for alternative methods such as divergence cleaning \citep{Toth2000,Balsara2004}. Such unreliable behavior is apparent both for simple and complex configurations \citep[see, e.g.,][]{Hopkins2016}, particularly for Powell divergence cleaning techniques \citep{Powell1999}. The finite discretization of the simulated domain introduces resistivity in our MHD solver \citep{Teyssier2006}. Since we set the physical magnetic diffusivity to $\eta = 0$, any magnetic diffusive effects in our models are purely numerical.

\subsection{Initial Conditions}
\label{ss:ICs}
Our MHD high-resolution cosmological simulations study the so-called \nut~galaxy \citep[previously studied in e.g.,][]{Martin-Alvarez2018,Martin-Alvarez2020}. The \nut~galaxy is a Milky Way-like spiral system forming in the center of an approximately spherical zoom region, generated using the initial conditions (ICs) originally presented by \citet{Powell2011}. The formation and evolution of the simulated galaxy have been studied in detail through the years to understand, e.g., its satellite population \citep{Geen2013}, angular momentum evolution \citep{Tillson2015}, hierarchical growth and star formation \citep{Kimm2017}, and the influence of CRs on its evolution \citep{Rodriguez-Montero2023}.

\begin{figure*}[ht!]
    \begin{center}
    \includegraphics[width=\textwidth]{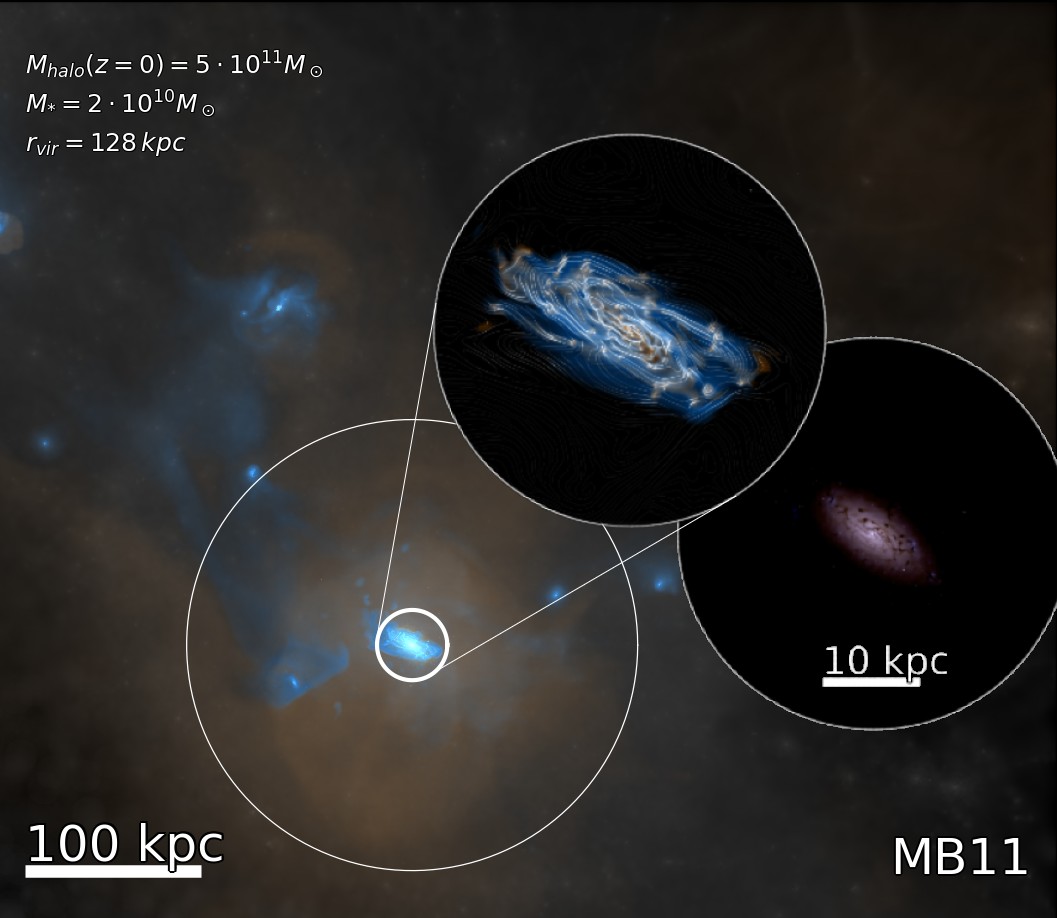}\\
    \end{center}
\caption{({\bf Large panel}) Colour composite image of the large-scale view of the environment around the simulated {\sc nut} galaxy. The galaxy is located in the center of the circles, which encompass the halo virial radius and the zoomed-in view in the inset panels. Colours correspond to dark matter density (grey), gas density (cyan), hot gas temperature (orange) and stellar density (gold). ({\bf Zoomed-in inset panels}) Zoom-in view of the galaxy, split into a combined view of the FIR (gold) and radio (blue) emission, and an SDSS-like color composite of the [u, g, r] filters. We include white streamlines indicating the orientation of the magnetic field within the projected plane.}
\label{fig:galaxy_front}
\end{figure*}

The \nut~galaxy is simulated in a cubic box with 12.5 comoving Mpc (cMpc) per side, with a spherical zoom region at its center that extends approximately $4.5$~cMpc across. We employ a mass resolution for the dark matter ($m_\text{DM}$) and stellar components ($m_{*}$) of $m_\text{DM} \simeq 5 \times 10^4~\Msun$ and $m_{*} \simeq 5 \times 10^3~\Msun$, respectively. Our simulations allow spatial refinement of the octree grid down to a minimum physical cell size of approximately $\sim10$~pc (equivalent to $\sim5-6$~pc radius for a particle-like treatment). At this resolution, \nut~showcases turbulent dynamo amplification \citep{Martin-Alvarez2018}, and displays turbulent properties similar (albeit not converged) to those found when using a uniform grid refinement of the galaxy \citep{Martin-Alvarez2022}. Within the resolved zoom region, the \nut~galaxy is the most massive system and is approximately located in the center of the region. The disk galaxy resides in a dark matter halo with virial mass $M_\text{vir} (\redshift = 0) \simeq 5 \times 10^{11}~\Msun$. All our simulations are generated according to the Wilkinson Microwave Anisotropy Probe (WMAP) 5 yr release cosmology \citep{Dunkley2009}. To provide some contextual information about the setup, in Figure~\ref{fig:galaxy_front}, we show a large-scale view of the studied galaxy in its cosmological environment with zoomed-in inset panels displaying the optical (Sloan Digital Sky Survey (SDSS)-like) and FIR+radio appearance of the system.

\subsection{Galaxy Formation Physics}
\label{ss:Subgrid}
To reproduce the formation of a realistic galaxy, we include various additional physical prescriptions, described in this section. 

We simulate metal-dependent gas cooling according to the thermodynamic properties of each gas cell. For temperatures above $10^4$~K, we interpolate precalculated {\sc cloudy} cooling tables \citep{Ferland1998}. In order to generate a realistic ISM, we also account for cooling below $10^4$~K following \citet{Rosen1995}. Explicit modeling of gas cooling at low temperatures is required for a self-consistent treatment of the cold gas distribution within the ISM, and will be fundamental to capture the small-scale structure that is responsible for the FIR emission properties. We include a standard ultraviolet (UV) background, activated at $\redshift = 10$ to replicate the impact of reionization \citep{Haardt1996}. 

We model star formation through a magneto-thermo-turbulent (MTT) star formation prescription\footnote{Originally implemented in \ramses~by {\it J. Devriendt}} presented in its thermo-turbulent form by \citet{Kimm2017, Trebitsch2017} and extended to its MHD form in \citet{Martin-Alvarez2020}. Star formation is allowed for cells at the highest level of resolution \citep{Rasera2006}, and only if the gravitational force overcomes the total support provided by the local turbulent, magnetic, and thermal pressures.

Cells allowed to form stars convert their gas content into stellar particles following a standard Schmidt law \citep{Schmidt1959}
\begin{equation}
    \dot{\rho}_\text{star} = \epsilon_\text{ff} \frac{\rho}{t_\text{ff}}\,,
    \label{eq:SchmidtLaw}
\end{equation}
where $\rho$ corresponds to the gas density, ${t_\text{ff}}$ to the freefall time, and $\epsilon_\text{ff}$ to the star formation efficiency. Instead of a fixed efficiency, $\epsilon_\text{ff}$ is a local quantity defined for each cell and depends on the MTT properties of the star forming cell and its close neighbors. The value of $\epsilon_\text{ff}$ follows the model by \citet{Padoan2011}, presented as {\it multi-freefall PN} in \citet{Federrath2012}. More details about our employed MTT star formation are provided in Appendix~B of \citet{Martin-Alvarez2020}.

All stellar particles in our simulation are allowed to generate supernova (SN) events following the mechanical stellar feedback prescription by \citet{Kimm2014}. Each SN injects mass, momentum and energy back to its hosting cell and its immediate neighbors. To determine the number of SN events for each stellar particle, we assume a Kroupa initial mass function \citep{Kroupa2001}, and for each explosion, we inject a total specific energy of $\varepsilon_\text{SN} = E_\text{SN} / M_\text{SN} \sim 10^{51} \erg / 10~\Msun$, with $E_\text{SN}$ and $M_\text{SN}$ the canonical SN explosion energy and progenitor mass. Each SN event returns a fraction of the progenitor mass $\eta_\text{SN} = 0.213$ to the ISM as baryonic gas, and a fraction of $\eta_\text{metals} = 0.075$ is assumed to be metal mass. 

\subsection{Magnetic Field Models and Simulation Suite}
\label{ss:Suite}
To quantify the influence of magnetization on the resulting observational properties, we analyze various simulations of the \nut~galaxy, each with a different model for magnetization. Some of our simulated galaxies are primarily magnetized by ab initio magnetic fields, seeded uniformly at the beginning of the simulations with comoving strength $B_0$. Our models \MBveinte, \MBdoce, \MBonce, and \MBdiez~explore various degrees of magnetization: $B_0 \sim 10^{-20}, 10^{-12}, 10^{-11}, 10^{-10}$ G, respectively. All models feature $B_0$ well below the current Planck upper limits on the primordial magnetic field strength, $B_{0}^{\rm{Planck}} < 10^{-9}$ G \citep{PlanckCollaboration2015}. The \MBveinte~and \MBdiez~models provide us with two theoretical edge cases: the former corresponds to negligible magnetization (i.e., magnetic fields in the fully kinematic regime), whereas the latter represents an extreme magnetization scenario. Importantly, \MBdiez, in particular, features extreme magnetizations that are disfavored by reionization constraints \citep{KMA2021} and galactic properties \citep{Martin-Alvarez2020}. Some of our findings in this work will further suggest a constraint of $B_0 < 10^{-10}$ G. We also investigate our \MBinj~model, where the magnetic fields in the galaxy are seeded through magnetized SN feedback. This is done by modifying the SN ejecta to also inject magnetic energy ($E_\text{inj} \sim 0.01\,E_\text{SN}$; \citealt{Beck2013a, Butsky2017, Vazza2017}). The injected magnetic fields are comparable to those found in supernova remnants (SNRs), with strengths of $\sim 10^{-5}$ G \citep[e.g.,][]{Parizot2006} when the injection is done at $\sim$10 pc scales. More details are provided in Appendix~A of \citet{Martin-Alvarez2021}. Overall, the \MBveinte, \MBdoce, \MBonce, and \MBdiez~models represent different degrees of magnetization in galaxies, whereas the \MBinj~simulation serves as an astrophysical, small-scale origin for the magnetic field in the galaxy \citep{Martin-Alvarez2021}. The simulations studied are summarized in Table~\ref{table:simulations}, where we also provide the resulting average magnetic field in the galaxy for each of the models. 

\begin{deluxetable*}{lccccl}
\centering
\tablecaption{Summary of the studied simulations.
\label{table:simulations} 
}
\tablecolumns{5}
\tablewidth{0pt}
\tablehead{\colhead{Simulation} & 	\colhead{$\dxmin$}  & \colhead{$B_0$} &  \colhead{$E_\text{inj}$} & \colhead{$\left<B_\text{gal}\right>$} & \colhead{Further Details} \\ 
 	&  \colhead{(pc)}	& \colhead{(G)}	& & \colhead{(G)}	
 \\
\colhead{(1)} & \colhead{(2)} & \colhead{(3)} & \colhead{(4)} & \colhead{(5)} & \colhead{(6)}}
\startdata
\MBdoce & 10 & $3 \times 10^{-12}$ & \xmark & $9 \times 10^{-6}$ & Regulated galactic magnetic field: intermediate magnetization\\
\MBonce & 10 & $3 \times 10^{-11}$ & \xmark & $3 \times 10^{-5}$ & Regulated galactic magnetic field: high magnetization\\
\MBinj  & 10 & $3 \times 10^{-20}$ & 0.01 $E_\text{SN}$  & $3 \times 10^{-5}$ & Small-scale regulated magnetic field: SNR seeding\\
\hline
\MBdiez & 10 & $3 \times 10^{-10}$ & \xmark & $2 \times 10^{-4}$ & Regulated galactic magnetic field: extreme magnetization\\
\MBveinte & 10 & $3 \times 10^{-20}$ & \xmark & $3 \times 10^{-11}$ & Negligible magnetization (does not affect evolution) 
\enddata
\tablenotetext{}{Notes. Column (1): model ID. Column (2): highest resolution full-cell physical size $\dxmin$. Column (3): comoving cosmic magnetic field strength $B_0$. Column (4): SN feedback magnetization. Column (5): mass-weighted magnetic field in a sphere containing the galaxy ($r < 8\,\kpc$). Column (6): additional details regarding the scenario explored by each model.}
\end{deluxetable*}

\subsection{Galaxy Centering and Tomographic Slicing}
\label{ss:HaloFinder}

Our {\it tomographic} analysis along the disk height coordinate relies on accurate centering of the galaxy position and the identification of the angular momentum direction. In order to identify the studied galaxy, we compute the position and characteristics of its dark matter halo using the {\sc HaloMaker} code \citep{Tweed2009}, applied exclusively to the dark matter component. We identify the system as its most massive progenitor at $\redshift = 10$, and determine its center by recursively applying a shrinking spheres algorithm \citep{Power2003} on the stellar component. From that redshift onward, we reapply this algorithm on each subsequent snapshot, initially repositioned to the updated center of mass from the galaxy's innermost 500 stellar particles in the previous output. The perpendicular direction to the plane of the disk is defined by the total angular momentum of the baryons of the galaxy, measured within a region with radius $r = 0.2\,r_\text{vir}$, with $r_\text{vir}$ the virial radius of the halo.

To investigate the scaling of intrinsic and observable quantities in our simulations, we generate projections where the LOS is aligned with the angular momentum of the galaxy (i.e., perpendicular to the disk plane). These projections employ the full AMR structure, are centered on the position of the galaxy and have a width of 30~kpc per physical side, with a resolution of 30~pc pixel-side (for guidance, each projection pixel has a width of approximately 3~cells at the finest level of refinement). To investigate the variation of the multiple physical parameters and observables studied as a function of the disk height, we vary the total thickness of the projections $\Delta \zcoord$, which span from $-0.5 \Delta \zcoord$ to $0.5 \Delta \zcoord$ along the perpendicular direction to the disk plane. We measure the quantities of interest in a wide ring with inner radius $r_\text{min} = 0.2\,\kpc$ and outer radius $r_\text{max} = 8\,\kpc$ to focus our investigation on the disk of the galaxy.

\section{Synthetic Observations and Telescope-like Observations: Very Large Array and SOFIA}
\label{s:Mocks}

In this section we describe our approach to generating synthetic observations. For each wavelength, we compute the Stokes parameters $I$, $Q$, and $U$, which are used to calculate the (linearly) polarized intensity, $PI$, and  polarization fraction, $p$, such as 
\begin{equation}
    PI = \sqrt{Q^2 + U^2},
    \label{eq:PI}
\end{equation}
\begin{equation}
    p = \frac{PI}{I}.
    \label{eq:p}
\end{equation}

Throughout this work, we display in various images the inferred plane-of-the-sky polarization through cyan quivers, rotated by $90^{\circ}$ to show the orientation of the plane-of-sky magnetic field. For our full-resolution observations, the quivers are sampled every 32~pixels, 2D gaussian smoothed and averaged inside a radius of 16~pixels to represent the inferred large-scale orientation of the magnetic field. While more polarimetric information is available, this depiction is selected for aesthetical reasons. In addition, we only display the polarization for pixels with a distance from the image center below 0.325 of the total image size (i.e., we only display quivers within a circle around the center of the image). We do not display quivers for pixels with a background value below the total average of the background map of each image.
We also show these quivers in telescope-like resolution observations; the magnetic field orientation is displayed in the Nyquist sampling (see Section~\ref{ss:Telescope}). To provide comparable information regarding the intrinsic magnetic field along the plane of the image, we also include thin white streamlines always generated with the full-resolution information of the magnetic field. These streamlines are calculated using the density-weighted magnetic field for the entire column displayed along the LOS.

\subsection{FIR Observations}
\label{ss:FIRMethod}
To obtain the total FIR and polarized emission, we employ a geometric approximation to estimate the dust emission \citep{LD1985,FP2000,PlanckXX2015,Chen2016,King2018,Lopez-Rodriguez2020}. The contributions to each of the Stokes parameters from a given cell are described by
\begin{equation}
    I_{i,\rm{FIR}} =  \ndust \left(1.0 - \pFIR \left(\frac{{B_x^2 + B_y^2}}{{B^2}} - \frac{2}{3}\right)\right) \dx,
    \label{eq:ImockFIRdust}
\end{equation}
\begin{equation}
    Q_{i,\rm{FIR}} = \pFIR  \ndust\ \left(\frac{{B_y^2 - B_x^2}}{{B^2}}\right) \dx,
    \label{eq:QmockFIRdust}
\end{equation}
\begin{equation}
   U_{i,\rm{FIR}} = \pFIR  \ndust\  \left( \frac{{2B_x B_y}}{{B^2}} \right) \dx.
    \label{eq:UmockFIRdust}
\end{equation}

In these equations, $\dx$ corresponds to the size of the cell, $\ndust$ is the dust number density, $\pFIR$ is the maximum polarization fraction, and $B$ is the magnetic field strength of the cell. We decompose the magnetic field into its components perpendicular to the projection LOS $B_x$ and $B_y$, with the total magnetic field given by $B^{2} = B_x^2 + B_{y}^2+B_{z}^2$. We set $\pFIR = 0.25$, according to the maximum polarization fraction observed by \citet{PlanckXII2020}. The total Stokes parameters $I$, $Q$, and $U$ are then computed as the integral of all the cells along the LOS. By employing these equations, we assume that dust properties and alignment remain unchanged at galactic scales, and that the polarized intensity is mainly depolarized, due to geometric depolarization along the LOS and within the studied beam size. We will study the effects of more detailed dust modeling in future work, through simulations accounting for on-the-fly dust formation and evolution \citep{Dubois2024}. For each cell, we estimate the dust number density following $\ndust = \left(\rho_\text{gas}\, Z\, \eta_\text{D/M}\right) m_\text{dust}^{-1} \, f_\text{cut} (T)$, with $\rho_\text{gas}$ the gas density, $Z$ the metallicity, $\eta_\text{D/M}=0.4$ a constant dust-to-metal ratio \citep{Dwek1998, Draine2007b}, a typical dust grain mass $m_\text{dust} = 1.26\times10^{-14}\,\g$  (with radius $0.1\,\mu m$ and density $3\,\g / \cm^{-3}$; \citealt{Zubko2004}), and $T$ the gas temperature. We set $f_\text{cut} (T) = \text{min}\left(1.0, \, \exp\left[1 - T / (1500\, \K)\right] \right)$, selected to avoid a sharp cut at $T = 1500\, \K$. To illustrate the decline in dust with temperature, $f_\text{cut} (T = 5 \times 10^3 \K) \sim 0.1$ and $f_\text{cut} (T = 8 \times 10^3 \K) \sim 0.01$, we repeat our analysis with two additional dust models in Appendix~\ref{ap:dust_models}: a no temperature cut model (i.e., $f_\text{cut} = 1$), and an ionization-based model \citep{Laursen2009}. Both yield virtually unchanged results for the quantities explored here, illustrating their independence from the treatment for the proportion of dust in the low gas density regime. Consequently, our employed geometric FIR modeling is dominated by the spatial distribution of the gas density and its metallicity rather than the gas temperature. 

Integrating Equations (\ref{eq:ImockFIRdust}), (\ref{eq:QmockFIRdust}), and (\ref{eq:UmockFIRdust}) leads to geometric estimates of the emitting FIR column, which will account for depolarization effects, due to incoherent magnetic field orientation along the LOS and within the beam size. Note that the final units of the Stokes parameters are per square centimeter, which can be converted to surface brightness after assuming a certain wavelength dependence of the dust temperature per LOS. Since this conversion is only a scaling factor, we decided to work in units of surface column density.

\subsection{Radio Observations}
\label{ss:RadioMethod}
We generate our synthetic synchrotron observations for radio frequencies using the {\sc polaris} code \citep{Reissl2019}. {\sc polaris} is a radiative transfer OpenMP parallelized code \citep{Reissl2016} that solves Stokes vector propagation along a given LOS through a Runge-Kutta solver for ray tracing. This accounts for absorption effects as well as Faraday rotation and depolarization, which may affect the polarized intensity in the outskirts of dense regions. To prepare our simulations for {\sc polaris}, we preprocess them with our {\sc ramses2polaris} code. We generate a {\sc polaris}-compatible octree grid, adaptively resolved with double the local AMR resolution of the simulation to account for displacements between the two grids and avoid aliasing. All native {\sc ramses} quantities required by {\sc polaris} are interpolated into this grid, and any disk thickness tomographic cuts are applied at this stage. While the limited resolution of cosmological simulations does not capture the small-scale structure of gas turbulence \citep{Kortgen2017} or the magnetic field, we provide {\sc polaris} with the magnetic field in the simulations without any modification, and do not apply any additional modeling to capture the sub-grid magnetic energy or magnetic field lines' substructure (e.g., \citealt{Reissl2019}).

To generate synthetic synchrotron observations, {\sc polaris} requires additional quantities not modeled by our simulations. These are all related to the distribution of electrons in the ISM. We compute the number density of the thermal electrons by interpolating the gas number density according to \citet{Pellegrini2020}. Particularly important for the synchrotron emission are the properties of CR electrons. Some simulations explicitly model CR energy density during their evolution \citep{Hanasz2013, Dubois2016, Pfrommer2017b, Hopkins2022}. This information can be used in the generation of synthetic radio observations \citep[e.g.,][]{Werhahn2021c,Werhahn2021a,Pfrommer2022}
or even employ on-the-fly modeling of the CR electrons and their energy spectrum \citep{Ponnada2023a}. As the simulations studied here do not explicitly model a CR component, we employ instead a simple post-processing static model based on that by \citet{Reissl2019}, which assumes a smooth distribution of CRs across the galaxy. This model and its caveats are further discussed in Appendix~\ref{ap:CR_models}, e.g. how this profile does not capture local effects and variations of MHD properties. We assume their energy spectrum to decrease with their increasing energy, following a power law with a fixed index $\pCR$. We set $\pCR = 2.3$, motivated by observations of late-type galaxies such as the system simulated for this work \citep{Lacki2013}. As done by \citet{Reissl2019}, we consider an energy range for the spectrum bounded by Lorentz factor $\gamma \in [4, 300]$, with minimum Lorentz factor $\gamma_\text{min}$ according to \citet{Webber1998}. The number density of electronic CRs $n_\text{CR}$ is set following the CR1 model by \citet{Reissl2019}, which assumes a Milky Way-like disk galaxy and sets:
\begin{equation}
    n_\text{CR}(r, z) = n_\text{CR,0} \exp\left(-\frac{r}{\Rgal}\right) \left[\cosh\left(\frac{z}{\hgal}\right)\right]^{-2},
    \label{eq:nCRdensity}
\end{equation}
where $n_\text{CR,0} = 1.74\, \times\, 10^{-4}~\text{e}^{-}\, \cc$ corresponds to the number density of CRs in the center of the Milky Way. This model is designed to match the observed number density of CR $\text{e}^{-}$ in the solar neighborhood \citep{Sun2008}. We select $\Rgal = 8$ kpc and $\hgal = 1$ kpc to adjust to the studied galaxy dimensions \citep[estimated through an approach similar to that by][]{Martin-Alvarez2020}. In Appendix~\ref{ap:CR_models}, we study how variations of the employed $\text{e}^-$ CR model influence our results, illustrating the resilience of this work in reaching the main conclusions of such modeling.

We generate synthetic observations using our modified version of {\sc polaris}\footnote{The modifications correspond to some minor revisions of the synchrotron module, made available on 2023 July 31 in the {\sc polaris} public repository \href{https://github.com/polaris-MCRT/POLARIS}{https://github.com/polaris-MCRT/POLARIS}. The public website for {\sc polaris} is \href{https://portia.astrophysik.uni-kiel.de/polaris/}{https://portia.astrophysik.uni-kiel.de/polaris/}}. While {\sc polaris} calculations employ the full expression for the contributed intensities per cell and their subsequent propagation, we note that the variations of the synchrotron emission between different regions are mostly governed by their different $n_\text{CR}$ and $B$ following
\begin{equation}
    I_{i,\rm{radio}} \propto n_\text{CR}\, B_\perp^{\frac{\pCR + 1}{2}} \dx \, f (\pCR,\, \gamma_\text{min},\, \gamma_\text{max}),
    \label{eq:Iradio}
\end{equation}
where $B_\perp$ is the local magnetic field strength perpendicular to the LOS, and $f (\pCR,\, \gamma_\text{min},\, \gamma_\text{max}, \lambda)$ encompasses all the factors in $I_{i,\rm{radio}}$ that only depend on these four quantities, assumed constant by our calculations. The total Stokes parameters $I$, $Q$, $U$ are the integral of all the cells along the LOS, accounting for the propagation of the Stokes vector through the medium of the galaxy and its circumgalactic medium. Details regarding the radiative transfer solver for synchrotron emission used by {\sc polaris}, the exact contribution from each cell to the Stokes parameters, and the complete computation process are provided by \citet{Reissl2016, Reissl2019}. The $\lambda = 6.2$ cm wavelength ($\sim 4.8$ GHz) strikes an optimal balance of a frequently employed wavelength and the highest signal-to-noise ratio for the observed galaxies at both radio and FIR wavelengths \citep{SALSAIV,SALSAV,Surgent2023}. Consequently, we limit our investigation to synchrotron emission at this wavelength (i.e., $6.2$ cm). 

\subsection{Telescope-like Observations}
\label{ss:Telescope}
To study telescope-like observations, we imitate the configurations of the SALSA survey \citep{SALSAIV} for the FIR, and the combined Very Large Array (VLA) and Effelsberg telescopes for the radio synchrotron observations \citep{BeckLive, Beck2019a}. A benefit of this combination of telescopes is their comparable angular resolutions ($\sim13\arcsec$). Our telescope-like observations account for three effects: (i) the distance to the observed system, (ii) the point-spread function (PSF) of the telescope, and (iii) the detector sampling pixelation. Notably, there are additional observational effects we do not include to facilitate interpretation, such as noise effects, calibration artifacts, or instrumental and detector aberrations. 

We position all our synthetic observations at a distance of 10 Mpc from the observer, selected to match the median distance to the sample observed by the SALSA survey \citep{SALSAIV}. We account for distance dimming in our synthetic radio maps, but maintain intrinsic column densities in the FIR. This distance is then taken into consideration when applying a convolution with the corresponding telescope PSF. We assume VLA $C$-band in its $D$ configuration (full width at half maximum (FWHM) of $12.6\arcsec$) and SOFIA in the $D$ band ($154~\mu$m, with FWHM of $13.6\arcsec$). Finally, we apply a Nyquist pixelation, resolving each PSF with $2\times2$ pixels.

\section{Results} \label{s:Results}
To facilitate the connection between observations and simulations, we start our analysis by reviewing the global appearance of the studied galaxy by comparing the intrinsic physical properties of the system with the synthetic observations. 

\subsection{Global Comparison of Synthetic FIR and Radio Observations}
\label{ss:GlobalReview}

\begin{figure*}[ht!]
\includegraphics[width=\textwidth]{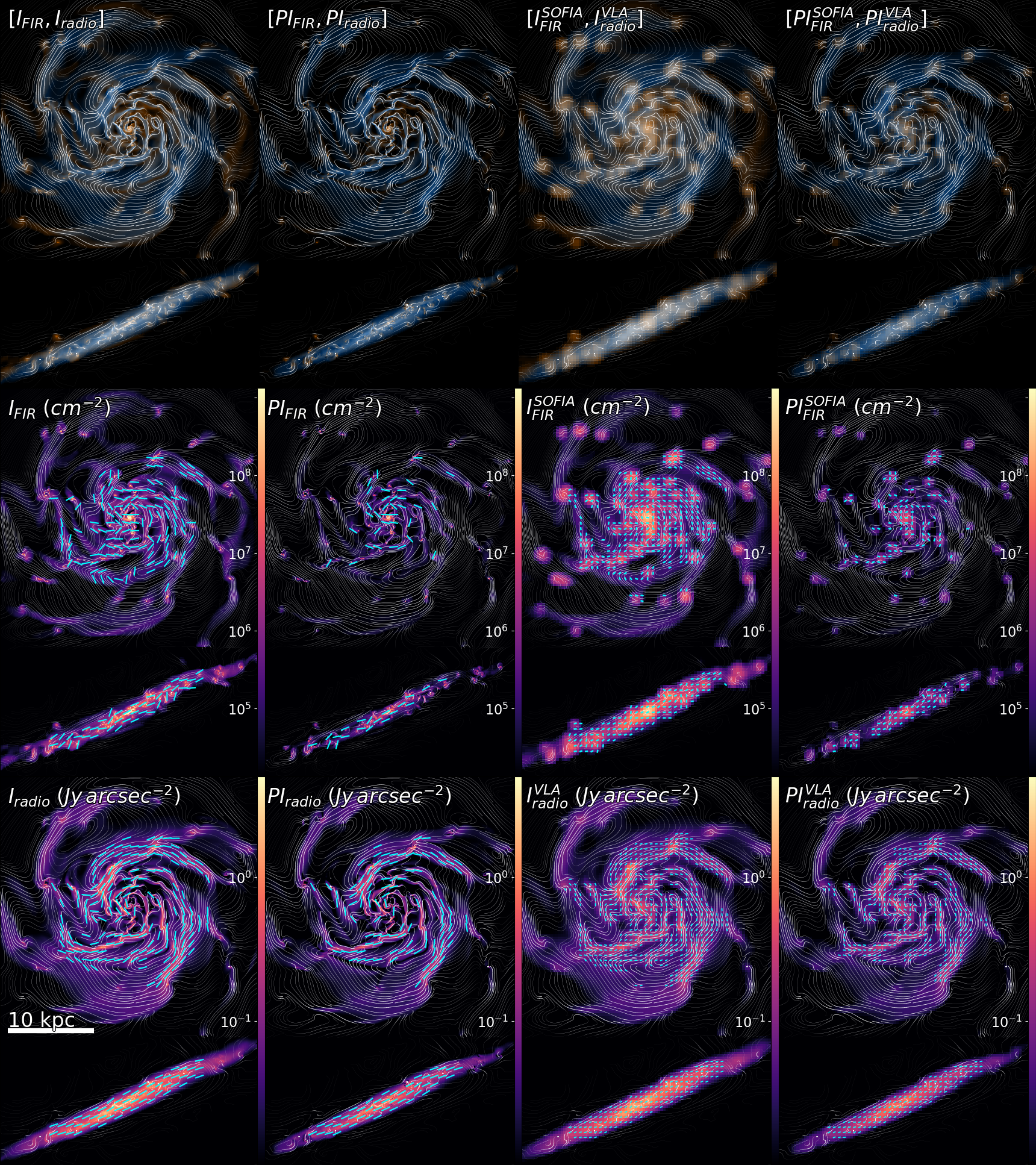}\\
\caption{FIR and radio emission comparison for the \MBonce~model, with each set of panels displaying a face-on and edge-on view of the disk galaxy. From left to right, the columns display full-resolution total intensity, full-resolution polarized intensity, telescope-like total intensity, and telescope-like polarized intensity. See Section~\ref{s:Mocks} for the details of
telescope-like synthetic observations. All panels have magnetic field lines overplotted as thin white streamlines. From top to bottom, each set of rows displays the following. ({\bf Top row}) Color composite view of the radio (blue) and FIR (gold) emission. ({\bf Middle row}) FIR geometric observation, with polarization measurements overlaid as cyan quivers rotated by $90^{\circ}$~to match the inferred orientation of the magnetic field. ({\bf Bottom row}) Same as the middle row, but now for the synthetic radio observations. FIR emission concentrates on dense clumps, whereas the radio emission is extended and diffuse. The radio magnetic field inferred from the polarized emission captures the large-scale magnetic field, and the FIR magnetic field captures the more turbulent and small-scale field.}
\label{fig:CombinedFIRradio}
\end{figure*}

Figure~\ref{fig:CombinedFIRradio} illustrates the main results of this work from a qualitative point of view. This figure compares our synthetic observations in radio and FIR for the \MBonce~model. Each panel combines a face-on and edge-on view of the galaxy, overlaid with streamlines representing the average magnetic field direction as thin white lines. These streamlines are slightly thickened in regions with stronger magnetic fields and removed from regions where their strength is negligible. Whenever present, cyan quivers display the orientation of the corresponding linear polarization, rotated by $90^{\circ}$ to align with the observationally inferred orientation of the local magnetic field in the plane of the sky. From left to right, in Figure~\ref{fig:CombinedFIRradio}, each of the displayed columns corresponds to full-resolution total intensity, full-resolution polarized intensity, telescope-like total intensity and telescope-like polarized intensity. Finally, from top to bottom, each of the rows displays the (top) combined color-composite view of the radio (blue) and FIR emission (gold), (middle) FIR emission, and (bottom) radio emission. 

The figure reveals a clustered distribution of the FIR intensity, typically concentrated in dense clumps of gas found within the spiral arms of the galaxy. These clumps have irregular shapes and extend for a few hundred parsecs in the simulated galaxy. However, the sizes of these clumps are below the resolution of the observations, so the clumps are vastly smoothed in the telescope-like observations to sizes comparable with the PSF and approximately circular shapes (i.e., unresolved sources). A similar result is found for the edge-on views, where the full-resolution images reveal a thin disk constituted by irregular cold gas clumps, mostly at low altitudes above and below the galactic plane. These clumps blend into a thicker appearance in the telescope-like observation, concealing the intrinsic width of the galactic disk. 

The total synchrotron emission has an extended and smooth distribution, following the spiral structure of the galaxy. The magnetic field strength approximately traces the thickness of the spiral arms, reflected by the magnetic field streamlines. Polarized synchrotron emission seems to show slightly thinner arms than the total synchrotron emission. This radio emission is not significantly modified by our telescope-like observation, suggesting that the scales for its coherence are comparable or larger than the size of the telescope PSF of approximately $300$ pc. Notably, our employed analytic smooth distribution for the CR electrons does not capture local deviations such as those found by more complex models, e.g., \citet{Werhahn2021a}. These could be significant for the full-resolution images and may propagate to their telescope-like counterparts. The more extended nature of the synchrotron emission, both in the face-on and edge-on orientations, is more apparent when directly comparing the two studied wavelength ranges, particularly reflected in the color composite panels.

The contrast between the large-scale and extended nature of the radio emission, and the small-scale and clumpy distribution of the FIR is clearly seen in the magnetic field inferred by their corresponding polarized intensities. The synchrotron quivers follow the orientation of the magnetic field across the plane of the sky almost perfectly, indicating that the radio synchrotron emission is a good tracer of the large-scale galactic magnetic field. This result suggests that the synchrotron emission is not significantly affected by structural anisotropies in the magnetic field at smaller scales. 

Conversely, the FIR magnetic field displays local deviations from the orientation of the intrinsic magnetic field at the scales depicted by the streamlines, which typically capture magnetic field structure at scales of approximately $\sim50 - 100\,\pc$. In those cases, the FIR magnetic field is frequently oriented perpendicular to the local dust density gradient. These deviations from the magnetic field are particularly clear in the telescope-like observation, where a non-negligible amount of beam depolarization is present. Examples are the clumps in the north-northwest direction above the galactic center, or in the eastern region of the image. Such beam depolarization results from the blending of diverse and unaligned linear polarizations contained within the beam of the observations. 

The spatial correlation between dust density and FIR magnetic field deviations from the intrinsic magnetic field depicted by the streamlines is a notable result when considering the limitations on grid size for present-day high-resolution galaxy formation simulations. An even higher amount of depolarization is likely to be recovered for enhanced spatial discretization that better captures the substructure of the gas clouds dominating the FIR intensity. We will show below that these high-density regions deviating from the local inferred magnetic field often correspond, as expected, to star forming regions. This association confirms the large angular dispersion of the FIR magnetic fields located in dense regions and the steep depolarization rate with increasing column density measured in nearby spiral galaxies  \citep{SALSAIV, SALSAV, Surgent2023}.

\begin{figure*}[ht!]
    \begin{center}
    \includegraphics[width=0.2\textwidth]{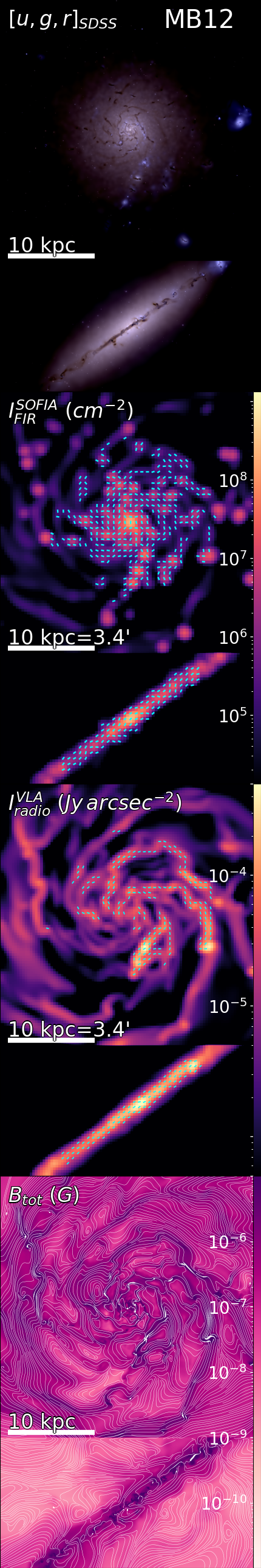}%
    \includegraphics[width=0.2\textwidth]{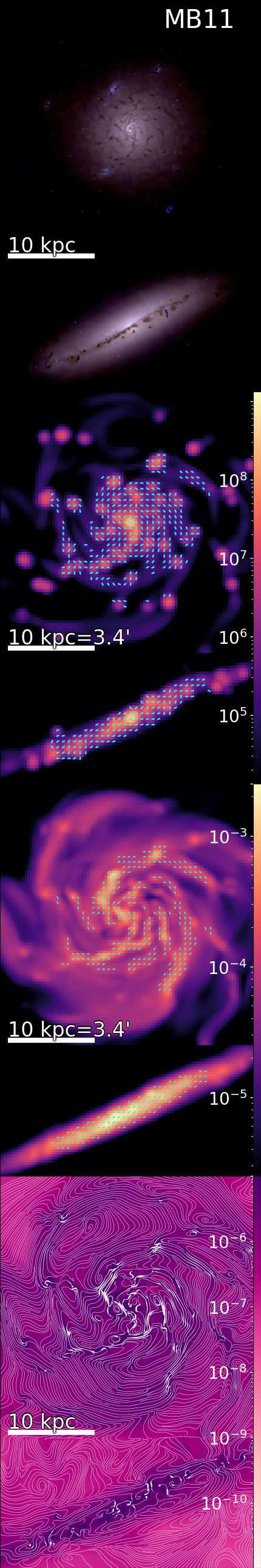}%
    \includegraphics[width=0.2\textwidth]{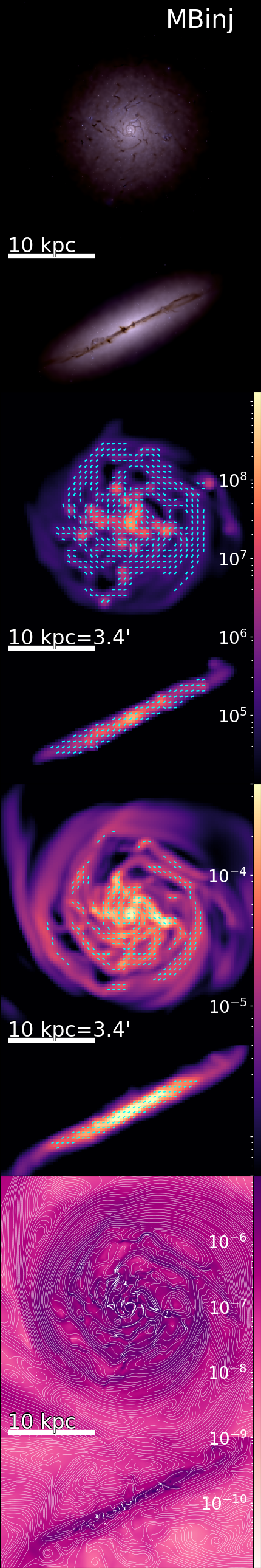}%
    \includegraphics[width=0.2\textwidth]{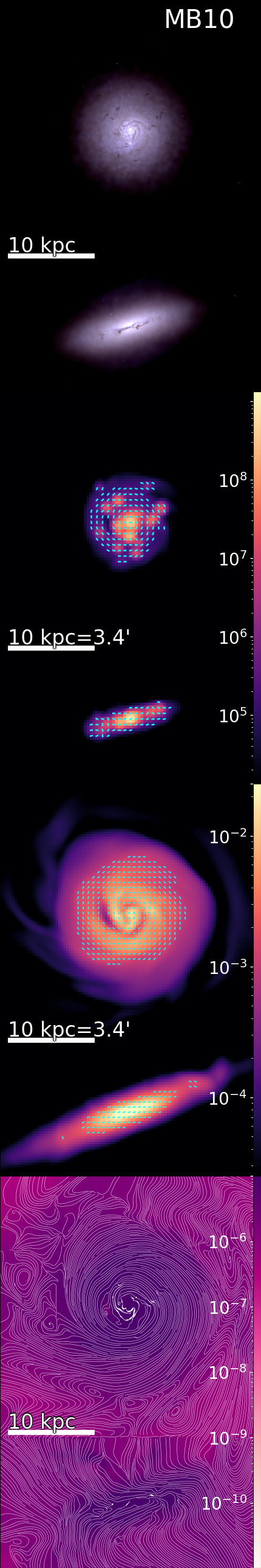}%
    \includegraphics[width=0.2\textwidth]{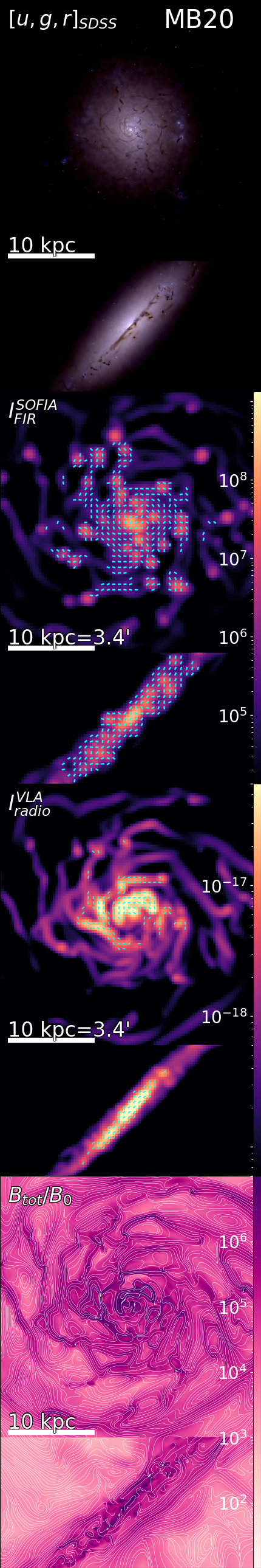}\\
    \end{center}
\caption{Comparison of observable intensities vs. projected intrinsic galaxy properties (continued in Figure~\ref{fig:Galaxies-p2}). From top to bottom, the rows display SDSS-like color composite of the [$u$, $g$, $r$] filters,  SOFIA-like FIR total intensity observation, VLA-like radio total intensity observation for synchrotron emission at $6.2$ cm, and magnetic field strength. On the FIR and radio emission panels, we overlay cyan quivers for the corresponding observed polarization rotated by $90^{\circ}$. The magnetic field strength includes white streamlines indicating the orientation of magnetic field lines within the projected plane. From left to right, each column displays the \MBdoce, \MBonce, \MBinj, \MBdiez, and \MBveinte~models. FIR emission shows some correlation with regions of ongoing star formation (displayed in Figure~\ref{fig:Galaxies-p2}), whereas the radio emission has a better resemblance to those of a stronger magnetic field.}
\label{fig:Galaxies-p1}
\end{figure*}

\begin{figure*}[ht!]
    \begin{center}
    \includegraphics[width=0.2\textwidth]{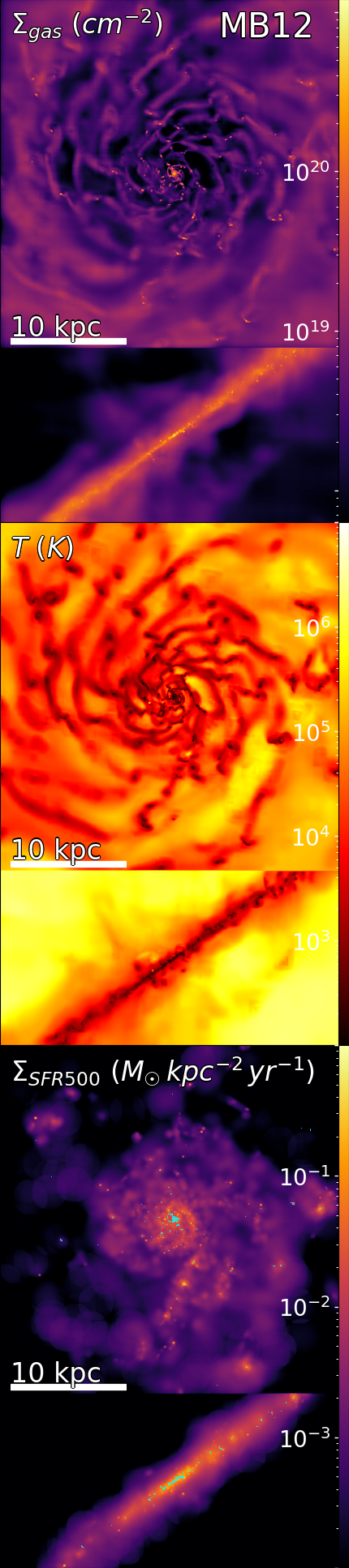}%
    \includegraphics[width=0.2\textwidth]{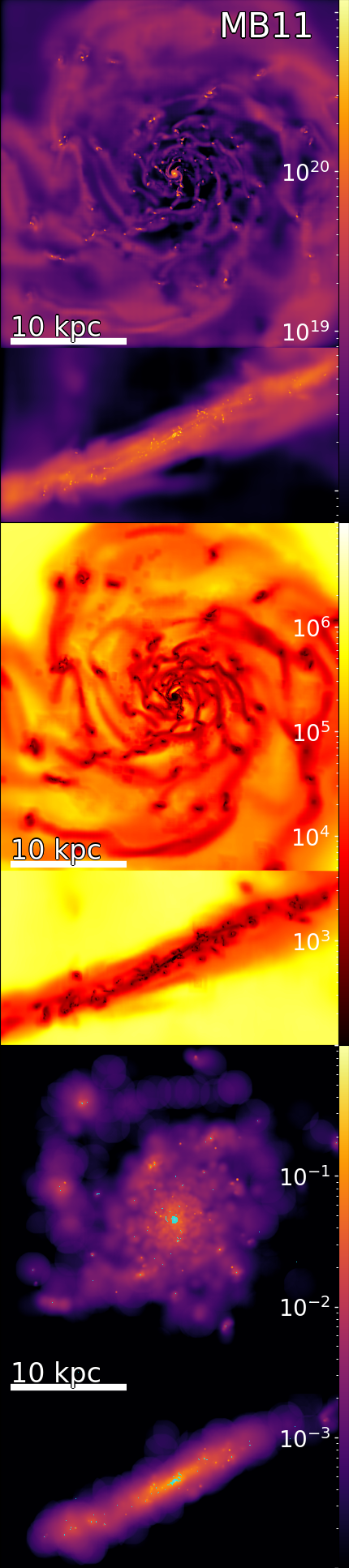}%
    \includegraphics[width=0.2\textwidth]{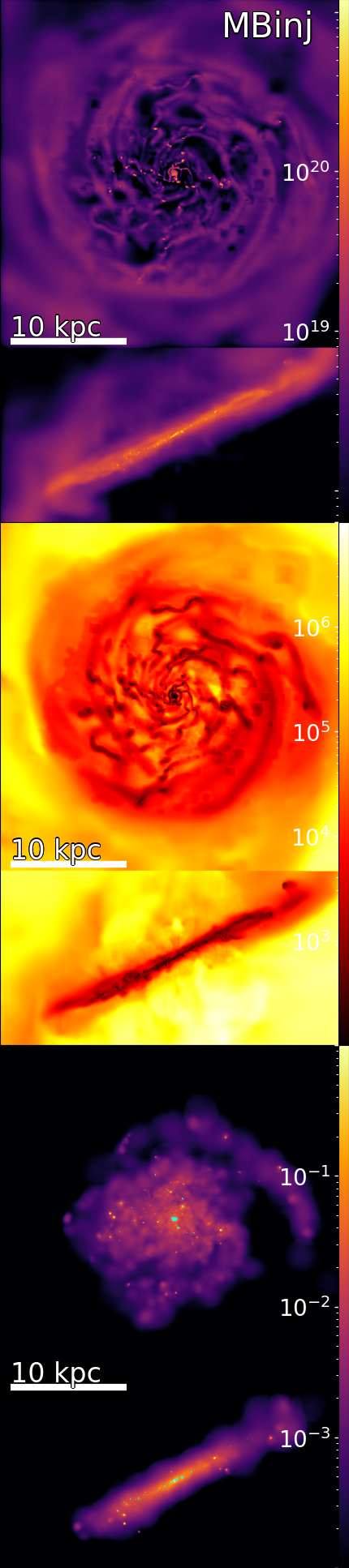}%
    \includegraphics[width=0.2\textwidth]{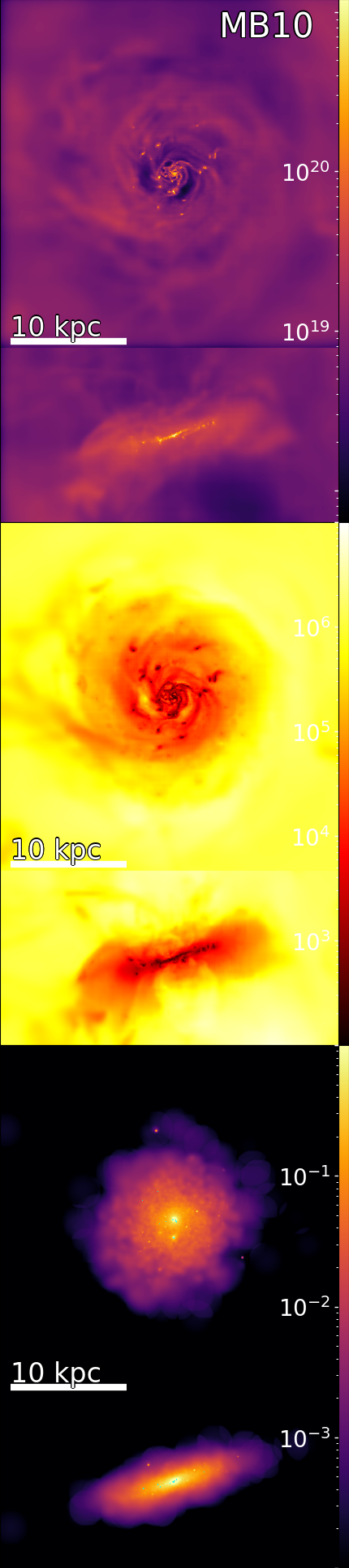}%
    \includegraphics[width=0.2\textwidth]{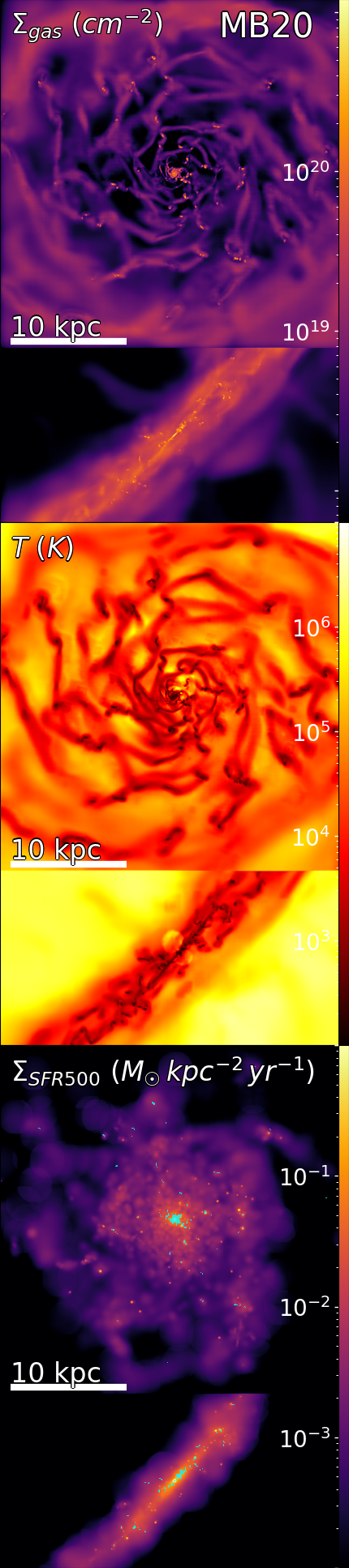}\\
    \end{center}
\caption{Figure~\ref{fig:Galaxies-p1} continued, now displaying projected intrinsic galaxy properties. From top to bottom, the rows show total gas surface density, density-weighted gas temperature, and SFR averaged over the last $500$ Myr with regions of ongoing star formation overplotted in cyan. From left to right, each column displays the \MBdoce, \MBonce, \MBinj, \MBdiez, and \MBveinte~models. FIR emission (displayed in Figure~\ref{fig:Galaxies-p1}) shows some correlation with regions of ongoing star formation.}
\label{fig:Galaxies-p2}
\end{figure*}

We present all the simulated galaxies investigated in this study in Figures~\ref{fig:Galaxies-p1} and \ref{fig:Galaxies-p2}. We compare their synthetic observations with various physical quantities of interest. From left to right, each column displays a different model: \MBdoce, \MBonce, \MBinj, \MBdiez, \MBveinte. From top to bottom, each of the rows in Figure~\ref{fig:Galaxies-p1} displays the observational quantities: SDSS filters' synthetic observation accounting for dust absorption\footnote{We model each stellar particle as a single stellar population, with spectral emission following \citet{Bruzual2003}. Dust absorption is modeled as an absorption screen.}, SOFIA-like FIR total intensity with polarization quivers, and the VLA-like radio total intensity with polarization quivers. In the last row, we include the density-weighted magnetic field strength with streamlines displaying the local orientation of the field in the projection plane. Figure~\ref{fig:Galaxies-p2} focuses on additional intrinsic MHD quantities. From top to bottom, each row displays gas surface number density, density-weighted gas temperature, and average star formation rate (SFR) surface density over the last $500$ Myr overplotted (in cyan) with regions of ongoing star formation according to our MTT model.

Some of the galaxies display similar structures, with \MBdiez~showing the most evident deviations as a result of its extreme magnetization shrinking the galaxy \citep{Martin-Alvarez2020}. Stronger magnetic fields appear to lead to tighter spiral structures, particularly for the arms observed in radio. The surface gas densities resemble the rippling pattern recently observed by JWST/MIRI in systems like the Whirpool galaxy (M51) by the program Feedback in Emerging Extrgalactic Star Clusers (ID: 1783; PI: Angela Adamo) and NGC\,628 \citep{Thilker2023} by Physics at High Angular Resolution in Nearby Galaxies-JWST (ID: 2107, PI: Janice Lee). Despite the similar appearance of the synthetic optical observations between our models and the density structures with the mid-infrared observations, different magnetizations lead to important variations in the synthetic FIR and radio observations. This illustrates the importance of these other wavelengths and observing techniques as independent probes for the ISM of galaxies. For example, the correlation of the FIR intensity with the SFR can be used as a proxy of the contribution of the small-scale magnetic field and the effect of the star formation activity. This work will provide an explanation of the proposed scenario by \citet{SALSAV} in which small-scale magnetic field and/or gas velocity dispersions, likely driven stellar feedback in regions of ongoing or recent star formation, are responsible for the depolarization of the FIR polarized intensity. This effect will be quantified in a follow-up work, as here we focus on the origin of measured magnetic fields at FIR and radio wavelengths. Furthermore, the FIR magnetic field structure of \MBdoce, \MBonce, and \MBinj ~have a striking resemblance to those of NGC~6946, M51, and M83, respectively \citep{SALSAIV,SALSAV}. Specifically, these three models as shown in Figure~\ref{fig:Galaxies-p1} have the following spatial correspondences with the FIR polarimetric observations: 
(a) NGC~6946 shows a very disordered and patchy magnetic field across the disk of the galaxy, with the FIR magnetic field mostly cospatial with the star forming region in the arms,
(b) M51 shows a well-ordered magnetic field in the central $\sim5$ kpc radius of the galaxy and more disordered in the outskirts, due to star formation activity and the total intensity is smooth and extended across the entire disk, and 
(c) M83 shows a large-scale ordered spiral magnetic field cospatial with the spiral arms with large angular dispersion of the field in the star-forming regions \citep[Figure~1 in][]{SALSAIV}.

With the exception of our two edge cases, \MBdiez~and \MBveinte, radio intensities across galaxies are comparatively similar, with the deviations in intensity correlated with the different magnetization of each system (see Equation~\ref{eq:Iradio}). Radio emission appears to somewhat trace a smoothed approximation of the gas surface density, filtered by regions of strong magnetic field. This correlation is somewhat less prominent for \MBveinte. The deviation of \MBveinte~and especially \MBdiez~from the usual appearance of galaxies in FIR (exclusively for \MBdiez) and radio observations indicate that magnetic fields are not only responsible for the observed emission, but also impact the underlying properties and distribution of the ISM, illustrated in Figure~\ref{fig:Galaxies-p2}. Similarly, we note that the total intensity of the radio of \MBdoce~shows more filamentary structures than the \MBonce~or \MBinj, which are smoother and better resemble radio observations \citep{Beck2015,Beck2019a}. These two models also show {\it magnetic arm-like} structures observed in spiral arms like NGC~6946 \citep{Beck2007}. Specifically, as shown in Figure~\ref{fig:Galaxies-p1} for \MBonce~and \MBinj~(and further reviewed below in Figures~\ref{fig:MB11_slice_telescope} and \ref{fig:MBinj_slice_telescope}), the polarized intensity has depolarized spiral arms that may be produced by the post-shock in the trailing part of the arm, as observed in NGC~6946 by \citet{Beck2007}. Thus, both \MBonce~and \MBinj~synthetic observations show features compatible with observations in both FIR and radio wavelengths. We will focus on these two models to discuss the main results of this manuscript.

\subsection{Integrated Slicing of the Polarized Emission of Galaxies}

\begin{figure*}[ht!]
\includegraphics[width=0.66\textwidth]{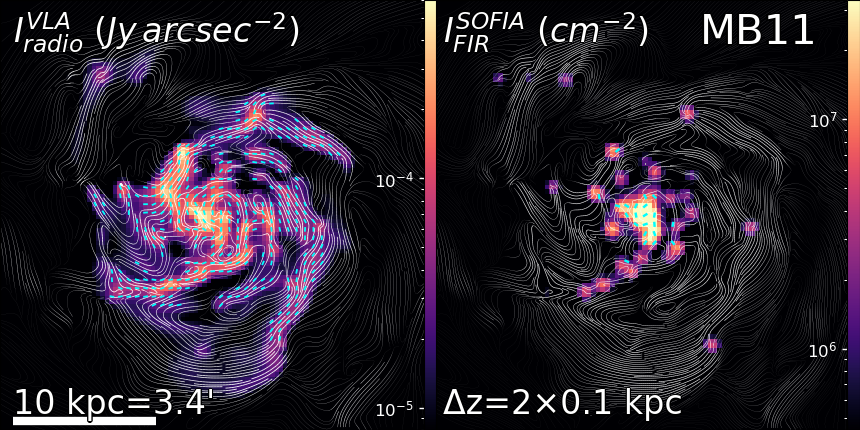}%
\includegraphics[width=0.33\textwidth]{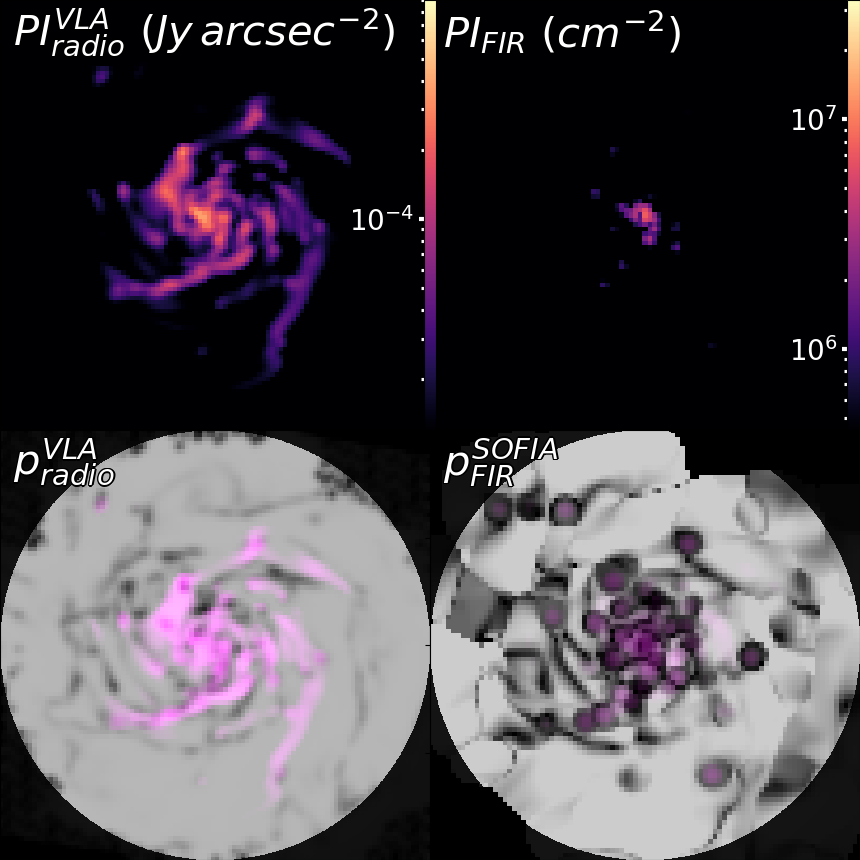}\\
\includegraphics[width=0.66\textwidth]{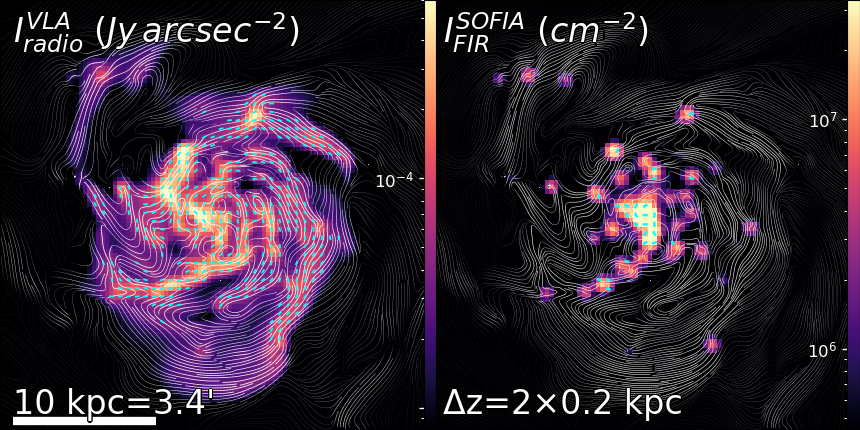}%
\includegraphics[width=0.33\textwidth]{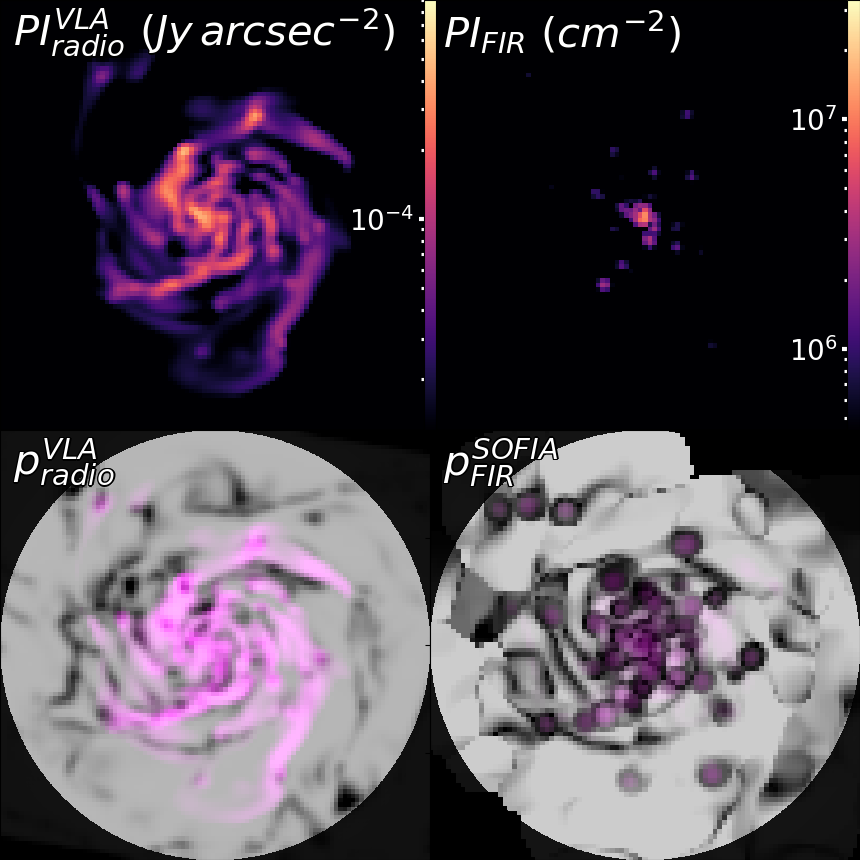}\\
\includegraphics[width=0.66\textwidth]{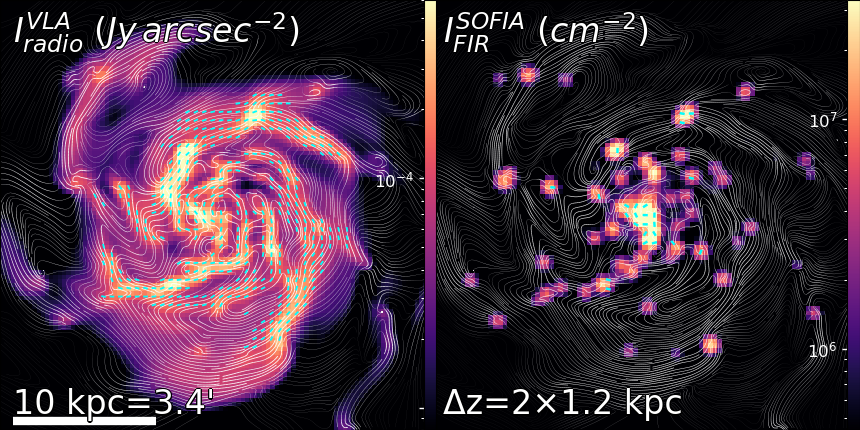}%
\includegraphics[width=0.33\textwidth]{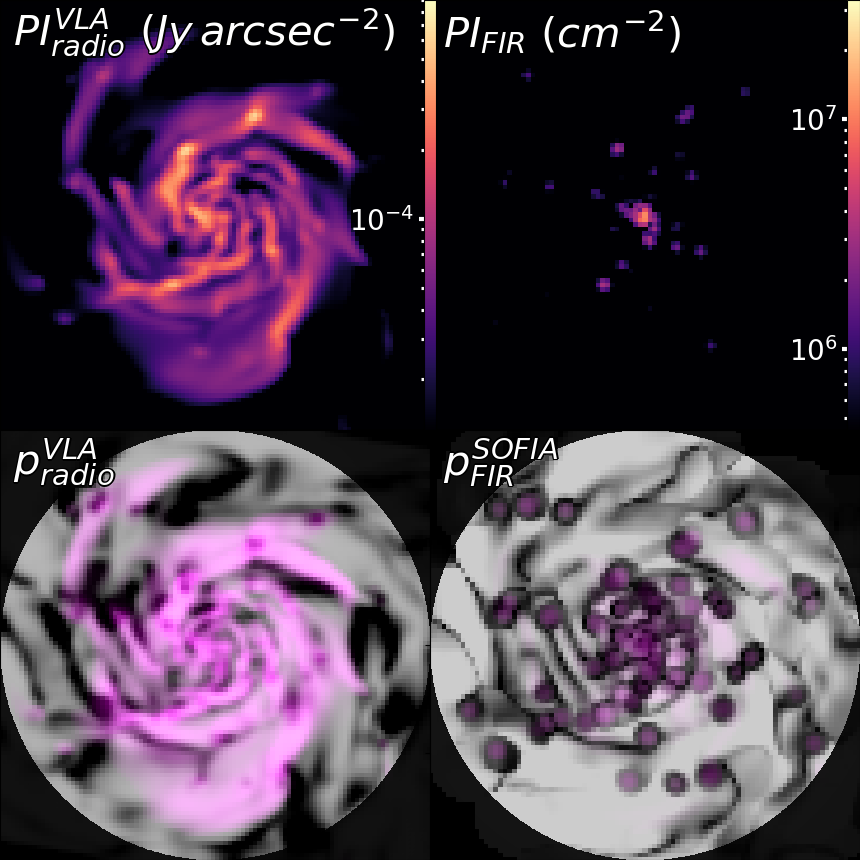}\\
\caption{Face-on projections centered on the galaxy for the vertical slicing of the observed telescope-like intensities for the \MBonce~model. From top to bottom, each set of panels displays disk half-thickness of 0.1, 0.2 and 1.2 kpc. The leftmost panel displays a VLA-like 6.2 cm synchrotron total intensity observation, whereas the second shows a SOFIA-like FIR total emission observation. The subset of smaller panels displays from top-left to bottom-right: VLA-like polarized intensity observation, SOFIA-like polarized intensity observation, VLA-like polarized fraction, and SOFIA-like polarized fraction. For the panels showing the polarized fractions, we employ a fixed color map that ranges from 0.3 to 0.6 (radio) and 0.05 to 0.1 (FIR), over which we overlay the corresponding total intensities in purple to guide the visual analysis of this relative quantity. Overall, synchrotron intensities at 6.2cm grow as the thickness of the projections is increased, whereas the FIR counterpart are approximately fixed within the inner 0.1 kpc. The polarization of the FIR is similarly fixed within this inner thickness, whereas the radio is somewhat depolarized as the disk thickness is increased.}
\label{fig:MB11_slice_telescope}
\end{figure*}

\begin{figure*}[ht!]
\includegraphics[width=0.66\textwidth]{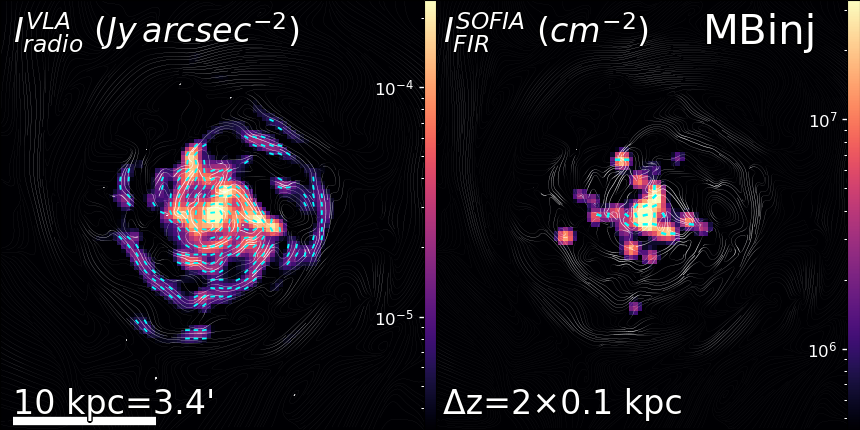}%
\includegraphics[width=0.33\textwidth]{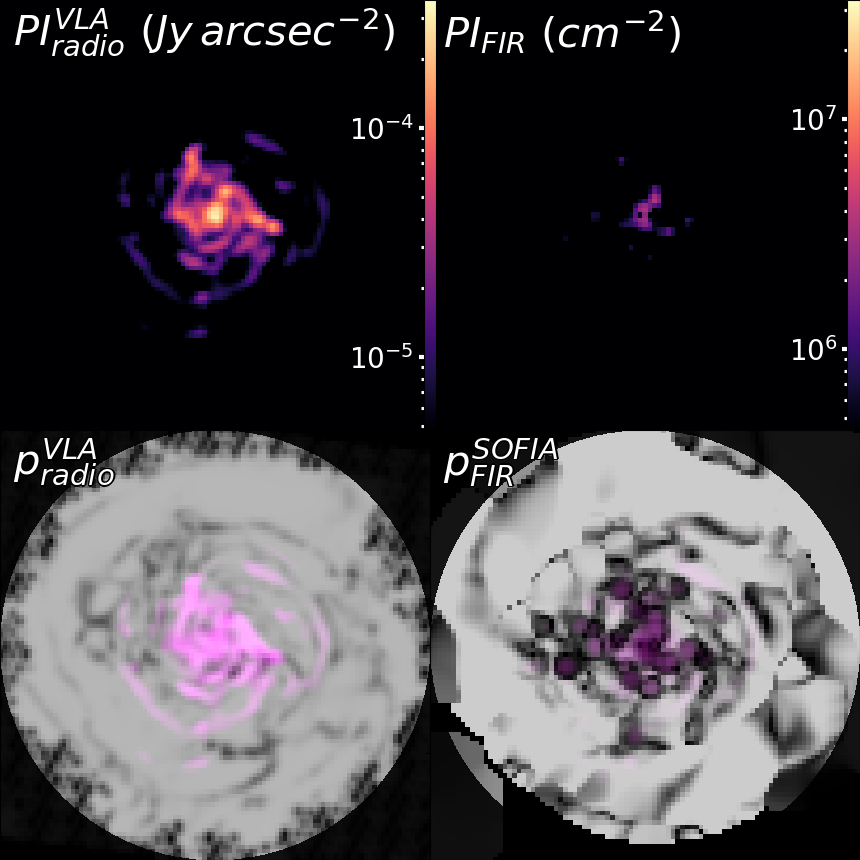}\\
\includegraphics[width=0.66\textwidth]{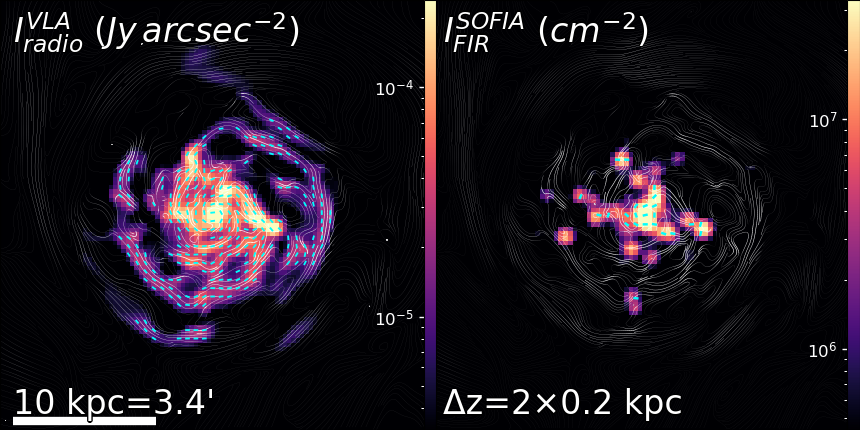}%
\includegraphics[width=0.33\textwidth]{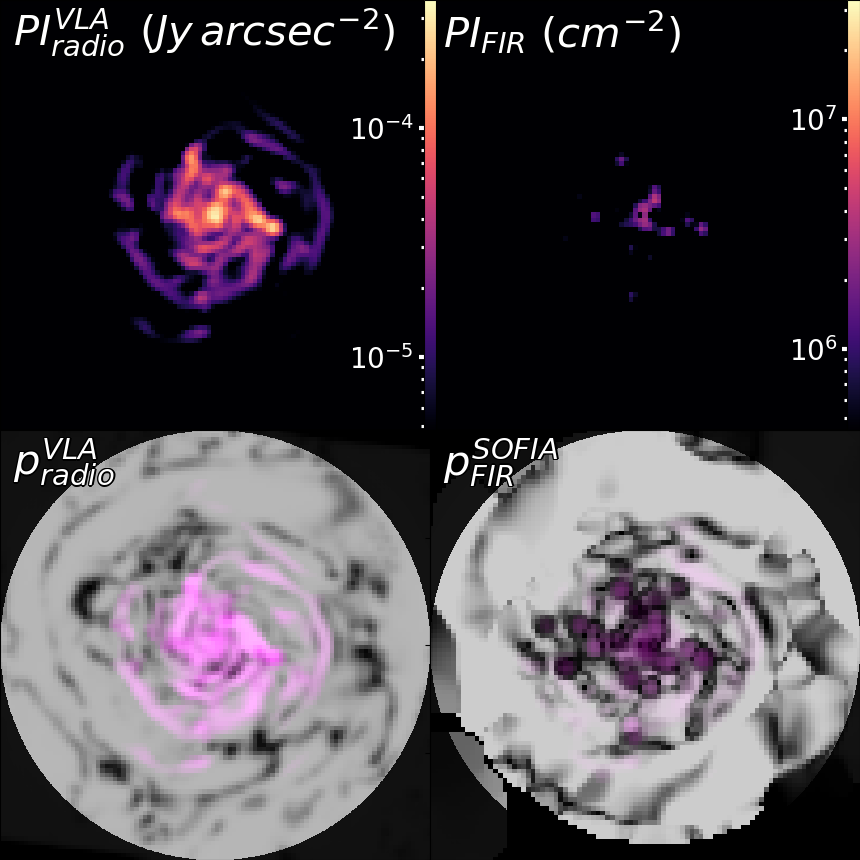}\\
\includegraphics[width=0.66\textwidth]{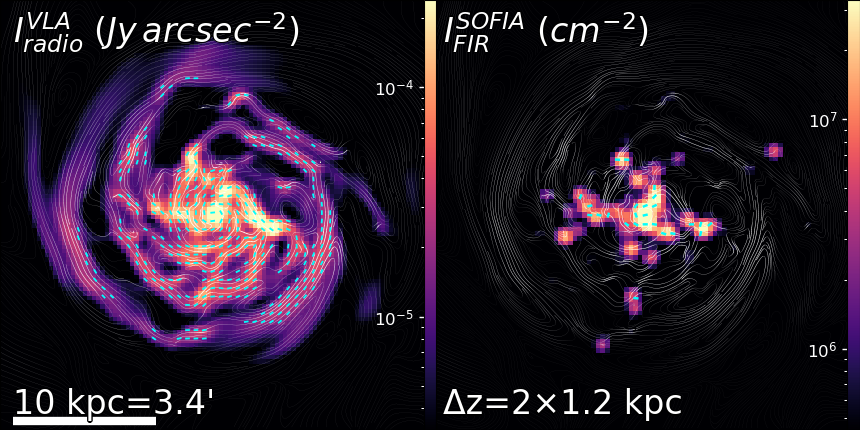}%
\includegraphics[width=0.33\textwidth]{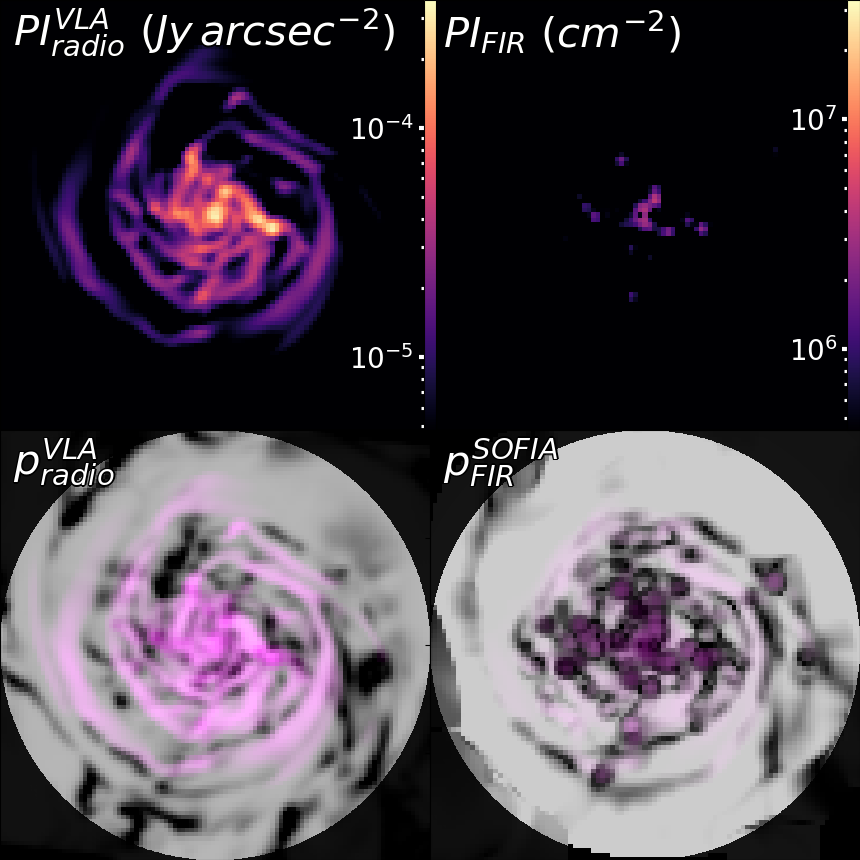}\\
\caption{Same as Figure~\ref{fig:MB11_slice_telescope}, but now for the \MBinj~model. As for the \MBonce~model, synchrotron intensities at 6.2 cm grow as the thickness of the projections is increased, with increasing radio depolarization when probing larger disk thicknesses. The FIR is fixed within $\Delta \zcoord \lesssim 0.1\,\kpc$, both for the total intensity and the observed depolarization.}
\label{fig:MBinj_slice_telescope}
\end{figure*}

To address the question of what is the source of polarized radio and FIR emission within the volume of galaxies, we review the distribution of their intensities along the disk height. Figures~\ref{fig:MB11_slice_telescope} and \ref{fig:MBinj_slice_telescope} display synthetic telescope-like observations of the \MBonce~and \MBinj~models for three different values of the disk half-thickness (0.5 $\Delta \zcoord$): 0.1, 0.2 and 1.2 kpc (i.e., integrated from -$0.5 \Delta \zcoord$ to +$0.5 \Delta \zcoord$). These are selected to capture the relative increase in intensity later explored in Section~\ref{ss:ScaleHeight}. The two leftmost columns correspond to the radio (first) and FIR (second) total intensities, overlaid with the inferred magnetic field orientation from their respective linearly polarized emissions. We also include in these panels the intrinsic orientation of the magnetic field within the disk slice as white streamlines (computed as described in Section \ref{s:Mocks}). The rightmost group of the four panels displays from top left to bottom right: radio-polarized intensity, FIR polarized intensity, radio-polarized fraction, and FIR polarized fraction. We fix the dynamic range displayed by the colors to a maximum-minimum-ratio of 75.0 for total and polarized intensities, and 2.0 for the polarization fractions so as to highlight differences in scaling. Linear polarization fractions range from 0.3 to 0.6 for radio and from 0.05 to 0.1 for FIR. To provide further clarity for these fractional quantities, we overlay the respective total intensity for each domain in purple.

The two figures reflect the more extended nature of the radio emission when compared with the FIR. A notable feature, however, is how this extended radio emission emerges for half-thicknesses $> 0.1\,\kpc$, indicating that the extended radio emission becomes increasingly prominent when we integrate up to higher altitudes above the midplane. The FIR emission experiences only minor changes, which will be quantified in Section \ref{ss:ScaleHeight}. Some FIR emitting clumps appear at radii $\gtrsim 8\,\kpc$ as the integrated thickness is increased but only contribute a small proportion of the total observed intensity. The appearance of these clumps are caused by deviations from a perfect cylindrical disk and by some gas clumps having orbits perturbed to some altitude above the midplane. The described trends are also reproduced by the polarized intensities. The polarization fractions provide further insight into how the intensities are distributed across disk heights. Based on the linear polarization maps, the polarization fraction of the FIR remains largely unchanged for altitudes $> 0.1\,\kpc$. The FIR polarized emission emerges mostly from the observed clumps, which have their polarization dominated by their central emission. These results will be quantified below, and are compatible with the observed clumpy polarized emission from spiral galaxies observed by the SALSA survey \citep{SALSAIV,SALSAV}. 

The synchrotron emission is approximately at its maximum average polarization fraction at $0.5 \Delta \zcoord \sim 0.1\,\kpc$ ($p_\text{radio} (\Delta \zcoord = 2 \times 0.1 \kpc) \sim 0.68$, $p_\text{radio} (\Delta \zcoord = 2 \times 1.2 \kpc) \sim 0.62$), indicating coherent emission at scales of $0.2\,\kpc$. Some star forming clumps display significant depolarization at low altitudes. As the thickness of the synthetic observations is increased, additional depolarization occurs, indicating that this radio emission is sensitive to depolarization effects along the LOS. Most of this depolarization with increasing thickness is localized trailing the synchrotron emission arms and the edges of regions with large polarized intensities. The radio-polarized emission maps show magnetic arm-like structures, which are more evident at low altitudes ($< 0.2\,\kpc$) but increase in intensity as the integration height increases. This result suggests that the measured {\it magnetic arms} observed in NGC\,6946 \citep{Beck2007} may be located close to the midplane of the galaxy, while the more diffuse radio-polarized emission arises from comparatively higher altitudes than its arm-related counterpart.

\subsection{Comparison of the Scale Height of FIR and Radio Emission}
\label{ss:ScaleHeight}

\begin{figure}[ht!]
\includegraphics[width=0.47\textwidth]{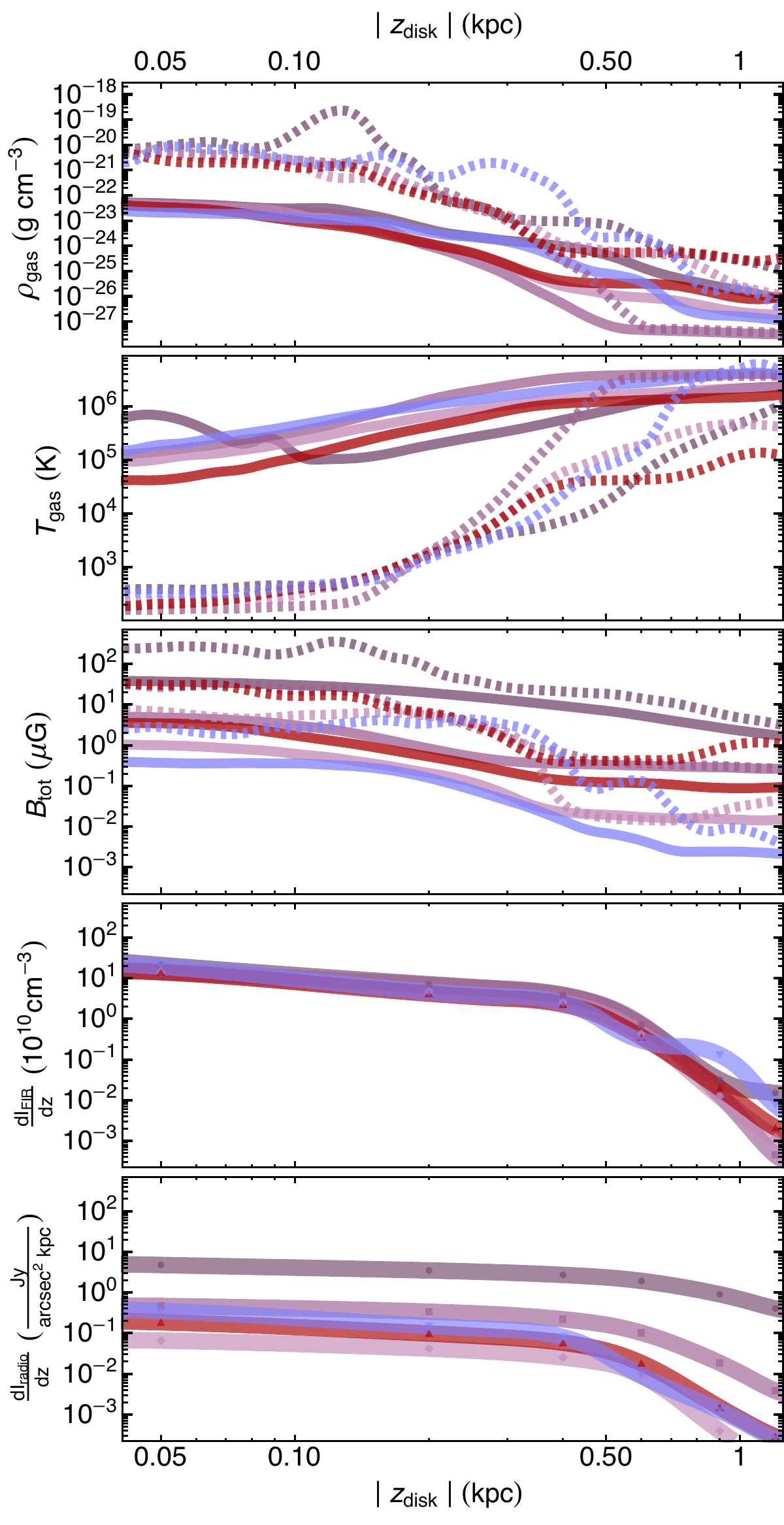}\\
\begin{center}
\vspace{-0.8cm}
\includegraphics[width=0.45\textwidth]{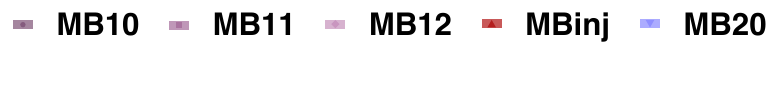}\\
\vspace{-0.9cm}
\end{center}
\caption{Vertical profiles as a function of height for the gas density (top panel), gas temperature (second panel), and magnetic field strength (third panel), as well as for the FIR (fourth panel) and radio (bottom panel) emission per unit height. Intrinsic quantities (i.e., gas density, gas temperature and magnetic field strength) are shown as both mass (dashed) and volume (solid) weighted. We increase the magnetic field of the \MBveinte~model by a factor of $10^7$ to include the simulation within the presented dynamic range. Magnetic fields approximately plateau at altitudes $h \gtrsim 0.6\,\kpc$. FIR displays the most pronounced decrease with height whereas radio is shallower.}
\label{fig:galaxy_profiles}
\end{figure}

To establish the typical emission profiles with respect to altitude above the galactic disk, we show them in Figure~\ref{fig:galaxy_profiles} combined with various intrinsic MHD quantities for comparison. The figure shows from top to bottom, gas density, gas temperature, and magnetic field strength profiles, as well as the intensities per unit height for the FIR (Figure~\ref{fig:galaxy_profiles}, fourth panel) and radio (Figure~\ref{fig:galaxy_profiles}, bottom panel) emissions. For comparison, we include both mass-weighted (dashed lines) and volume-weighted (solid lines) average measurements of the intrinsic MHD quantities (i.e., gas density, gas temperature and magnetic field strength).

For the MHD quantities, volume-weighted profiles are comparable across all our models, except perhaps for \MBdiez. The temperature profile for this model reflects a shrunken galaxy that is somewhat hotter in its innermost region, due to a more radially concentrated disk, and colder at intermediate altitudes ($|\zcoord| \sim 0.2\,\kpc$), due to a wider thick disk (see Figure~\ref{fig:Galaxies-p2}; also \citealt{Martin-Alvarez2020}). Despite this, the mass-weighted temperature profile of \MBdiez, which traces the properties of the thin disk and the denser phases of the ISM, is similar to those of the other models. Mass-weighted profiles have a more complicated structure, with deviations driven by massive and dense gas clumps (e.g., \MBdiez~at $\sim 0.15\,\kpc$ or \MBveinte~at $\sim 0.2\,\kpc$), leading to the substructure observed in the magnetic field profile. 
Volume-weighted gas densities and magnetic field strength, as well as the two intensities studied, display an approximately power-law-like decrease with a break at an altitude of $|\zcoord| \sim 0.6\,\kpc$, with the change in slope somewhat less prominent in the radio. Similarly, the FIR emission also displays the largest decrease, whereas the radio has a shallower reduction with altitude. Magnetic field strengths approximately flatten out above altitudes of $|\zcoord| \gtrsim 0.4 - 0.6\,\kpc$ comparable to the gas density profiles. The magnetic field strength inner profile decreases with an approximate power law of $-1.0 \pm 0.2$, which we estimate is comparable to that in the profiles presented by \citet{Pakmor2017}. Similarly, the gas density profiles have an approximate power-law index of $-2.8 \pm 0.2$. For the intensities, the inner part of the profile ($|\zcoord| < 0.6\,\kpc$) decreases with $\alpha$ power-law indices ($\text{dI} / \text{d}|\zcoord| \propto |\zcoord|^{\alpha}$) of $\alpha_\text{radio} = - 0.56 \pm 0.11$, and  $\alpha_\text{FIR} = -0.99 \pm 0.17$.

\begin{figure*}[ht!]
\includegraphics[width=0.9\textwidth]{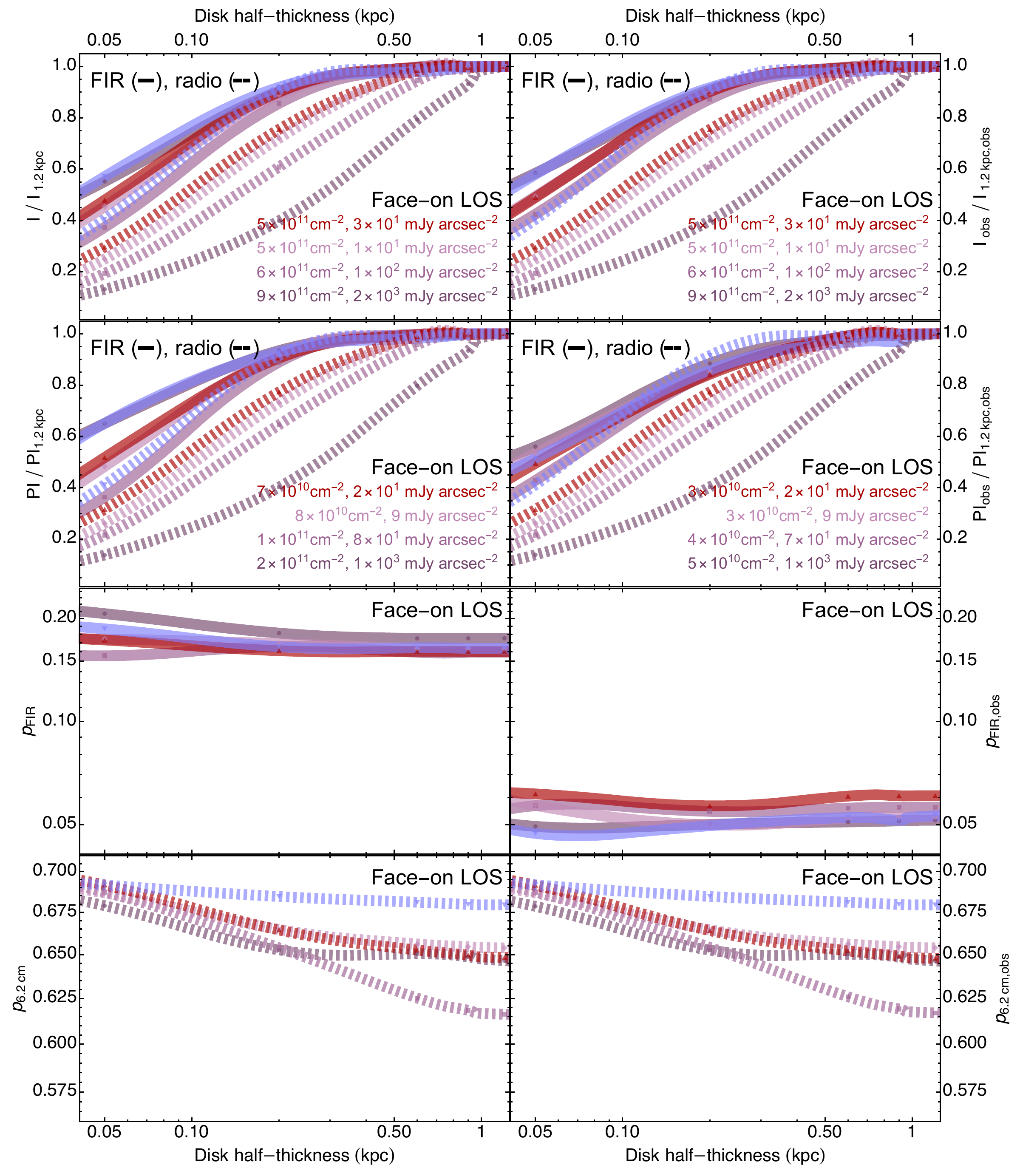}\\
\begin{center}    
\vspace{-0.8cm}
\includegraphics[width=0.4\textwidth]{rowlegend.pdf}\\
\vspace{-0.9cm}
\end{center}
\caption{Relative contribution as a function of galactic altitude for multiple quantities integrated between $-0.5 \Delta \zcoord$ and $+0.5 \Delta \zcoord$, and normalized to their integral at $0.5 \Delta \zcoord = 1.2\,\kpc$. From top to bottom, the rows display total intensity, polarized intensity, FIR polarization fraction, and radio synchrotron polarization fraction. All quantities are computed using face-on synthetic observations. The left and right columns correspond to the full-resolution and telescope-like observations, respectively. FIR emission is shown using solid lines whereas the radio is represented with dashed lines. $\ge80\%$ of the FIR emission is concentrated below $0.2\,\kpc$, whereas radio emission has similar integrated contributions at heights of $0.3 - 0.4\,\kpc$.}
\label{fig:HeightScale}
\end{figure*}

We now quantify how the relative intensity at FIR and radio wavelengths changes as a function of the altitude above the galactic disk, and the corresponding scale heights for each of the observables. For this, we compute the relative growth of multiple quantities. This corresponds to their integral between $-0.5 \Delta \zcoord$ and $+0.5 \Delta \zcoord$ normalized to the value of this integral for $0.5 \Delta \zcoord = 1.2\,\kpc$. Figure~\ref{fig:HeightScale} shows this relative growth with respect to disk half-thickness of $0.5 \Delta \zcoord$ for the normalized total intensity (top row), normalized polarized intensity (second row), FIR polarization fraction (third row), and radio polarization fraction (last row). These are shown both for full-resolution (left column) and telescope-like observations (right column). The panels combine the results for FIR (solid lines) and radio (dashed lines) emission. Intensities are normalized to their value at a 1.2 kpc disk half-thickness, but their integrated values employed for this are provided inside the corresponding panels for completeness.

We find the FIR intensity to increase rapidly within the central $0.2\,\kpc$ of the half-thickness, with approximately $80\%$ of the total and polarized emission concentrated within this layer. Radio intensity only reaches such percentages at scales of $0.3 - 0.4\,\kpc$, and approximately follows its power-law increase ($\text{dI}_\text{radio} \propto |\zcoord|^{-0.56 \pm 0.11} \text{d}|\zcoord|$) up to heights of $\sim 0.5-0.8\,\kpc$.
This is true except for the edge-case models: \MBveinte~and \MBdiez. The negligible magnetization field case presented by \MBveinte~has a more concentrated FIR and radio emission. On the other hand, \MBdiez~has a considerably more extended radio emission and FIR with its most significant contribution coming from scales comparable to those of \MBveinte. This suggests that stronger fields lead to a thicker distribution above the disk midplane of the underlying magnetic energy \citep{Martin-Alvarez2020}, and is reflected by the radio. As expected, the total intensities are unchanged when our telescope-like observation method is applied. Full-resolution polarized intensities also follow a relative growth similar to their total intensity counterparts, with some variation at the innermost heights likely driven by the degree of magnetic field anisotropy. The overall depolarization within the explored altitude range is small, approximately 3\% for the FIR and 5\% for the radio. When considering the depolarization occurring in dense gas clouds, as shown by Figures~\ref{fig:MB11_slice_telescope} and \ref{fig:MBinj_slice_telescope}, this suggests that for face-on observations, the magnetic field is, except in localized regions (dense gas clumps for FIR and sometimes radio emissions; and trailing emission in magnetic arms in radio), generally ordered along the LOS above heights of $\sim100$ pc. The synchrotron emission scarcely changes when comparing the polarized intensities and polarization fractions between our full-resolution and telescope-like synthetic observations. This further supports that the emission (and magnetic field) structures probed by these lower frequencies are large scale in nature, with sizes comparable to or larger than the physical resolution of our telescope (i.e., 300 parsecs). 

Comparatively, the FIR polarized intensity clusters considerably within $0.1\,\kpc$, with virtually all models featuring $70\%$ of their FIR emission within this half-thickness. This suggests that the small-scale structure variations seen in the full-resolution observations are now dominated by depolarization, due to the SOFIA beam. This is further reflected by a reduced polarized fraction from about $15\%$ in the intrinsic scenario, down to $\sim5\%$ in the telescope-like scenario. This supports that the FIR emission is a probe of magnetic fields at scales $\lesssim 100$ parsecs, and supports the case for future FIR polarimetric missions with even higher resolutions as a powerful window to probe into the deep and small-scale magnetic fields of galaxies. 

\begin{figure}[ht!]
\includegraphics[width=0.47\textwidth]{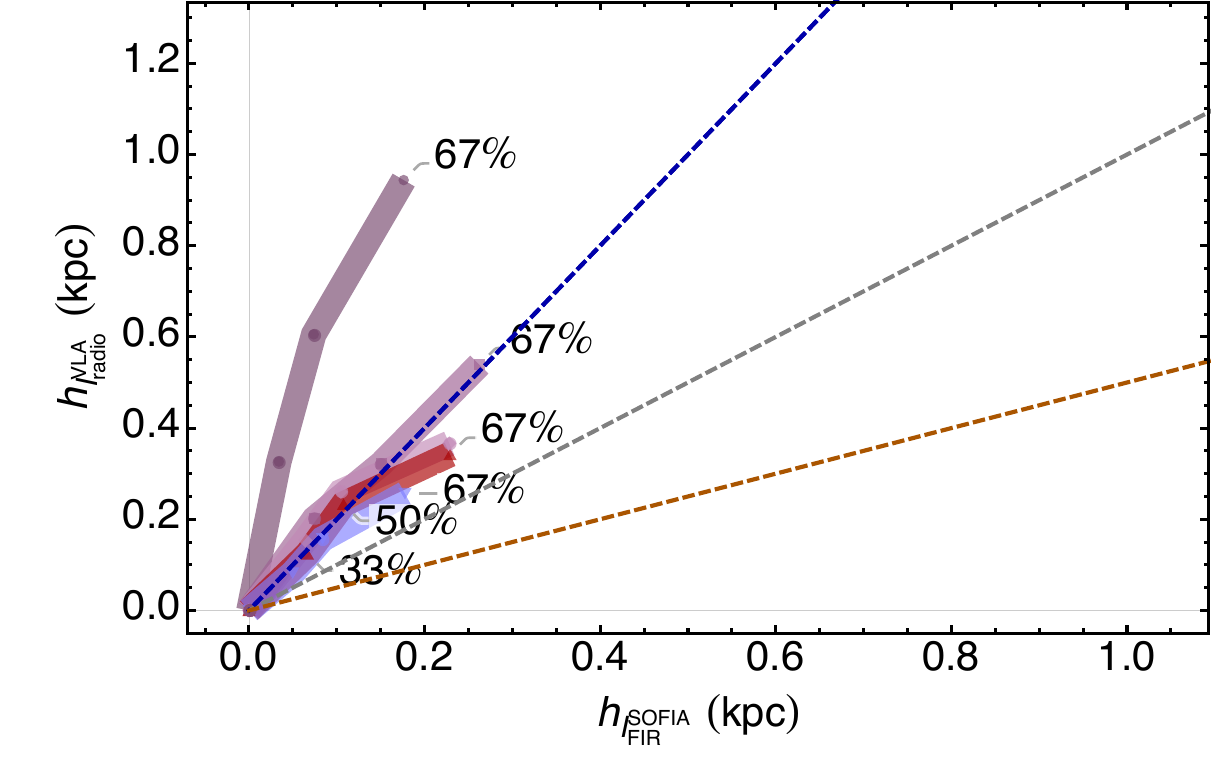}\\
\begin{center}    
\vspace{-0.8cm}
\includegraphics[width=0.4\textwidth]{rowlegend.pdf}\\
\vspace{-0.9cm}
\end{center}
\caption{Comparison of height scale for radio ($y$-axis) vs. FIR ($x$-axis). Different lines show the altitudes at which a given percentage of the integrated emission is reached for each of the reviewed quantities. We include three dashed lines to aid visual inspection indicating the one-to-one (gray), two-to-one (blue), and half-to-one (orange) relations. Lines scaling along the blue dashed line have quantities for the $y$-axis that proportionally extend twice the height of the $x$-axis, and inversely for those scaling along the orange dashed line. Total radio synchrotron intensity has approximately double the scale height of the FIR emission.}
\label{fig:FIRvsRadio_panel}
\end{figure}

\begin{figure*}[ht!]
\includegraphics[width=\textwidth]{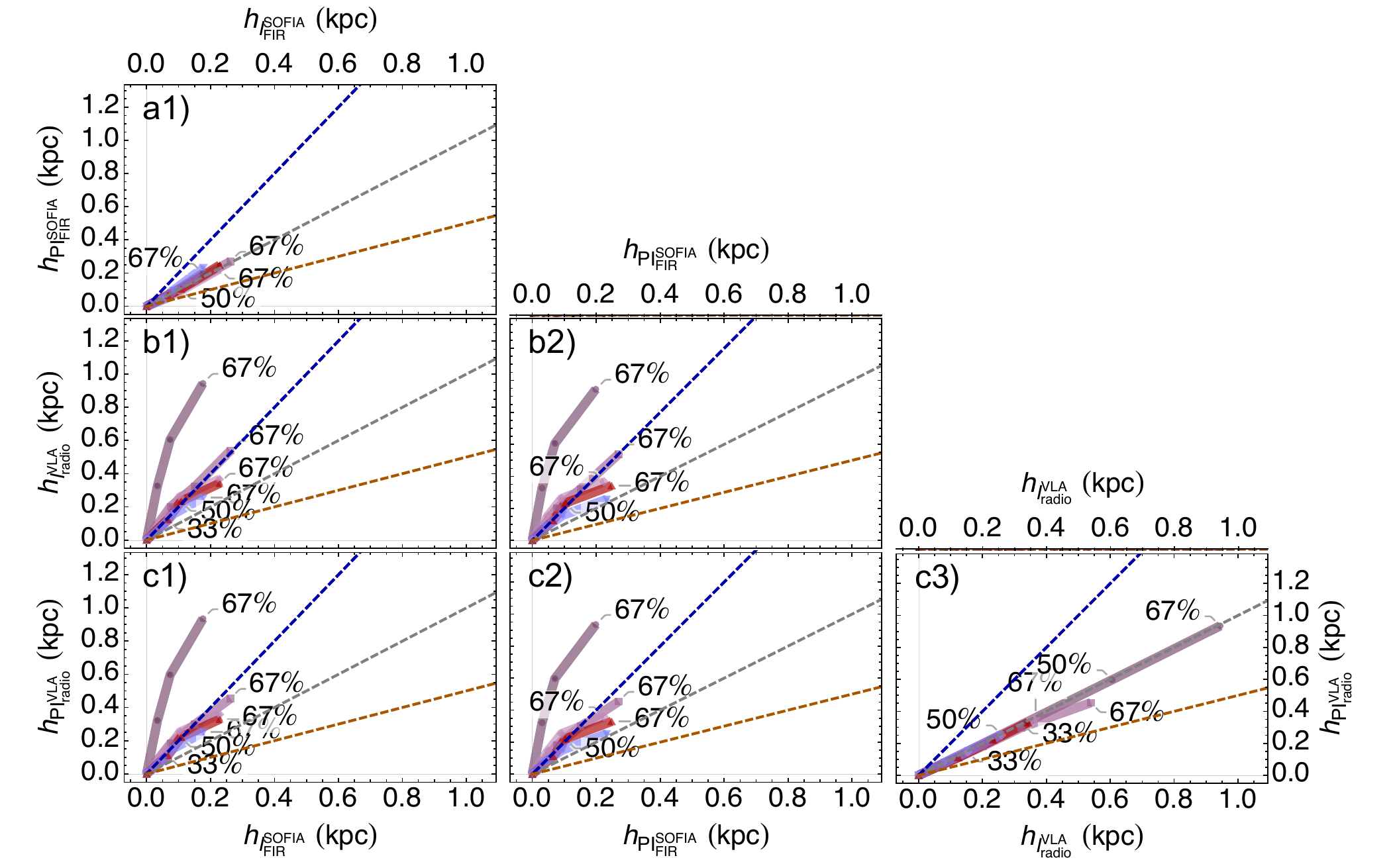}\\
\begin{center}    
\vspace{-0.8cm}
\includegraphics[width=0.4\textwidth]{rowlegend.pdf}\\
\vspace{-0.9cm}
\end{center}
\caption{Comparison of height scale for various combinations of observational quantities. From left to right, the $x$-axes of the columns correspond to SOFIA-like FIR total intensity, SOFIA-like FIR polarized intensity, and VLA-like radio synchrotron total intensity. From top to bottom, the rows display the SOFIA-like FIR polarized intensity, VLA-like radio synchrotron total intensity, and VLA-like radio synchrotron polarized intensity. We include three dashed lines to aid visual inspection indicating the one-to-one (gray), two-to-one (blue), and half-to-one (orange) relations. Lines scaling along the blue dashed line have quantities for the $y$-axis that proportionally extend twice the height of the $x$-axis, and inversely for those scaling along the orange dashed line. Both total and polarized radio synchrotron intensity have height scales approximately double to their FIR counterparts.}
\label{fig:FIRvsRadio_triangle}
\end{figure*}

To compare the altitudes reached by the FIR and radio emissions, we review in Figure~\ref{fig:FIRvsRadio_panel} the scale height at which a given percentage of the emission is obtained for the FIR ($x$-axis) and radio ($y$-axis) intensities. To aid in the interpretation of this type of graph, we include three dashed lines: the one-to-one (gray), two-to-one (blue) and half-to-one (gold) relations between the two scale heights. An important result from this figure is the approximate two-to-one scaling between most of the radio and FIR intensities. We estimate that $\ge50$\% of the FIR intensity arises from a vertical scale of $\le0.11$ kpc, while at radio wavelengths, the vertical scale responsible for half of the emission is $\le0.24$ kpc (Table~\ref{table:scale_summary}). This result indicates that most of the radio intensities arise from disk regions with thickness approximately twice in size as that of the FIR intensities, i.e., $\frac{\hIradio}{\hIFIR} = 2.16\pm0.13$ The averaged vertical scales were estimated using the combination of the models  \MBinj, \MBdoce~and \MBonce, labeled as {\bf Combined} (i.e. excluding \MBveinte~and \MBdiez) in Table \ref{table:scale_summary}.

We further compare total and polarized intensities in Figure~\ref{fig:FIRvsRadio_triangle}. Overall, the comparison of total to polarized intensities for both FIR and radio wavelengths scales approximately along the one-to-one relation (a1 and c3 panels). This indicates that polarized emission globally probes approximately the same column as total intensities, at least at scale heights larger than the resolution elements of the simulations. In combination with the trend for depolarization discussed above, this suggests that regions where depolarization is important have a comparatively higher contribution to their final polarization from emission layers at higher altitudes, which are less likely to be perfectly aligned with emission emerging from lower altitudes. This is reflected in the linear polarization fraction shown in Figures~\ref{fig:MB11_slice_telescope} and \ref{fig:MBinj_slice_telescope}. The two-to-one scaling between the radio and FIR intensities is preserved for the polarized intensity: $\frac{\hPIradio}{\hPIFIR} = 2.11\pm0.24$. Depolarization along the LOS may become more important for galaxies observed under some degree of inclination. Such effects of inclination are studied in Section  \ref{section:InclinationEffects}.

In addition, all the results above are approximately unchanged when comparing the full-resolution and telescope-like observations. This is shown in Appendix~\ref{ap:TelescopeVsIntrinsic}, and it indicates that telescope observations at their currently attainable resolutions succeed in probing the intrinsic distribution of the emission along different disk heights down to at least a few times the resolution of our simulations ($\lesssim$ 30 pc). Consequently, we limit our analysis to the telescope-like case, but note that our results are unchanged vis-\`a-vis the full-resolution case.

\begin{deluxetable*}{lcccccc}
\centering
\tablecaption{Summary of the measured scale heights of our simulations.
\label{table:scale_summary} 
}
\tablecolumns{7}
\tablewidth{0pt}
\tablehead{\colhead{Model} & 	\colhead{$\hIFIR$}  & \colhead{$\hPIFIR$} &  \colhead{$\hIradio$} & \colhead{$\hPIradio$} & \colhead{$\frac{\hIradio}{\hIFIR}$} & \colhead{$\frac{\hPIradio}{\hPIFIR}$} \\ 
 & \colhead{(kpc)} & 	\colhead{(kpc)}  & \colhead{(kpc)}  & \colhead{(kpc)}  \\
\colhead{(1)} & \colhead{(2)} & \colhead{(3)} & \colhead{(4)} & \colhead{(5)} & \colhead{(6)} & \colhead{(7)} }
\startdata
\multicolumn{7}{c}{Face-on orientation}\\
\hline
\MBinj  &  0.11 &  0.10 &  0.23 &  0.22 &  2.2 &  2.1 \\ 
\MBdoce &  0.11 &  0.09 &  0.26 &  0.25 &  2.5 &  2.8 \\ 
\MBonce &  0.15 &  0.15 &  0.32 &  0.30 &  2.1 &  2.0 \\ 
{\bf Combined} &  $ 0.11 \pm 0.01 $ &  $ 0.11 \pm 0.01 $ &  $ 0.24 \pm \
0.03 $ &  $ 0.23 \pm 0.03 $ &  $ 2.2 \pm 0.1 $ &  $ 2.1 \pm \
0.2 $ \\ 
\MBveinte &  0.08 &  0.10 &  0.16 &  0.15 &  1.9 &  1.6 \\ 
\MBdiez &  0.07 &  0.07 &  0.61 &  0.59 &  8.2 &  8.4 \\ 
\hline
\multicolumn{7}{c}{Inclined orientation}\\
\hline
\MBinj &  0.13 &  0.12 &  0.27 &  0.26 &  2.1 &  2.2 \\ 
\MBdoce &  0.14 &  0.08 &  0.30 &  0.29 &  2.2 &  3.6 \\ 
\MBonce &  0.21 &  0.22 &  0.39 &  0.36 &  1.8 &  1.6 \\ 
{\bf Combined} &  $ 0.16 \pm 0.03 $ &  $ 0.14 \pm 0.04 $ &  $ 0.32 \pm \
0.04 $ &  $ 0.30 \pm 0.03 $ &  $ 2.0 \pm 0.1 $ &  $ 2.5 \pm \
0.5 $ \\ 
\MBveinte &  0.10 &  0.09 &  0.17 &  0.16 &  1.7 &  1.9 \\ 
\MBdiez &  0.08 &  0.04 &  0.63 &  0.55 &  8.0 &  15.7 
\enddata
\tablenotetext{}{Notes. 
Column (1): model name.
Column (2): scale height of the $50$\% integrated FIR total intensity.
Column (3): scale height of the $50$\% integrated FIR polarized intensity. 
Column (4): scale height of the $50$\% integrated radio total intensity. 
Column (5): scale height of the $50$\% integrated radio-polarized intensity. 
Column (6): ratio between the radio and FIR total intensity scale heights.
Column (7): ratio between the radio and FIR polarized intensity scale heights.
We include a {\bf Combined} scale height estimated as the average of the models \MBinj, \MBdoce\, and \MBonce, excluding the edge cases \MBveinte\, and \MBdiez.}
\end{deluxetable*}

\subsection{FIR as a Tracer of the Cold Medium and Radio Synchrotron as a Tracer of Warm Neutral and Cold Phases}

A different height scaling for each of the studied wavelengths is somewhat expected when considering that the FIR is a dust-generated emission (Equation~\ref{eq:ImockFIRdust}), whereas the radio intensity is proportional to some power of the magnetic field strength and the CR electron distribution (Equation~\ref{eq:Iradio}), with both further affected by the magnetic field structure. Stronger magnetic fields are theoretically expected and measured by observations probing denser gas phases \citep{Crutcher2012, Roche2018}. Despite having weaker magnetic fields, due to the large fraction of galaxy volume in the warm and hot gas phases and the volume-filling distribution of CR electrons in the galaxy, the synchrotron radiation emerging from these phases will constitute a significant fraction of the observed extended radio emission.

\begin{figure*}[ht!]
    \begin{center}
    \includegraphics[width=0.2\textwidth]{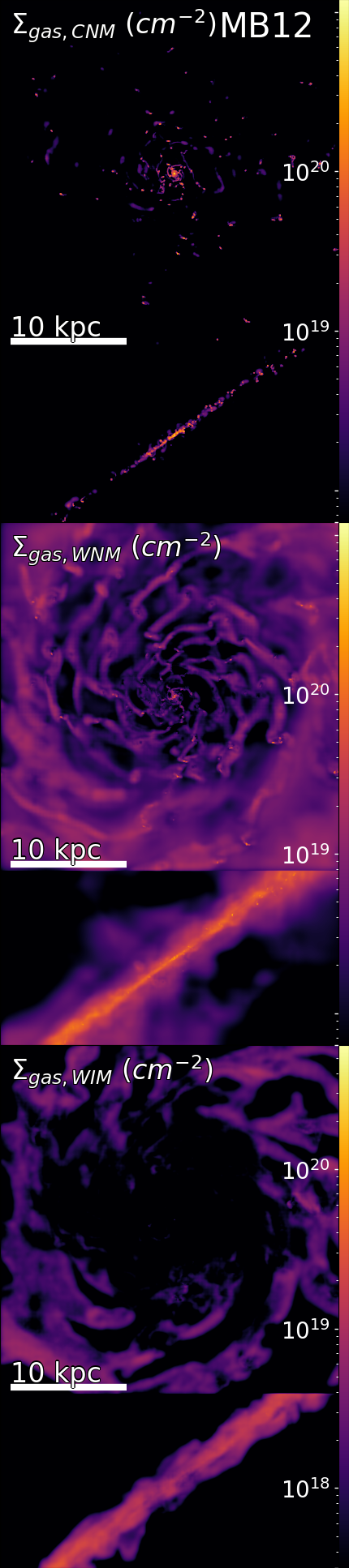}%
    \includegraphics[width=0.2\textwidth]{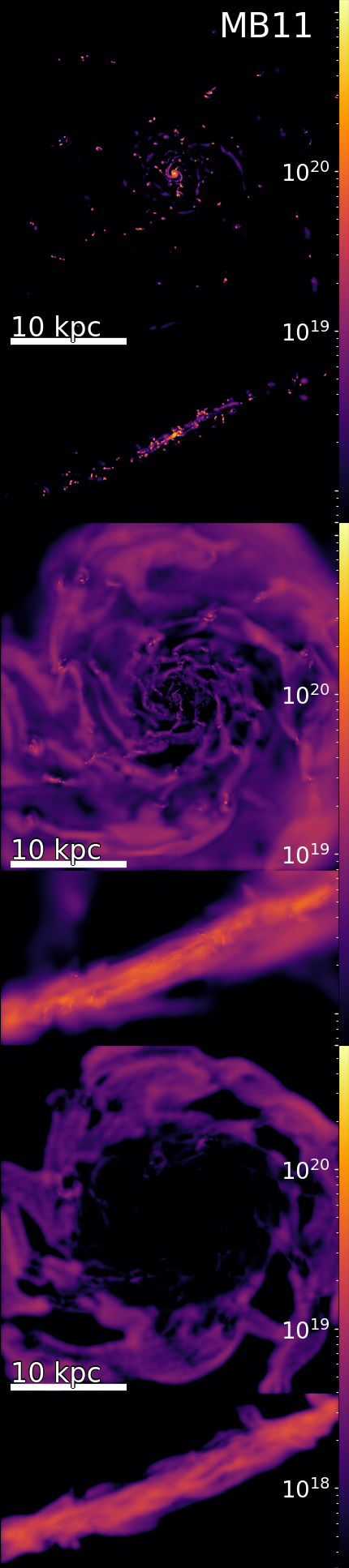}%
    \includegraphics[width=0.2\textwidth]{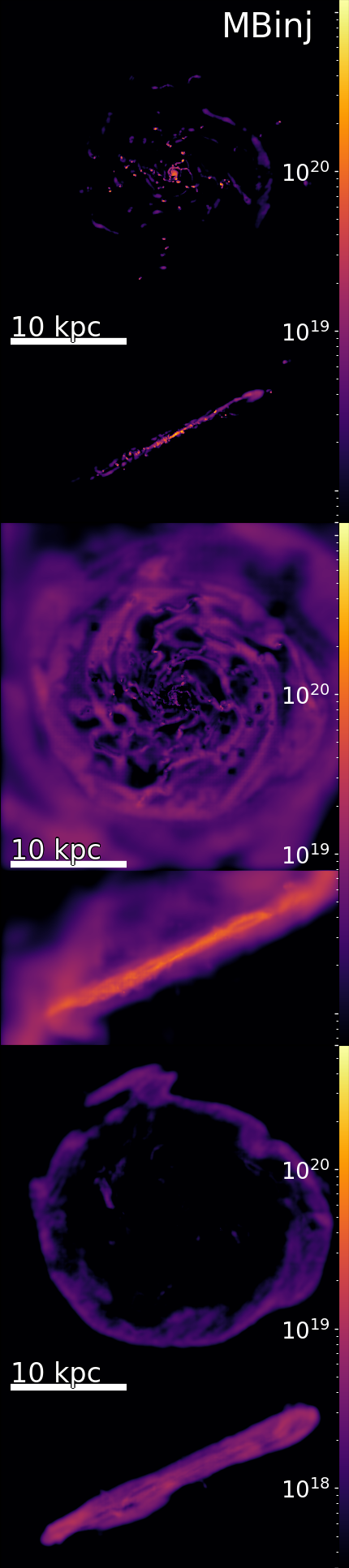}%
    \includegraphics[width=0.2\textwidth]{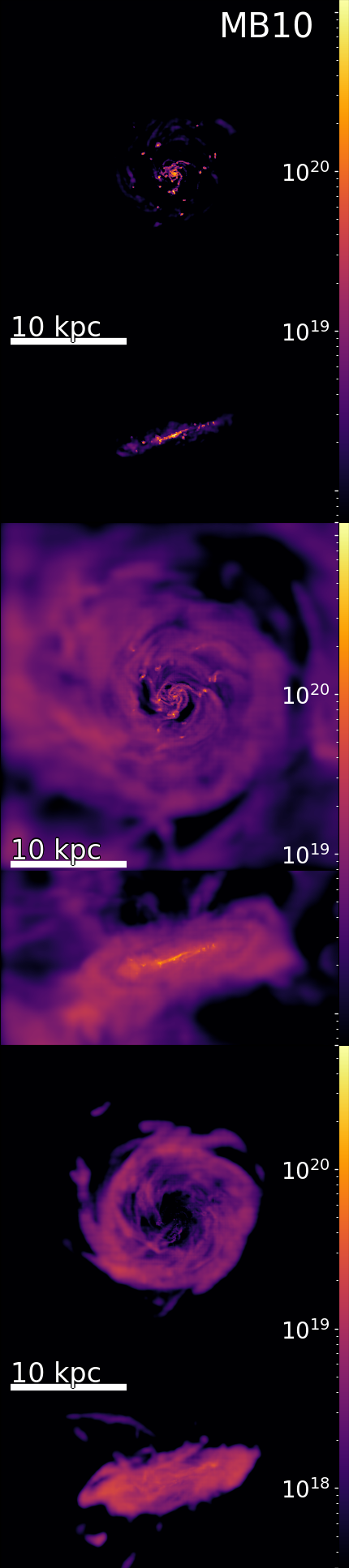}%
    \includegraphics[width=0.2\textwidth]{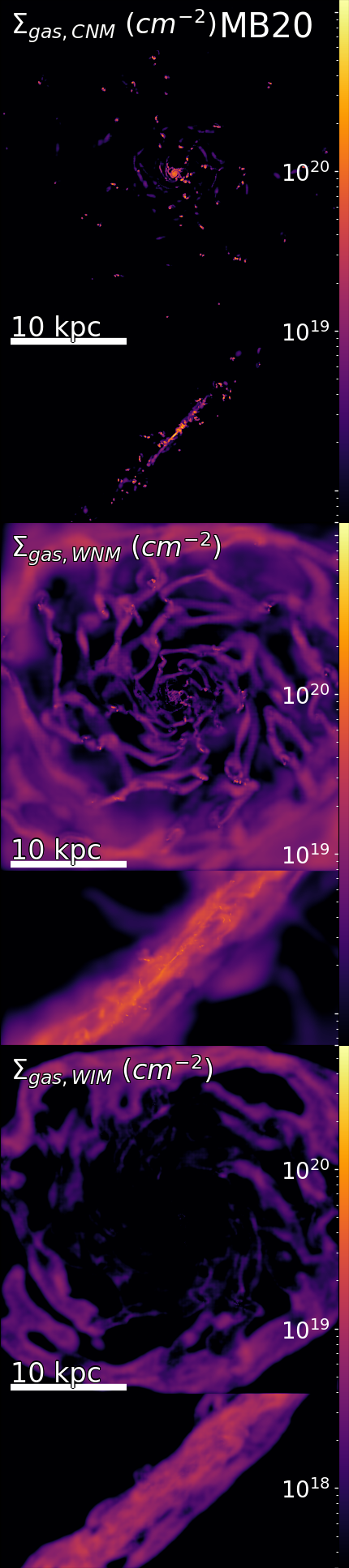}\\
    \end{center}
\caption{Projected views of the galaxies (same as shown in Figures~\ref{fig:Galaxies-p1} and \ref{fig:Galaxies-p2}) now showing galaxy ISM phase surface gas densities. From top to bottom, the rows display CNM phase gas surface density, WNM gas surface density, and WIM gas surface density. From left to right, each column displays \MBdoce, \MBonce, \MBinj, \MBdiez, and \MBveinte~models.}
\label{fig:Galaxies-p3}
\end{figure*}

To investigate the correlation between the two studied wavelengths and the ISM phases, we divide the ISM into four different phases. Each gas cell is classified, according to its temperature, $T$, and hydrogen ionization, $x_\text{HII}$, into
\begin{itemize}
    \item Cold neutral medium (CNM): $T \le 200\,\K$,
    \item Warm neutral medium (WNM): $ 200\,\K < T \le 10^6\,\K$, with its WNM gas density $\rho_\text{gas, WNM} = \rho_\text{gas, warm}\,(1 - x_\text{HII})$,
    \item Warm ionized medium (WIM): $ 200\,\K < T \le 10^6\,\K$, with its WIM gas density $\rho_\text{gas, WIM} = \rho_\text{gas, warm}\,x_\text{HII}$,
    \item Hot medium (HM): $10^6\,\K < T$,
\end{itemize}
where $x_\text{HII}$ is the hydrogen ionization fraction for each cell computed according to the gas' local properties. The quantity $\rho_\text{gas, warm}$ is the gas density $\rho_\text{gas}$ for cells with temperature fulfilling $T \in \left[200, 10^6\right]\,\K$, and set to $\rho_\text{gas, warm} = 0$ otherwise. We note that modeling a self-consistent ionization of the gas, which requires radiative transfer \citep[e.g.,][]{Rosdahl2015a,Katz2022a}, would modify the total amount of ionized hydrogen, its distribution, and its dynamics \citep{Rosdahl2015b,Martin-Alvarez2023,Yuan2024}. To provide further insight into the distribution of gas across the ISM phases we defined, we display in each subsequent row of Figure~\ref{fig:Galaxies-p3} the surface density of the CNM, WNM and WIM phases. The figure displays the same galaxies and orientations as Figures~\ref{fig:Galaxies-p1} and \ref{fig:Galaxies-p2}, with columns from left to right corresponding to the \MBdoce, \MBonce, \MBinj, \MBdiez~and \MBveinte~models. Overall, the distributions of CNM and WNM gas surface densities are similar across simulations, and especially for our three main models studied (i.e., \MBdoce, \MBonce~and \MBinj). As discussed by \citet{Martin-Alvarez2020}, the \MBdiez~shows a somewhat more radially concentrated galaxy, and with a thicker disk (particularly evident for its WNM). For all models, most of the mass in the WIM is concentrated at radii of $r \gtrsim 10\,\kpc$, with a distribution comparable to the WNM at those radii. In this section, we study the correlations between each of the ISM phases and the FIR and radio emission. We find these to be preserved despite any ISM phase variations that may be introduced by the different studied magnetizations, except perhaps for the larger deviations of the extreme \MBdiez~case. In future work, we will review in more detail the impact of the magnetic field on the ISM and its different phases.

\begin{figure*}[ht!]
\includegraphics[width=\textwidth]{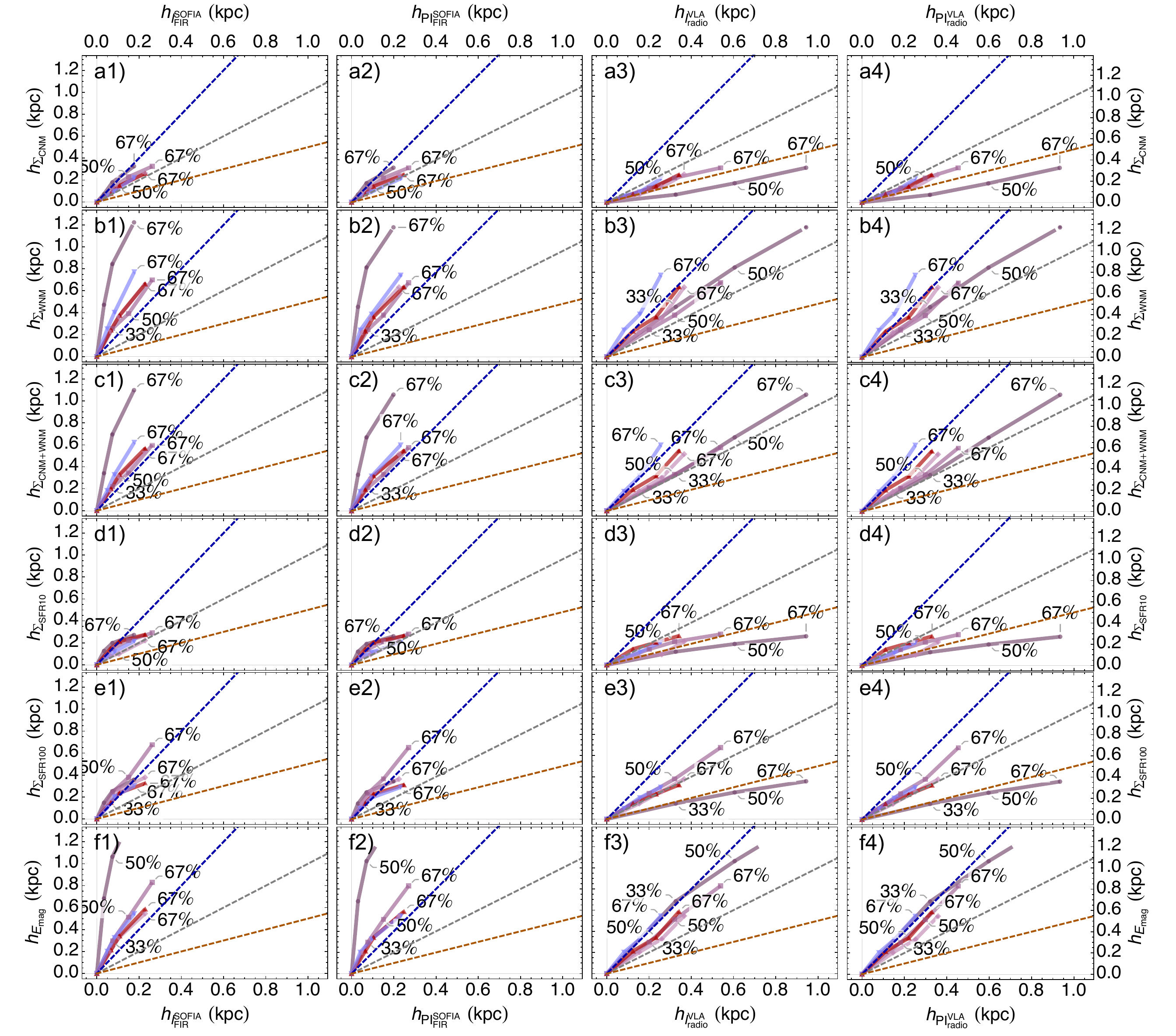}\\
\begin{center}    
\vspace{-0.8cm}
\includegraphics[width=0.4\textwidth]{rowlegend.pdf}\\
\vspace{-0.9cm}
\end{center}
\caption{Same as Figure~\ref{fig:FIRvsRadio_triangle}, now comparing telescope-like intensities with multiple intrinsic quantities of interest for which we found or expected correlations. From top to bottom, these correspond to CNM column density $\Sigma_\text{CNM}$, WNM column density $\Sigma_\text{WNM}$, neutral hydrogen column density $\Sigma_\text{CNM}+\Sigma_\text{WNM}$, 10 Myr SFR column density $\Sigma_\text{SFR10}$, 100 Myr SFR column density $\Sigma_\text{SFR100}$, and integrated total magnetic energy $E_\text{mag}$. Both FIR and radio emission are more concentrated than the integrated magnetic energy (f panels). We find that the CNM is traced by the FIR emission (panels (a1), (a2)), whereas the radio synchrotron emission is an approximate tracer for the combination of the WNM and CNM phases instead (except for \MBveinte; panels (c3), (c4)).}
\label{fig:emission_vs_phase}
\end{figure*}

To provide a more quantitative relation between the gas phases and the different types of emission, we repeat our scale height comparison, now between the two wavelengths studied against the ISM phases, magnetic energy, and SFR. Due to the approximately one-to-one scaling between telescope-like and full-resolution synthetic observations, we only review comparisons for the former. In Figure~\ref{fig:emission_vs_phase}, we show a comparison of the scale height between the radio and FIR intensities and the intrinsic quantities with which we found or expected interesting correlations. The $x$-axis in each column in Figure~\ref{fig:emission_vs_phase} shows from left to right SOFIA-like FIR  total intensity, SOFIA-like FIR polarized intensity, VLA-like radio total intensity, and VLA-like radio-polarized intensity. The quantities of interest are shown for each row's respective $y$-axis, and correspond from top to bottom to CNM surface density, WNM surface density, neutral gas (CNM and WNM) combined surface density, ongoing SFR (over the last 10 Myr), recent SFR (over the last 100 Myr), and total magnetic energy. The average ratios between the quantities found to display correlations are summarized in Table~\ref{table:phases_summary}. We reviewed various other quantities not shown in this work, and found no significant correlations. These are listed in the bottom of Table~\ref{table:phases_summary}.

We find some important correlations between the emission and some of the ISM phases. In particular, FIR emission scales comparably to the CNM surface density ($\frac{\hCNM}{\hIFIR} \sim \frac{\hCNM}{\hPIFIR} \sim 1.3\pm0.1$; panels (a1), (a2)), whereas the radio emission roughly matches the neutral gas surface density scaling with height ($\frac{\hCNMWNM}{\hIradio} \sim \frac{\hCNMWNM}{\hPIradio} \sim 1.3\pm0.1$; panels (c3), (c4)), except for the \MBveinte~with a negligible magnetic field. This latter correlation is driven by the combined mass, as exclusively considering the WNM leads to a somewhat more extended scale height than that from the synchrotron (panels (b3), (b4)), and was recently also found by \citet{Ponnada2023a}. This supports the consideration of the FIR emission as a density probe, while the radio can be comparatively interpreted as a volume-filling \textbf{mass} probe. 

We also explore correlations with the scaling of SFR, both at timescales indicative of ongoing star formation ($10$ Myr; $H_{\alpha}$) and recent activity ($100$ Myr, UV). FIR is a better tracer of the ongoing star formation ($\frac{\hSFR}{\hIFIR} = 1.5\pm0.2$; panels (d1), (d2)), although we note that our FIR emission has a higher contribution from the central-most altitudes (i.e., a rapid increase up to $\sim50\%$ of the emission) up to $\sim 0.1\,\kpc$. Both radio and FIR become comparable in scale at heights of $\sim 0.2\,\kpc$. On the other hand, the radio intensity matches well the scale heights of star formation on longer timescales ($100$ Myr), i.e., $\frac{\hSFRlong}{\hPIradio} = 1.1\pm0.1$ (panels (e3), (e4)). It shows approximately one-to-one scaling for all models except for \MBdiez, which features a concentrated star formation profile driven by its extreme magnetization \citep{Martin-Alvarez2020}. 

Finally, we display the scaling of the two types of emission with magnetic energy (bottom row). These relations show that the magnetic energy in all our galaxies extends significantly further into higher altitudes than the intensities in our synthetic observations. This result may suggest that these two wavelength ranges do not capture a considerable amount of magnetic energy, likely due to the decrease of number density with height of both electronic CRs and dust. Furthermore, with magnetic field lines above the galactic disk likely oriented along the direction of winds and outflows \citep{Lopez-Rodriguez2021,Lopez-Rodriguez2023}, synchrotron, and dust emission will now be preferentially oriented toward LOSs with some degree of inclination. 

\begin{deluxetable*}{lcccccc}
\centering
\tablecaption{Total integrated projected ratio between phase surface densities and the ISM phases.
\label{table:phases_summary} 
}
\tablecolumns{8}
\tablewidth{0pt}
\tablehead{\colhead{Model} & 	\colhead{$\frac{\hCNM}{\hIFIR}$}  & \colhead{$\frac{\hCNM}{\hPIFIR}$} &  \colhead{$\frac{\hCNMWNM}{\hIradio}$} & \colhead{$\frac{\hCNMWNM}{\hPIradio}$} & \colhead{$\frac{\hSFR}{\hIFIR}$} & \colhead{$\frac{\hSFRlong}{\hPIradio}$} \\ 
\colhead{(1)} & \colhead{(2)} & \colhead{(3)} & \colhead{(4)} & \colhead{(5)} & \colhead{(6)} & \colhead{(7)} }
\startdata
\multicolumn{7}{c}{Face-on orientation} \\
\hline
\MBinj &  $ 1.23 \pm 0.01 $ &  $ 1.22 \pm 0.01 $ &  $ 1.47 \pm \
0.01 $ &  $ 1.57 \pm 0.01 $ &  $ 1.83 \pm 0.04 $ &  $ 1.03 \pm \
0.01 $ \\
\MBdoce & $ 1.21 \pm 0.01 $ &  $ 1.41 \pm 0.01 $ &  $ 1.23 \pm \
0.01 $ &  $ 1.27 \pm 0.01 $ &  $ 1.02 \pm 0.01 $ &  $ 1.00 \pm \
0.01 $ \\ 
\MBonce &  $ 1.56 \pm 0.02 $ &  $ 1.54 \pm 0.02 $ &  $ 1.09 \pm \
0.01 $ &  $ 1.20 \pm 0.01 $ &  $ 1.53 \pm 0.03 $ &  $ 1.24 \pm \
0.01 $ \\  
{\bf Combined}  &  $ 1.33 \pm 0.11 $ &  $ 1.39 \pm 0.09 $ &  $ 1.26 \pm 0.11 $ &  $ 1.34 \pm 0.11 $ &  $ 1.5 \pm 0.2 $ &  $ 1.09 \pm 0.07 $ \\ 
\MBveinte &  $ 1.26 \pm 0.01 $ &  $ 1.08 \pm 0.01 $ &  $ 2.20 \pm \
0.02 $ &  $ 2.25 \pm 0.02 $ &  $ 1.42 \pm 0.01 $ &  $ 1.29 \pm \
0.01 $ \\ 
\MBdiez &  $ 2.30 \pm 0.02 $ &  $ 2.38 \pm 0.04 $ &  $ 1.13 \pm 
0.01 $ &  $ 1.15 \pm 0.01 $ &  $ 2.77 \pm 0.07 $ &  $ 0.42 
\pm 0.01 $ \\
\hline
\multicolumn{7}{c}{Inclined orientation ($66^{\circ}$ with respect to the face-on orientation) }\\
\hline
\MBinj  &  $ 1.22 \pm 0.01 $ &  $ 1.27 \pm 0.01 $ &  $ 1.28 \pm \
0.01 $ &  $ 1.34 \pm 0.01 $ &  $ 1.66 \pm 0.04 $ &  $ 0.87 \
\pm 0.01 $ \\ 
\MBdoce &  $ 1.12 \pm 0.01 $ &  $ 1.87 \pm 0.02 $ &  $ 1.05 \pm \
0.01 $ &  $ 1.10 \pm 0.01 $ &  $ 0.86 \pm 0.01 $ &  $ 0.93 \
\pm 0.01 $ \\
\MBonce &  $ 1.15 \pm 0.01 $ &  $ 1.09 \pm 0.01 $ &  $ 1.47 \pm \
0.01 $ &  $ 1.65 \pm 0.01 $ &  $ 1.00 \pm 0.01 $ &  $ 0.40 \pm \
0.01 $ \\ 
{\bf Combined} &  $ 1.16 \pm 0.03 $ &  $ 1.41 \pm 0.2 $ &  $ 1.3 \pm \
0.1 $ &  $ 1.4 \pm 0.2 $ &  $ 1.2 \pm 0.2 $ &  $ 0.7 \pm \
0.2 $ \\ 
\MBveinte &  $ 1.39 \pm 0.02 $ &  $ 1.77 \pm 0.04 $ &  $ 4.28 \pm \
0.02 $ &  $ 4.56 \pm 0.03 $ &  $ 1.05 \pm 0.01 $ &  $ 1.23 \pm \
0.01 $ \\  
\MBdiez  &  $ 2.05 \pm 0.03 $ &  $ 4.5 \pm 0.1 $ &  $ 1.26 \pm \
0.01 $ &  $ 1.41 \pm 0.01 $ &  $ 2.51 \pm 0.04 $ &  $ 0.41 \
\pm 0.01 $  
\enddata
\tablenotetext{}{Notes. 
Column (1): model name.
Column (2): average ratio between the CNM surface density and FIR total intensity.
Column (3): average ratio between the CNM surface density and FIR polarized intensity,.
Column (4): average ratio between the neutral gas surface density and radio total intensity.
Column (5): average ratio between the neutral gas surface density and radio-polarized intensity.
Column (6): average ratio between the ongoing SFR over the last 10 Myr and FIR total intensity.
Column (7): average ratio between the ongoing SFR over the last 100 Myr and radio total intensity.
We include a {\bf Combined} scale height estimated as the average of the models \MBinj, \MBdoce~and \MBonce~between 0.3 and 0.66 of the total integrated distribution, excluding the edge cases \MBveinte~and \MBdiez. For completeness, we list but do not show additional quantities studied that did not display any significant correlations: WIM surface density, HM surface density, volume, CNM volume, WNM volume, CNM+WNM volume, WIM volume, hot medium volume, as well as magnetic energy separated into the energy contained in the CNM, WNM, CNM+WNM, WIM, and hot medium phases.}
\end{deluxetable*}

\subsection{The Effects of Inclination on the Vertical Distribution of FIR and Radio Emission}
\label{section:InclinationEffects}

\begin{figure*}[ht!]
\includegraphics[width=0.9\textwidth]{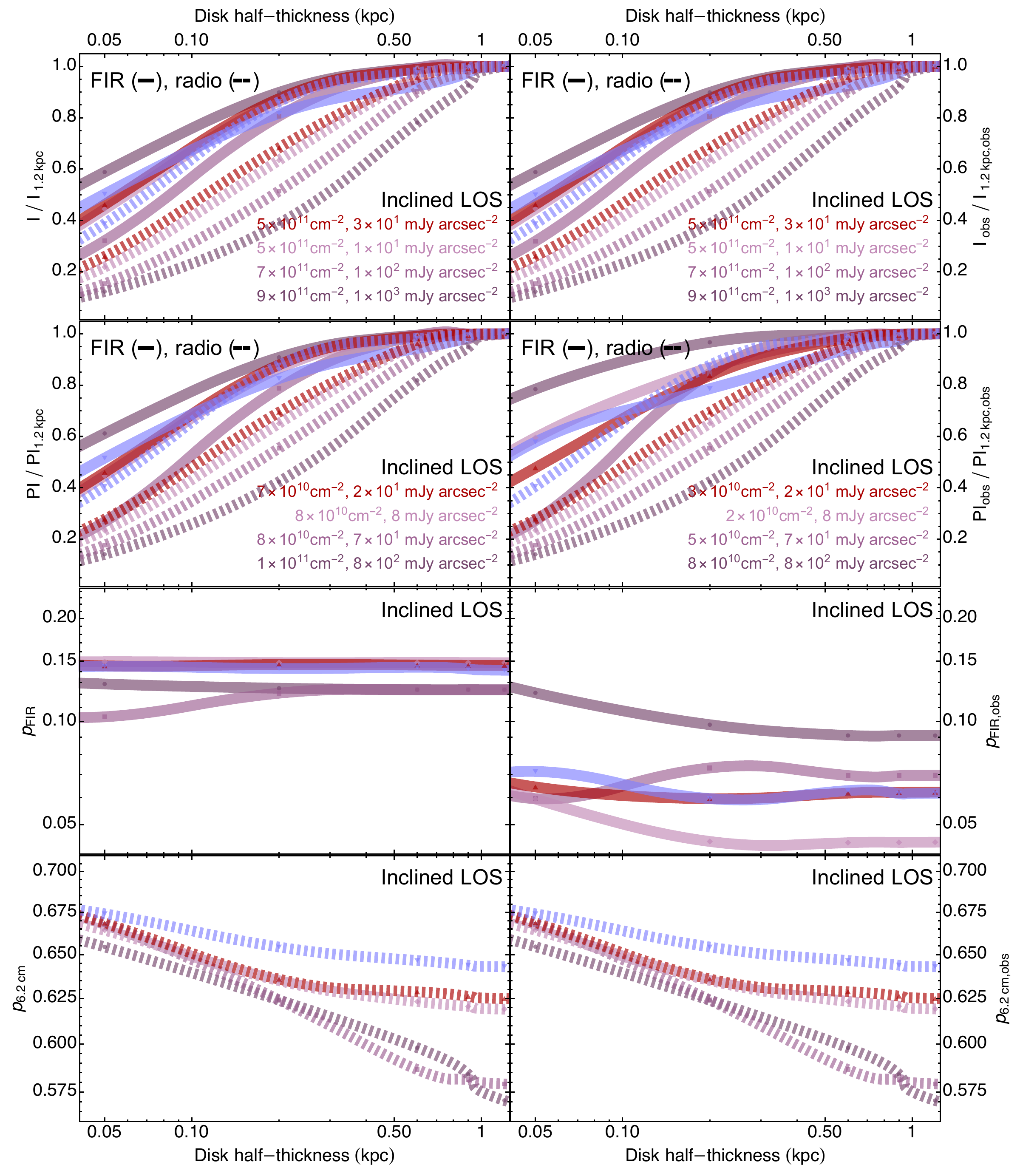}\\
\begin{center}    
\vspace{-0.8cm}
\includegraphics[width=0.4\textwidth]{rowlegend.pdf}\\
\vspace{-0.9cm}
\end{center}
\caption{Same as Figure~\ref{fig:HeightScale}, now for an inclined LOS (66$^{\circ}$ inclination with respect to the face-on orientation). Total intensities are unaffected by inclination effects. However, more considerable depolarization is found, especially for the FIR emission, when accounting for a telescope-like configuration.}
\label{fig:HeightScaleInclined}
\end{figure*}

The reviewed height scalings and their interrelations were studied under the most straightforward scenario of a fully face-on orientation for the observed galaxy. To generalize our results to the broader context of observations, we review whether they hold for inclined observations. Throughout this section, we assume a LOS inclined $66^{\circ}$ with respect to the face-on orientation. An intermediate inclination of $45^{\circ}$ was also studied, with all the studied scalings and correlations closely resembling our face-on orientation case. This considerable inclination of $66^{\circ}$ is selected with the objective of reviewing whether our results hold when including a non-negligible ISM density column in each LOS pixel while avoiding a fully edge-on LOS. As our objective is to determine relations with respect to altitude over the midplane of the galaxy, we continue to modify the integrated thickness along the direction perpendicular to the disk plane, as before. In Figure~\ref{fig:HeightScaleInclined}, we repeat the same comparison shown in Figure~\ref{fig:HeightScale} for the relative contribution to the integrated intensities as a function of disk thickness, now for the inclined LOS. The half-intensity scales and scale ratios for our inclined observations of each emission are summarized in Table~\ref{table:scale_summary}.

As expected from simply LOS-integrated quantities, the total intensities are approximately unchanged. The measured scale heights of the polarized intensities increase with inclination, featuring a more prominent displacement of the profiles toward higher half-thickness for the telescope-like synthetic observations (Figure~\ref{fig:HeightScaleInclined}, right column). This effect is found for both the radio and FIR emission, with a larger relative increase for the synchrotron emission. Polarized fraction maps (Figures~\ref{fig:MB11_slice_telescope} and \ref{fig:MBinj_slice_telescope}) reveal LOS depolarization effects with an overall decrease in the polarization fraction in FIR from $\sim0.18$ to $\sim0.14$ (Figure~\ref{fig:HeightScaleInclined}). A flat profile with increasing disk thickness in the full-resolution case indicates that the intrinsic depolarization occurs close to the disk plane. However, the more significant variations in the telescope-like observations show how resolution-limited observations will include important beam depolarization effects with varying contributions along the LOS. While inclination depolarization is also found for the radio emission, its reduction of approximately $\sim0.02$ represents a small proportional decrease. Radio observations appear less sensitive to both inclination and beam depolarization, likely due to synchrotron emission emerging from larger galactic scales and consequently probing more ordered magnetic fields. 

\begin{figure*}[ht!]
\includegraphics[width=\textwidth]{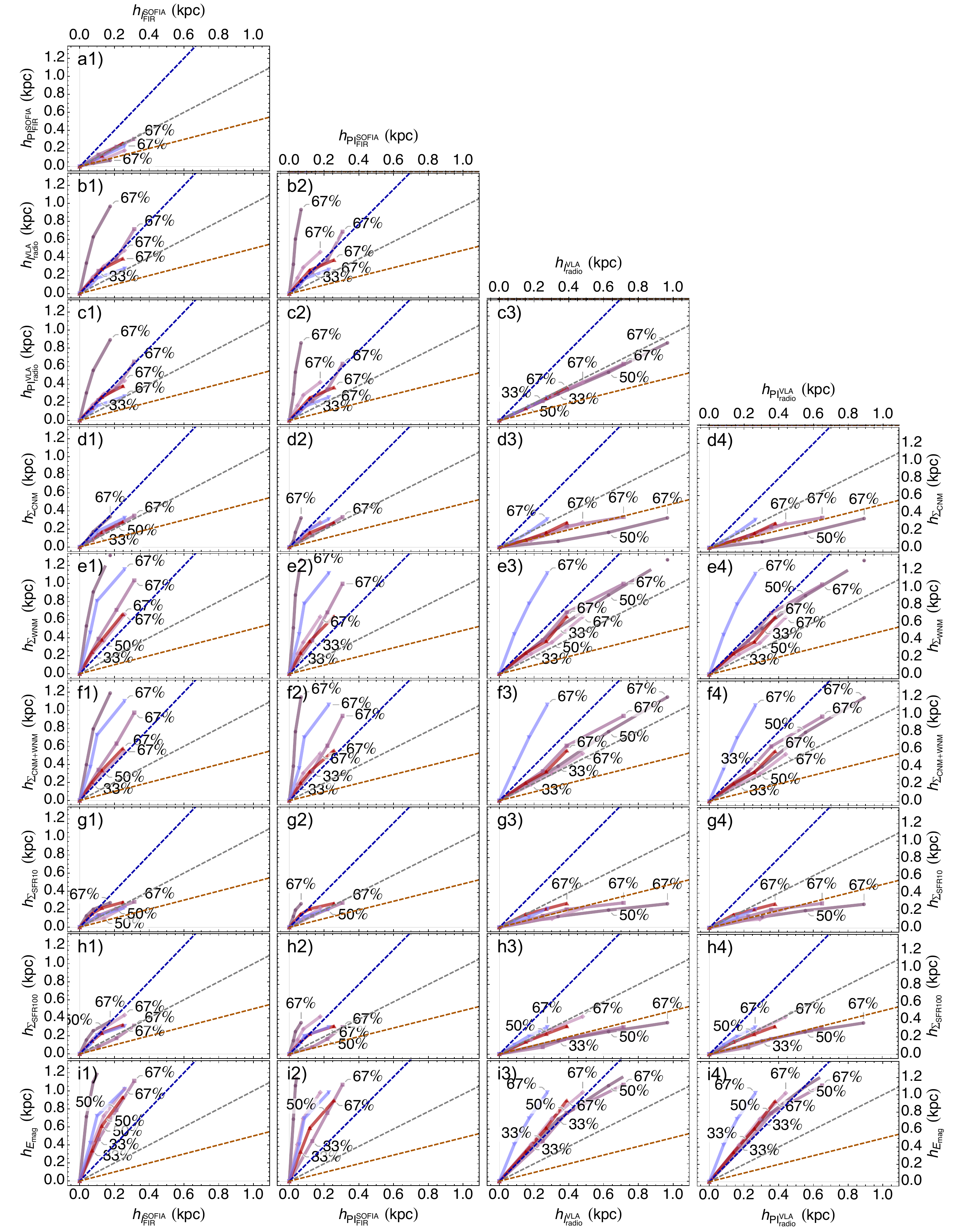}\\
\begin{center}    
\vspace{-0.8cm}
\includegraphics[width=0.4\textwidth]{rowlegend.pdf}\\
\vspace{-0.9cm}
\end{center}
\caption{Same as Figures~\ref{fig:FIRvsRadio_triangle} and \ref{fig:emission_vs_phase}, now for an inclined LOS. We include the one-to-one (gray), two-to-one (blue), and half-to-one (orange) scaling relations to aid visual inspection.}
\label{fig:triangle_inclined}
\end{figure*}

Finally, we review the main correlations studied on our face-on configuration for the inclined observations in Figure~\ref{fig:triangle_inclined}. Here, we compare the telescope-like intensities with each other in the upper triangle, and the correlations between emission, magnetic energy, star formation, and phases in the bottom $6 \times 4$ set of panels in Figure~\ref{fig:triangle_inclined}. As before, we include dashed lines for the one-to-one (gray), two-to-one (blue) and half-to-one (orange) scalings between the $y-$ and $x$-axis variables. 

Polarized intensities now have more concentrated scale heights than their total intensity counterparts (Figure~\ref{fig:triangle_inclined}, panels (a1), (c3)). This reveals that observed polarizations are likely fixed at the disk midplane and undergo depolarization as they traverse an increasing column of the ISM. Depolarization effects for radio intensities remain small, and as shown in Figures~\ref{fig:MB11_slice_telescope} and \ref{fig:MBinj_slice_telescope}, are limited to localized regions of the galaxy. In our inclined observations, we continue to find the synchrotron emission to probe approximately double the scale of the FIR (Figure~\ref{fig:triangle_inclined}, panel (b1)), with the ratio becoming slightly higher for polarized intensities (Figure~\ref{fig:triangle_inclined}, panel (c2)), due to the scale reduction of the polarized FIR, i.e., $\frac{\hIradio}{\hIFIR} = 2.04\pm0.11$, $\frac{\hPIradio}{\hPIFIR} = 2.48\pm0.59$. The aforementioned correlations with different phases and quantities remain approximately unchanged, as shown in the remaining panels. 

This reinforces our conclusions for the FIR as a probe of small-scale magnetic fields in the cold dense gas typical at altitudes $\lesssim 0.2\,\kpc$, whereas the radio provides information about large-scale magnetic fields typically entrained in the neutral gas of the galaxy and within a thicker disk of scale height $\sim 0.4\,\kpc$.

\section{Conclusions} 
\label{s:Conclusions}

We have studied the different scales and ISM phases probed by FIR and radio synchrotron polarimetric intensities employing our high-resolution cosmological MHD simulations of a Milky Way-like galaxy. We generated our simulations employing our own modified version of the {\sc ramses} code \citep{Teyssier2002}, in its CT MHD implementation \citep{Fromang2006,Teyssier2006} featuring numerical precision divergence-less magnetic fields. Our suite of simulations spans different magnetization levels and mechanisms, which reflect the resilience of our results toward magnetic field variations in galaxies, only showing some deviations for the edge cases (negligible magnetization \MBveinte, and extreme magnetization \MBdiez).

We generate our synthetic observations through geometrical approximations for the FIR and with a modified version of the {\sc polaris} code \citep{Reissl2019} for the synchrotron emission in radio. We explore the approximate appearance of these two emissions and quantitatively determine their scale heights above the disk midplane. Our main results can be summarized as follows:
\begin{itemize}
    \item Overall, the appearance of our synthetic observations in both FIR and radio emission closely resembles observations. For instance, we observe striking similarities between \MBdoce, \MBonce, and \MBinj~with NGC~6946, M51, and M83, respectively. Furthermore, we identify structures in the polarized radio emission maps that resemble the observed magnetic arms of NGC~6946 (see Section \ref{ss:GlobalReview}).
    \item Synchrotron emission is more extended and pervasive than FIR emission. The former covers virtually the entirety of the galaxy surface and disk thickness, whereas the FIR emerges from a number of resolved dusty filaments and clumps, unresolved in the telescope-like observation. While both radio and FIR polarization trace the large-scale magnetic field of the galaxy, the FIR deviates locally, due to its higher sensitivity to the small-scale magnetic field structure.
    \item In addition to a more extended surface coverage in the face-on observations, the synchrotron emission emerges from larger disk thicknesses of $\sim 0.4\,\kpc$ doubling those of the FIR emission ($\lesssim 0.2\,\kpc$), respectively. This supports the consideration of the FIR emission as a density probe, while the radio can be comparatively interpreted as a volume-filling mass probe.
    \item  We find FIR to trace the CNM, whereas synchrotron emission scales with neutral gas (warm and cold neutral phases) column density. Both exhibit some degree of positive correlation with star formation, with the FIR emission a better match for short timescales ($\sim 10$ Myr) and the radio for longer ones ($\sim 100$ Myr).
    \item Radio emission correlates spatially with regions of high magnetic energy. However, we find magnetic energy to extend toward higher altitudes with a vertical profile featuring a scale height approximately double to that of the synchrotron emission, and extending into the halo. We attribute this deviation from the magnetic energy profile to a comparatively steeper vertical profile of the CR electrons and dust number densities modulating the synchrotron and dust emission, as well as a poloidal magnetic field configuration.
    \item The polarized radio intensity experiences depolarization with increasing disk thickness, but only in localized regions trailing magnetic arms. On the other hand, FIR emission is rapidly depolarized within thin disk scales. Radio emission is relatively insensitive to telescope beam depolarization (at scales of $\sim 300$ pc) whereas this is the dominant source of depolarization for the FIR. This is in agreement with the suggested large galactic scale nature of the synchrotron emission and a smaller scale for the FIR emission, possibly below the resolutions of our studied simulations.
    \item All our results are approximately unchanged for inclined LOS observations, with the most notable effect being a greater depolarization of the FIR emission.
\end{itemize}

Overall, these results reveal the complementarity of FIR and radio polarimetric observations, with the FIR emission tracing small-scale magnetic fields in the cold phase, closer to the midplane of the disk, and sensitive to ongoing star formation ($\sim$ 10 Myr). We independently confirm that synchrotron emission traces the neutral gas (warm and cold neutral phases; \citealt{Ponnada2023a}) column density. We also find that radio emission provides information about large-scale magnetic fields, more ordered on galactic scales, corresponding with regions of recent star formation ($\sim 100$ Myr), and pervasive across the surface of the galaxy.

Through the inclusion of additional components, such as radiative transfer to self-consistently model the ionization of the neutral gas or CRs to better capture galactic outflows and modeling of synchrotron emission \citep{Werhahn2021c, Pfrommer2022, Martin-Alvarez2023, Ponnada2023a}, galaxy formation simulations will feature (pending caveats to the accuracy of the simulations and their models) sophisticated and self-consistent modeling for the intrinsic physical quantities of galaxies \citep[e.g.][]{Werhahn2021a,Hopkins2022} as well as their synthetic observations across the electromagnetic spectrum. Our study shows the potential for MHD galaxy formation simulations to reproduce radio and FIR observations and their observables, as we delve into the Square Kilometre Array era and the next generation of FIR observatories such as the Probe far-Infrared Mission for Astrophysics \citep[PI: J. Glenn; e.g.,][]{Moullet2023}. 

\section{Acknowledgments}
We kindly thank the referee for insightful comments and suggestions that improved the quality of this paper. This work is part of the Horizon-UK project, which used the DiRAC Complexity system, operated by the University of Leicester IT Services, which forms part of the STFC DiRAC HPC Facility (\href{www.dirac.ac.uk}{www.dirac.ac.uk}). This equipment is funded by BIS National E-Infrastructure capital grant ST/K000373/1 and STFC DiRAC Operations grant ST/K0003259/1. The equipment was funded by BEIS capital funding via STFC capital grants ST/K000373/1 and ST/R002363/1 and STFC DiRAC Operations grant ST/R001014/1. DiRAC is part of the National e-Infrastructure. The authors would like to acknowledge the use of the University of Oxford Advanced Research Computing (ARC) facility in carrying out this work. \href{http://dx.doi.org/10.5281/zenodo.22558}{http://dx.doi.org/10.5281/zenodo.22558}. This work of S.M.A. and E.L.-R. is supported by the NASA/DLR Stratospheric Observatory for Infrared Astronomy (SOFIA) under the 08\_0012 Program.  SOFIA is jointly operated by the Universities Space Research Association, Inc. (USRA), under NASA contract NNA17BF53C, and the Deutsches SOFIA Institut (DSI) under DLR contract 50OK0901 to the University of Stuttgart. S.M.A. also acknowledges support from the Kavli Institute for Particle Astrophysics and Cosmology
(KIPAC) Fellowship. I.M.C acknowledges support from the Spanish Ministry of Universities ref. UNI/551/2021-May 26 (EU Next Generation funds), and from ANID program FONDECYT Postdoctorado 3230653. A.B. was supported by an appointment to the NASA Postdoctoral Program at the NASA Ames Research Center, administered by Oak Ridge Associated Universities under contract with NASA. A.B. is supported by the program Hubble Archival Research project AR 17041, and Chandra Archival Research project ID No. 24610329, provided by NASA through a grant from the Space Telescope Science Institute and the Center for Astrophysics Harvard \& Smithsonian, operated by the Association of Universities for Research in Astronomy, Inc., under NASA contract NAS 5-03127.
M.J.J. acknowledges support from the STFC consolidated grant [ST/W000903/1], a UKRI Frontiers Research Grant [EP/X026639/1], which was selected by the European Research Council (ERC) and generous support from the Hintze Family Charitable Foundation through the Oxford Hintze Centre for Astrophysical Surveys.
K.T. acknowledges support from the ERC (grant agreement No. 771282), and from
the Foundation of Research and Technology - Hellas Synergy Grants Program through project POLAR, jointly implemented by the Institute of Astrophysics and the Institute of Computer Science.
E.N. is funded by the "ENIGMA" PNRR project under the "Young Researchers" call of the Italian Ministry of Universities and Research.

%





\appendix

\section{The Effect of Alternative Dust Models on the Synthetic FIR Emission}
\label{ap:dust_models}
In the absence of on-the-fly dust modeling, a post-processing prescription is required to estimate a dust number density for each cell in our simulations. In this appendix, we review three different dust models, showing how they all yield approximately unchanged results for the quantities explored in this work. The three dust models studied are

\noindent{\bf Model 1:} fixed metal-to-dust ratio with high-temperature fading
\begin{equation}
\ndust = \frac{\rho_\text{gas}\, Z}{m_\text{dust}} \, \eta_\text{D/M} \, f_\text{cut} (T),
\label{eq:dust1}
\end{equation}
\noindent{\bf Model 2:} fixed metal-to-dust ratio with no temperature cut
\begin{equation}
\ndust =  \frac{\rho_\text{gas}\, Z}{m_\text{dust}} \, \eta_\text{D/M},
\label{eq:dust2}
\end{equation}
\noindent{\bf Model 3:} ionization-dependent metal-to-dust ratio \citep[based on][]{Laursen2009} 
\begin{equation}
\ndust = \frac{\rho_\text{gas}\, Z\, }{m_\text{dust}} \, \eta_\text{D/M}
\left(x_\text{HI} + f_\text{ion}\,x_\text{HII}\right).
\label{eq:dust3}
\end{equation}
In these expressions, $\rho_\text{gas}$ is the gas density, $Z$ is the gas metallicity, $\eta_\text{D/M}=0.4$ represents a constant ratio of dust-to metal mass \citep{Draine2007b}, and $m_\text{dust}$ is our assumed typical dust mass $m_\text{dust} = 1.26\times10^{-14}\,\g$ (see the main text). Model 1 includes a smooth gas temperature cut above $T > 1500\,\K$ following the functional form $f_\text{cut} (T) = \text{min}\left(1.0, \, \exp\left[1 - T / (1500\, \K)\right] \right)$. Model 2 reviews the effect of removing this temperature cut, and model 3 explores instead an ionization-dependent model as proposed by \citet{Laursen2009}, with $x_\text{HI}$ and $x_\text{HII}$ corresponding to the neutral and ionized hydrogen mass fractions. $f_\text{ion}$ represents the constant fraction of the entrained metal mass in ionized hydrogen that is assumed to be in the form of dust $f_\text{ion} = 0.01$.

\begin{figure}[ht!]
\includegraphics[width=0.72\textwidth]{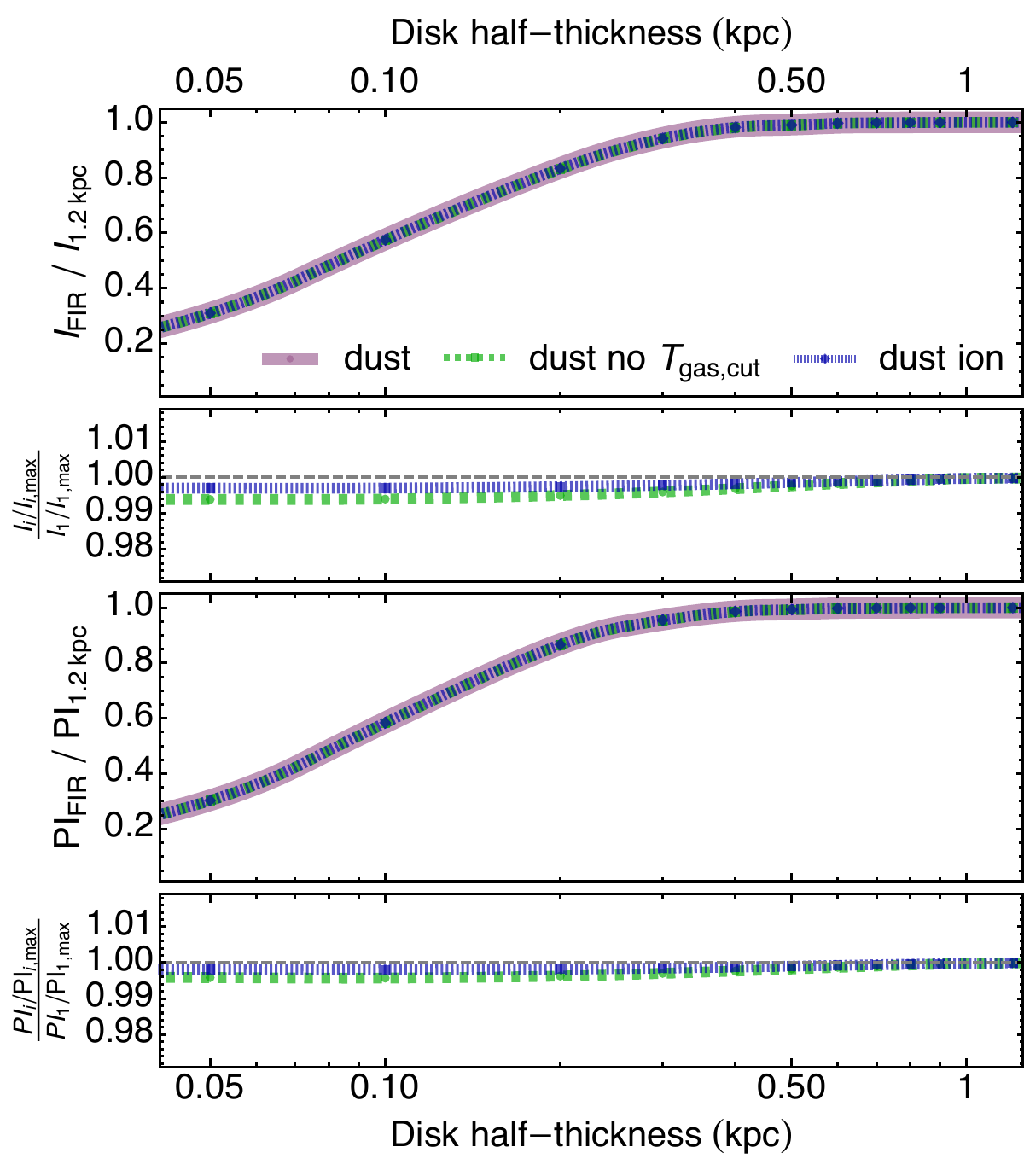}%
\includegraphics[width=0.28\textwidth]{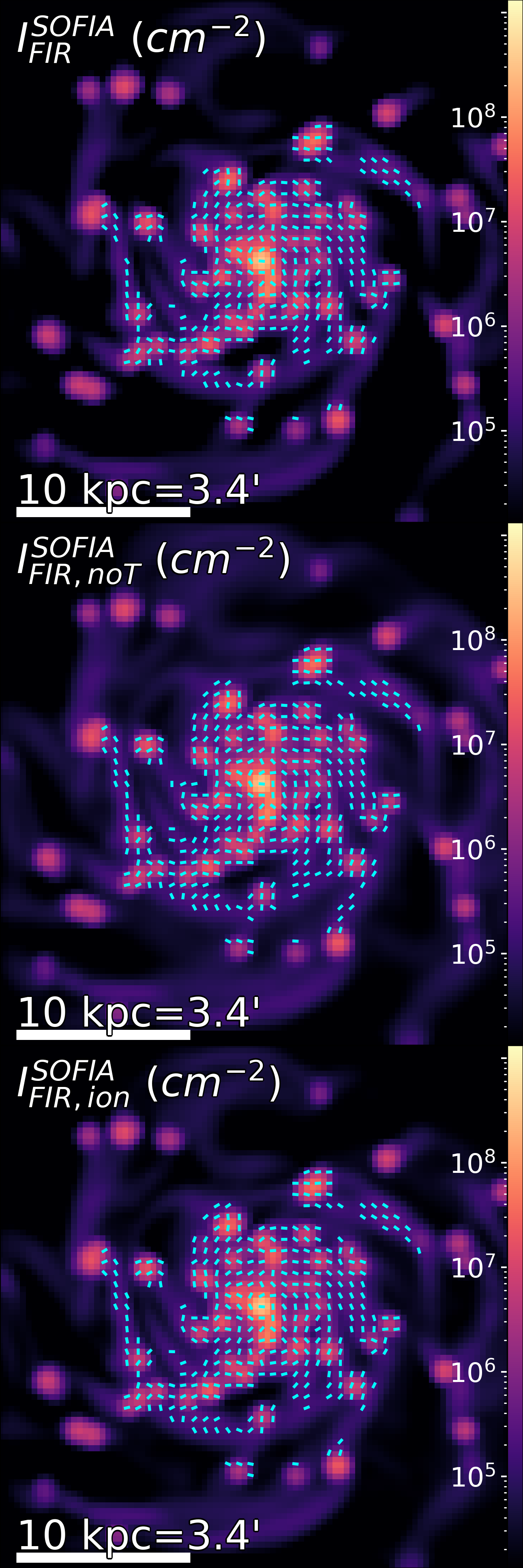}\\
\caption{Comparison of the three studied dust models for the \MBonce~model. ({\bf Left column}) Relative contribution as a function of galactic altitude to the integrated total intensity (top row) and integrated polarized intensity (third row). The second and fourth rows show the ratio of relative contributions between models 2 and 3 with respect to the model 1 reference, for the total and polarized emission, respectively. ({\bf Right column}) SOFIA-like FIR geometric magnetic field observation, with polarization measurements overlaid as cyan quivers rotated by
90$^{\circ}$ to match the inferred orientation of the magnetic field. From top to bottom, the panels display models 1, 2, and 3. Overall, varying dust models do not significantly affect the relative contribution as a function of disk half-thickness nor the appearance of our synthetic observations.}
\label{fig:three_dust}
\end{figure}

We compare these three models in Figure~\ref{fig:three_dust}. The left column of Figure~\ref{fig:three_dust} shows the relative intensity contribution as a function of galactic altitude for the \MBonce~model, both for the total (first row) and the polarized (third row) FIR emission. All three dust models display approximately equal scaling with increasing disk altitude. We also show the total (second row) and polarized (fourth row) emission ratios of model 2 and model 3 with respect to our reference dust calculation (i.e., model 1). The approximate equivalence is unsurprising: the main deviation between the three models is the fraction of the total mass assumed to be dust in the low-density regime. Deviations between models 1 and 2 are dependent on the distribution of high gas temperatures (through $f_\text{cut}$), Model 3 is primarily dependent on the distribution of high gas temperatures at low gas densities (reducing $x_\text{HI} / x_\text{HII}$). However, these variations are subdominant when compared to the effect of higher gas densities and gas metallicity, which will generally correspond to regions of low temperature and low ionization. For qualitative comparison, in the right column of Figure~\ref{fig:three_dust}, we show projections for each of the three dust models, which are approximately unchanged. Variations are only notable in regions of low $I_\text{FIR}$, which contribute insignificantly to the integrated emission budget. This further illustrates how our results are robust with respect to simple post-processing dust models. We reserve more detailed dust modeling for future work.

\section{The Influence of CR Electron Variation in the Synchrotron Emission Extension}
\label{ap:CR_models}
\begin{figure}[ht!]
\includegraphics[width=0.72\textwidth]{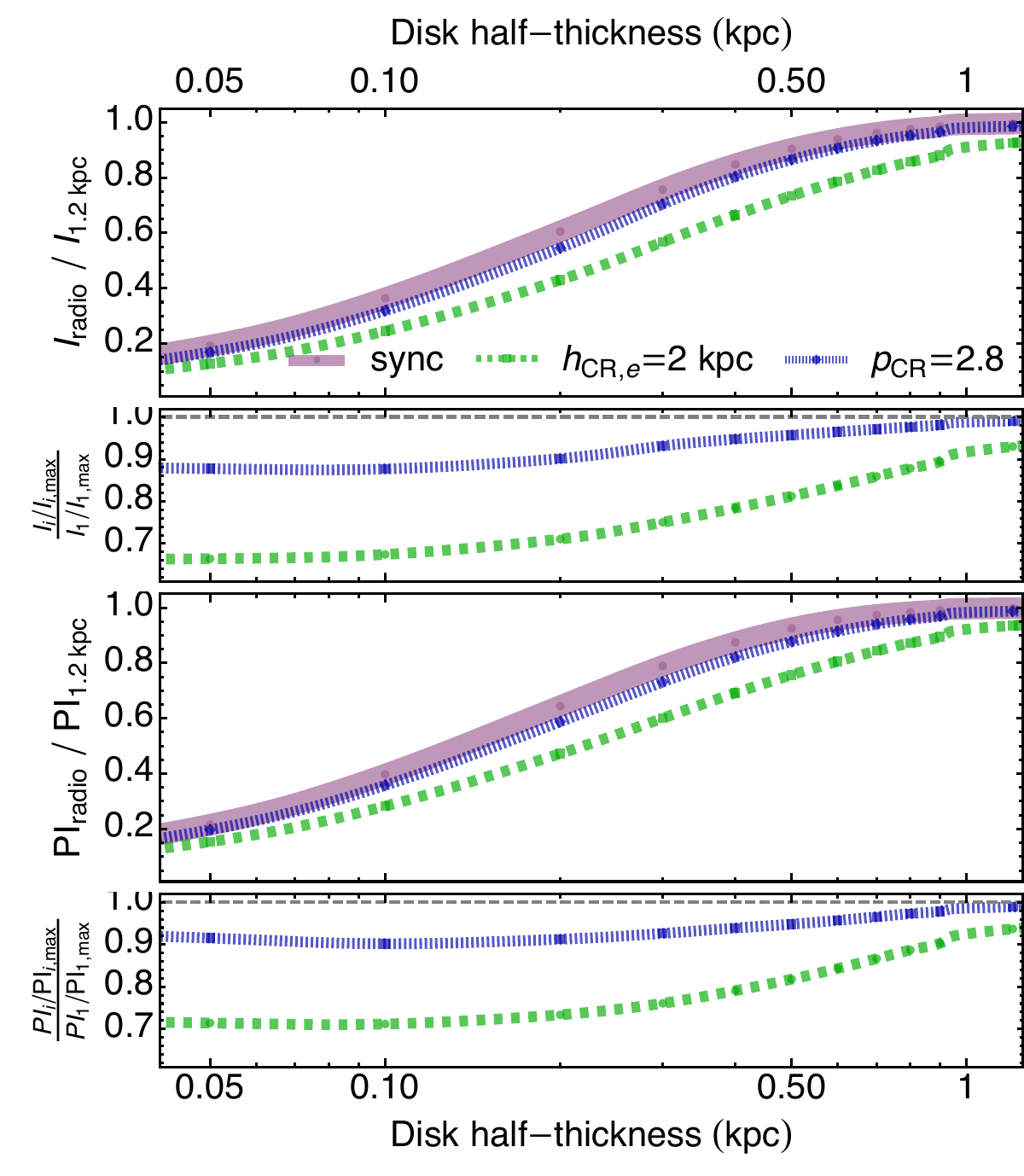}%
\includegraphics[width=0.28\textwidth]{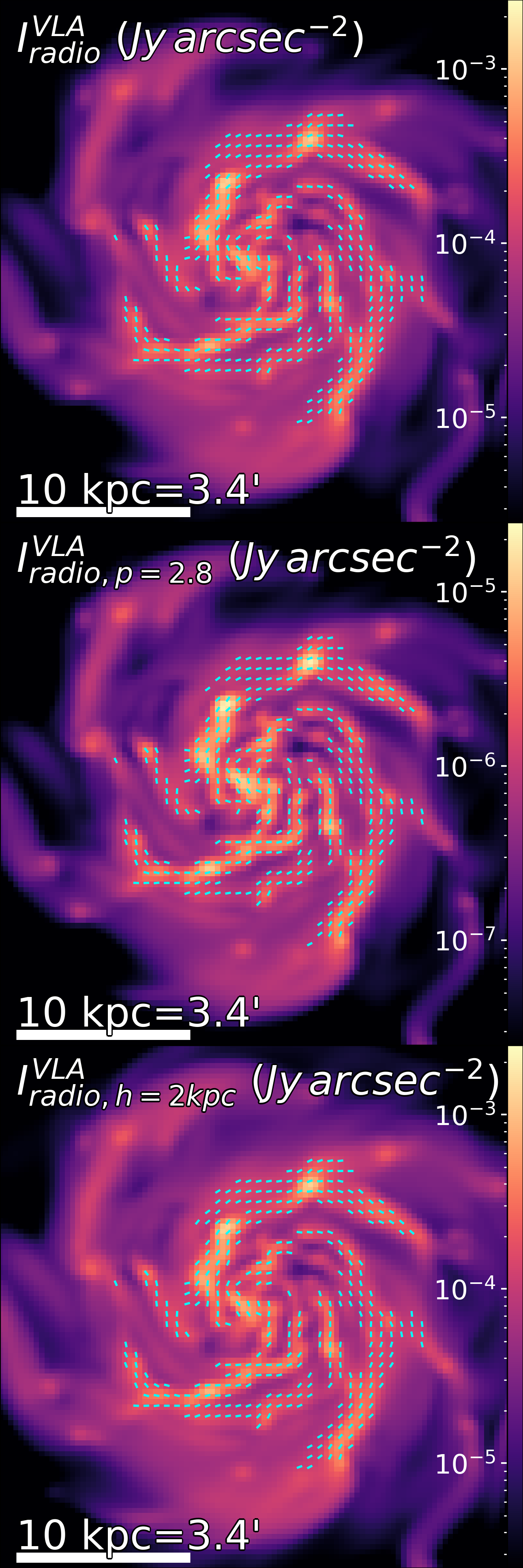}\\
\caption{\label{fig:CR_models} Comparison of different CR $e^-$ configurations for the synchrotron emission of the \MBonce~model. ({\bf Left column}) Relative contribution as a function of galactic altitude to the integrated total intensity (top row) and integrated polarized intensity (third row). The second (total intensity) and fourth (polarized intensity)
rows show the ratio of relative contributions for the extended thickness CR $e^-$ disk and the steeper CR $e^-$ spectrum models  with respect to our fiducial CR $e^-$ distribution. ({\bf Right column}) VLA-like synchrotron observations with inferred plane-of-the-sky magnetic field overlayed as cyan quivers. From top to bottom, the panels display the fiducial CR $e^-$ configuration, a sharper $\pCR = 2.8$ configuration, and a CR $e^-$ configuration with $\hgal = 2\,\kpc$. Exploring large deviations from the expected CR $e^-$ configuration only leads to intensities more extended with altitude, and approximately unchanged maps of the observations.}
\end{figure}

Our CR electron configuration assumes a post-processing, analytic, and smooth distribution (see Equation~\ref{eq:nCRdensity}), based on observations of these energetic particles in the Milky Way \citep[e.g.,][]{Webber1998,Sun2008}. While the distribution employed in the main study represents our best estimate for the properties of CRs in such disk galaxies, there are two possible main variations of this model that may affect our results. We review their general effects in Figure~\ref{fig:CR_models}, and provide their resulting scale heights in Table~\ref{table:CR_models}. The first is our selection of the CR energy spectrum index $\pCR = 2.3$. This shallower intrinsic spectrum is expected in galaxies with a younger CR population. Nonetheless, observations frequently observe synchrotron spectral indices closer to $\alpha \sim 1$ (corresponding to $\pCR \lesssim 3.0$; 
\citealt{Heesen2014,Tabatabaei2017}). While the observed spectral index does not correspond to the intrinsic index of the underlying CR population, for the \MBonce~simulation, we review the influence of a steeper spectrum with $\pCR = 2.8$ in Figure~\ref{fig:CR_models}. The relative contribution per disk half-thickness is somewhat more extended than in our fiducial case, with a half-intensity height increase factor $1.14$. The synthetic observation map for $\pCR = 2.8$ has an appearance extremely similar to our fiducial case, but we note a dramatic reduction of the resulting intensity (of $\sim 2$ dex). Finally, edge-on synchrotron observations from disk galaxies detect their total radio emission at altitudes of $h \gtrsim 1\,\kpc$. \citet{Krause2018} quote scale heights of $\left(1.1 \pm 0.3\right)\,\kpc$ and $\left(1.4 \pm 0.7\right)\,\kpc$ for the scale heights of total radio intensity in their $C$-band and $L$-band, respectively. These appear to correspond to simple averages of the scale height across their studied population of galaxies. When we incorporate their provided information for the associated uncertainty in the average calculation, we retrieve $\left(0.91 \pm 0.15\right)\,\kpc$ and $\left(1.07 \pm 0.04\right)\,\kpc$. More importantly for this work, we also provide here the weighted median $\left(0.73 \pm 0.35\right)\,\kpc$ and $\left(1.06 \pm 0.08\right)\,\kpc$ (and simple median $\left(1.08 \pm 0.42\right)\,\kpc$ and $\left(1.09 \pm 0.45\right)\,\kpc$) values for the $C$ and $L$-bands, respectively. Notably, total radio intensities include a contribution from thermal emission, and thus provide only lower limits for the synchrotron scale heights. Overall, these observations suggest that the electron distribution should, for a magnetic field naturally expected to decrease above the midplane, extend beyond the altitude of $\hgal \gtrsim 1\,\kpc$. While the extension of the CR electrons above the midplane is bound to be affected by galaxy properties, we leave their exploration to future work. Here, we compare the best estimate for our galaxy of $\hgal = 1\,\kpc$ with a considerably larger $\hgal = 2\,\kpc$ in Figure~\ref{fig:CR_models}. Assuming $\hgal = 2\,\kpc$ leads to a more extended distribution of synchrotron intensities that still tapers off at $\sim 1\,\kpc$, the resulting half-intensity height is increased by a factor of $1.6$. As for the previous case, the synthetic observation map has a remarkably similar appearance, and displays somewhat higher intensities. Overall, we emphasize here that uncertainty associated with the best estimates for the CR $e^{-}$ configuration employed is smaller than the variations explored in this appendix, and that the two variations explored here both lead to somewhat higher scales and a more extended distribution of the emission, further reinforcing our central conclusion of the radio exploring larger scales than the FIR emission.

Nonetheless, our employed configuration relies on a simple analytical model that assumes a smooth distribution of CR electrons across the simulated galaxies. More sophisticated models will account for further details of CR physics, where some simulations are now self-consistently modeling the evolution of CR energy density \citep[e.g.][]{Dubois2016, Pfrommer2017b, Hopkins2020}, or even a direct self-consistent model of the evolution of the CR electrons \citep{Hopkins2022}. Accounting for such CR physics will affect the resulting synchrotron emission \citep{Werhahn2021a,Werhahn2021c,Ponnada2023b}. Some of these effects are the distribution of CR electrons modified by the intrinsic MHD properties, CR energy losses through cooling processes, CR acceleration or propagation according to the local magnetic field structure. Accounting for some of these, \citet{Werhahn2021a} find some small-scale substructures for the distribution of CR energy density. This would influence the full-resolution observations, and may be significant for the beam-smoothed telescope-like synthetic observations studied here. Similarly, the model by \citet{Werhahn2021a} appears to find, at times, larger CR energy densities at higher altitudes above the midplane than at the midplane for intermediate radii of the disk. This could lead to a more extended synchrotron emission with altitude. Analogously, \citet{Ponnada2023b} show how different CR transport physics may leave an imprint on radio synchrotron emission. Due to these considerations, the conclusions presented in this manuscript will benefit from further review through more sophisticated modeling of the configuration of CR electrons in the future.

\begin{deluxetable*}{lcccccc}
\centering
\tablecaption{Summary of the measured scale heights of the studied variations of the CR $e^-$ configuration. 
\label{table:CR_models} 
}
\tablecolumns{6}
\tablewidth{0pt}
\tablehead{\colhead{$\hIradio$} & \colhead{$\hPIradio$} & 
\colhead{$h_\text{I radio, 2.8}$} & \colhead{$h_\text{PI radio, 2.8}$} & 
\colhead{$h_\text{I radio, 2kpc}$} & \colhead{$h_\text{PI radio, 2kpc}$} \\ 
 \colhead{(kpc)} & \colhead{(kpc)} & \colhead{(kpc)} & \colhead{(kpc)} & \colhead{(kpc)} & \colhead{(kpc)}}
\startdata
\hline
0.31 &  0.28 &  0.35 &  0.32 &  0.50 &  0.44 \\ 
\enddata
\tablenotetext{}{Notes:
From left to right, each pair of columns displays the total and polarized intensities scale heights for 
(i) our fiducial CR configuration, 
(ii) a steeper $\pCR = 2.8$ spectrum, and 
(iii) a more extended CR disk with $\hgal = 2\,\kpc$.}
\end{deluxetable*}

\section{Telescope versus intrinsic scale height measurements}
\label{ap:TelescopeVsIntrinsic}

\begin{figure*}[ht!]
\includegraphics[width=\textwidth]{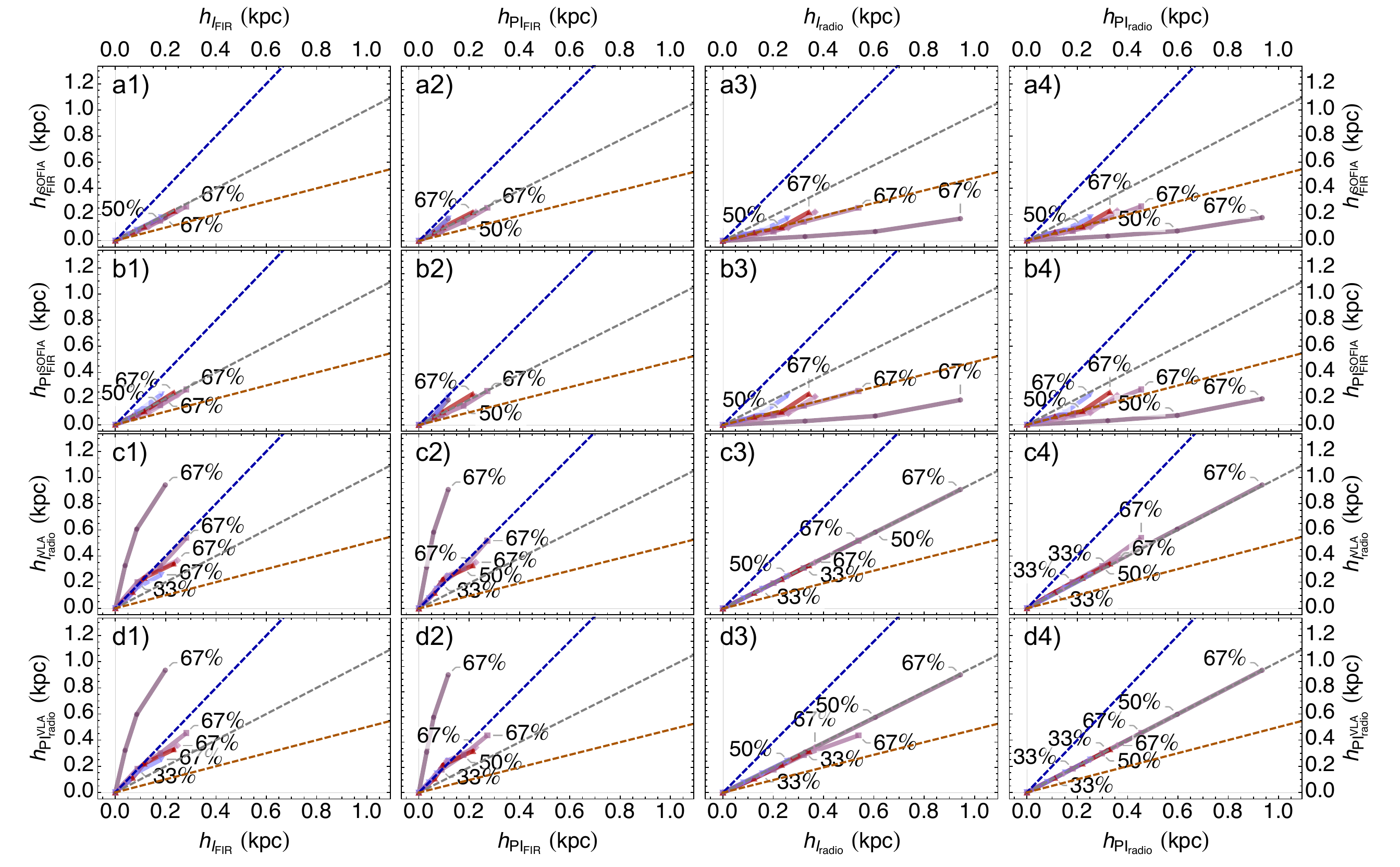}\\
\begin{center}    
\vspace{-0.8cm}
\includegraphics[width=0.4\textwidth]{rowlegend.pdf}\\
\vspace{-0.9cm}
\end{center}
\caption{Comparison of height scale for various combinations of observational quantities. From left to right (top to bottom), the columns (rows) correspond to the FIR total intensity, FIR polarized intensity, radio synchrotron total intensity, and radio synchrotron polarized intensity for their full-resolution (telescope-like) case. We include three dashed lines to aid visual inspection, indicating the one-to-one (gray), two-to-one (blue), and half-to-one (orange) relations. Telescope-like intensities scale approximately as their full-resolution equivalents and are used for our main analysis.}
\label{fig:TelescopeVsIntrinsic}
\end{figure*}

For the sake of clarity and brevity within the main text of this manuscript, we limit our comparison of altitude scales for the FIR versus radio emissions to their telescope-like observations. 

This is motivated by the approximately equal scaling of the telescope-like and full-resolution cases with altitude. This is illustrated by their direct comparison in Figure~\ref{fig:TelescopeVsIntrinsic}. This figure features, as in Figure~\ref{fig:FIRvsRadio_panel}, the scaling of two quantities for different percentages of their total integrated emission. Here, we show all the intrinsic measurements on the $x$-axis and the telescope-like ones on the $y$-axis. Overall, in the instrinsic versus telescope-like counterpart set of panels in Figure~\ref{fig:TelescopeVsIntrinsic} (main diagonal: panels (a1), (b2), (c3), (d4)), the relations are tighter for the radio emission, further reinforcing the idea that synchrotron in galaxies emerges from comparable scales, whereas the FIR is more sensitive to beam-related effects, particularly for its polarized intensity. All the remaining relative scalings explored in this work are preserved between intrinsic and telescope-like, but not explicitly displayed.


\bibliographystyle{aasjournal}
\bibliography{main.bib}

\begin{thebibliography}{}
\expandafter\ifx\csname natexlab\endcsname\relax\def\natexlab#1{#1}\fi
\providecommand{\url}[1]{\href{#1}{#1}}
\providecommand{\dodoi}[1]{doi:~\href{http://doi.org/#1}{\nolinkurl{#1}}}
\providecommand{\doeprint}[1]{\href{http://ascl.net/#1}{\nolinkurl{http://ascl.net/#1}}}
\providecommand{\doarXiv}[1]{\href{https://arxiv.org/abs/#1}{\nolinkurl{https://arxiv.org/abs/#1}}}

\bibitem[{Balsara \& Kim(2004)}]{Balsara2004}
Balsara, D.~S., \& Kim, J. 2004, The Astrophysical Journal, 602, 1079,
  \dodoi{10.1086/381051}

\bibitem[{Beck {et~al.}(2013)Beck, Dolag, Lesch, \& Kronberg}]{Beck2013a}
Beck, A.~M., Dolag, K., Lesch, H., \& Kronberg, P.~P. 2013, Monthly Notices of
  the Royal Astronomical Society, 435, 3575, \dodoi{10.1093/mnras/stt1549}

\bibitem[{Beck(2007)}]{Beck2007}
Beck, R. 2007, Astronomy and Astrophysics, 470, 539,
  \dodoi{10.1051/0004-6361:20066988}

\bibitem[{Beck(2015)}]{Beck2015}
---. 2015, Astronomy {\&} Astrophysics, 578, A93,
  \dodoi{10.1051/0004-6361/201425572}

\bibitem[{Beck {et~al.}(2019{\natexlab{a}})Beck, Berkhuijsen,
  Gie{\ss}{\"{u}}bel, \& Mulcahy}]{Beck2019a}
Beck, R., Berkhuijsen, E.~M., Gie{\ss}{\"{u}}bel, R., \& Mulcahy, D.~D.
  2019{\natexlab{a}}, \dodoi{10.1051/0004-6361/201936481}

\bibitem[{Beck {et~al.}(2019{\natexlab{b}})Beck, Chamandy, Elson, \&
  Blackman}]{Beck2019b}
Beck, R., Chamandy, L., Elson, E., \& Blackman, E.~G. 2019{\natexlab{b}},
  Galaxies, 8, 4, \dodoi{10.3390/GALAXIES8010004}

\bibitem[{Beck \& Wielebinski(2013)}]{BeckLive}
Beck, R., \& Wielebinski, R. 2013, 215,
  \dodoi{https://doi.org/10.48550/arXiv.1302.5663}

\bibitem[{Bernet {et~al.}(2008)Bernet, Miniati, Lilly, Kronberg, \&
  Dessauges-Zavadsky}]{Bernet2008}
Bernet, M.~L., Miniati, F., Lilly, S.~J., Kronberg, P.~P., \&
  Dessauges-Zavadsky, M. 2008, Nature, 454, 302, \dodoi{10.1038/nature07105}

\bibitem[{{Borlaff} {et~al.}(2023){Borlaff}, {Lopez-Rodriguez}, {Beck},
  {Clark}, {Ntormousi}, {Tassis}, {Martin-Alvarez}, {Tahani}, {Dale}, {del
  Moral-Castro}, {Roman-Duval}, {Marcum}, {Beckman}, {Subramanian},
  {Eftekharzadeh}, \& {Proudfit}}]{SALSAV}
{Borlaff}, A.~S., {Lopez-Rodriguez}, E., {Beck}, R., {et~al.} 2023, \apj, 952,
  4, \dodoi{10.3847/1538-4357/acd934}

\bibitem[{Brandenburg \& Subramanian(2005)}]{Brandenburg2005}
Brandenburg, A., \& Subramanian, K. 2005, {Astrophysical magnetic fields and
  nonlinear dynamo theory},  North-Holland,
  \dodoi{10.1016/j.physrep.2005.06.005}

\bibitem[{Bruzual \& Charlot(2003)}]{Bruzual2003}
Bruzual, G., \& Charlot, S. 2003, Monthly Notices of the Royal Astronomical
  Society, 344, 1000, \dodoi{10.1046/j.1365-8711.2003.06897.x}

\bibitem[{Buck {et~al.}(2020)Buck, Pfrommer, Pakmor, Grand, \&
  Springel}]{Buck2020}
Buck, T., Pfrommer, C., Pakmor, R., Grand, R.~J., \& Springel, V. 2020, Monthly
  Notices of the Royal Astronomical Society, 497, 1712,
  \dodoi{10.1093/mnras/staa1960}

\bibitem[{Buie {et~al.}(2022)Buie, Scannapieco, \& Mark~Voit}]{Buie2022}
Buie, E., Scannapieco, E., \& Mark~Voit, G. 2022, The Astrophysical Journal,
  927, 30, \dodoi{10.3847/1538-4357/ac4bc2}

\bibitem[{Butsky {et~al.}(2017)Butsky, Zrake, Kim, Yang, \& Abel}]{Butsky2017}
Butsky, I., Zrake, J., Kim, J.-h., Yang, H.-I., \& Abel, T. 2017, The
  Astrophysical Journal, 843, 113, \dodoi{10.3847/1538-4357/aa799f}

\bibitem[{{Chen} {et~al.}(2016){Chen}, {King}, \& {Li}}]{Chen2016}
{Chen}, C.-Y., {King}, P.~K., \& {Li}, Z.-Y. 2016, \apj, 829, 84,
  \dodoi{10.3847/0004-637X/829/2/84}

\bibitem[{Cottle {et~al.}(2020)Cottle, Scannapieco, Bruggen, Banda-Barragan, \&
  Federrath}]{Cottle2020}
Cottle, J., Scannapieco, E., Bruggen, M., Banda-Barragan, W., \& Federrath, C.
  2020.
\newblock \url{http://arxiv.org/abs/2002.07804}

\bibitem[{{Crutcher}(1999)}]{Crutcher1999}
{Crutcher}, R.~M. 1999, \apj, 520, 706, \dodoi{10.1086/307483}

\bibitem[{Crutcher(2012)}]{Crutcher2012}
Crutcher, R.~M. 2012, Annual Review of Astronomy and Astrophysics, 50, 29,
  \dodoi{10.1146/annurev-astro-081811-125514}

\bibitem[{Draine {et~al.}(2007)Draine, Dale, Bendo, Gordon, Smith, Armus,
  Engelbracht, Helou, Kennicutt, Li, Roussel, Walter, Calzetti, Moustakas,
  Murphy, Rieke, Bot, Hollenbach, Sheth, \& Teplitz}]{Draine2007b}
Draine, B.~T., Dale, D.~A., Bendo, G., {et~al.} 2007, The Astrophysical
  Journal, 663, 866, \dodoi{10.1086/518306}

\bibitem[{Dubois \& Commer{\c{c}}on(2016)}]{Dubois2016}
Dubois, Y., \& Commer{\c{c}}on, B. 2016, Astronomy {\&} Astrophysics, 585,
  A138, \dodoi{10.1051/0004-6361/201527126}

\bibitem[{Dubois \& Teyssier(2010)}]{Dubois2010}
Dubois, Y., \& Teyssier, R. 2010, Astronomy {\&} Astrophysics, 523, A72,
  \dodoi{10.1051/0004-6361/200913014}

\bibitem[{{Dubois} {et~al.}(2024){Dubois}, {Rodr{\'\i}guez Montero}, {Guerra},
  {Trebitsch}, {Han}, {Beckmann}, {Yi}, {Lewis}, \& {Jang}}]{Dubois2024}
{Dubois}, Y., {Rodr{\'\i}guez Montero}, F., {Guerra}, C., {et~al.} 2024, arXiv
  e-prints, arXiv:2402.18515, \dodoi{10.48550/arXiv.2402.18515}

\bibitem[{Dunkley {et~al.}(2009)Dunkley, Komatsu, Nolta, Spergel, Larson,
  Hinshaw, Page, Bennett, Gold, Jarosik, Weiland, Halpern, Hill, Kogut, Limon,
  Meyer, Tucker, Wollack, \& Wright}]{Dunkley2009}
Dunkley, J., Komatsu, E., Nolta, M.~R., {et~al.} 2009, Astrophysical Journal,
  Supplement Series, 180, 306, \dodoi{10.1088/0067-0049/180/2/306}

\bibitem[{Dwek(1998)}]{Dwek1998}
Dwek, E. 1998, The Astrophysical Journal, 501, 643, \dodoi{10.1086/305829}

\bibitem[{Federrath \& Klessen(2012)}]{Federrath2012}
Federrath, C., \& Klessen, R.~S. 2012, Astrophysical Journal, 761, 156,
  \dodoi{10.1088/0004-637X/761/2/156}

\bibitem[{Ferland {et~al.}(1998)Ferland, Korista, Verner, Ferguson, Kingdon, \&
  Verner}]{Ferland1998}
Ferland, G.~J., Korista, K.~T., Verner, D.~A., {et~al.} 1998, Publications of
  the Astronomical Society of the Pacific, 110, 761, \dodoi{10.1086/316190}

\bibitem[{{Fiege} \& {Pudritz}(2000)}]{FP2000}
{Fiege}, J.~D., \& {Pudritz}, R.~E. 2000, \apj, 544, 830,
  \dodoi{10.1086/317228}

\bibitem[{{Fletcher}(2010)}]{Fletcher2010}
{Fletcher}, A. 2010, in Astronomical Society of the Pacific Conference Series,
  Vol. 438, The Dynamic Interstellar Medium: A Celebration of the Canadian
  Galactic Plane Survey, ed. R.~{Kothes}, T.~L. {Landecker}, \& A.~G. {Willis},
  197, \dodoi{10.48550/arXiv.1104.2427}

\bibitem[{Fromang {et~al.}(2006)Fromang, Hennebelle, \& Teyssier}]{Fromang2006}
Fromang, S., Hennebelle, P., \& Teyssier, R. 2006, Astronomy {\&} Astrophysics,
  457, 371, \dodoi{10.1051/0004-6361:20065371}

\bibitem[{Geach {et~al.}(2023)Geach, Lopez-Rodriguez, Doherty, Chen, Ivison,
  Bendo, Dye, \& Coppin}]{Geach2023Polarized2.6}
Geach, J.~E., Lopez-Rodriguez, E., Doherty, M.~J., {et~al.} 2023,
  \dodoi{10.1038/s41586-023-06346-4}

\bibitem[{Geen {et~al.}(2013)Geen, Slyz, \& Devriendt}]{Geen2013}
Geen, S., Slyz, A., \& Devriendt, J. 2013, Monthly Notices of the Royal
  Astronomical Society, 429, 633, \dodoi{10.1093/mnras/sts364}

\bibitem[{Gent {et~al.}(2023)Gent, Mac~Low, \& Korpi-Lagg}]{Gent2023}
Gent, F.~A., Mac~Low, M.-M., \& Korpi-Lagg, M.~J. 2023.
\newblock \url{https://arxiv.org/abs/2306.07051v2
  http://arxiv.org/abs/2306.07051}

\bibitem[{Haardt \& Madau(1996)}]{Haardt1996}
Haardt, F., \& Madau, P. 1996, The Astrophysical Journal, 461, 20,
  \dodoi{10.1086/177035}

\bibitem[{Hanasz {et~al.}(2013)Hanasz, Lesch, Naab, Gawryszczak, Kowalik, \&
  W{\'{o}}lta{\'{n}}ski}]{Hanasz2013}
Hanasz, M., Lesch, H., Naab, T., {et~al.} 2013, Astrophysical Journal Letters,
  777, 38, \dodoi{10.1088/2041-8205/777/2/L38}

\bibitem[{Heesen {et~al.}(2014)Heesen, Brinks, Leroy, Heald, Braun, Bigiel, \&
  Beck}]{Heesen2014}
Heesen, V., Brinks, E., Leroy, A.~K., {et~al.} 2014, Astronomical Journal, 147,
  103, \dodoi{10.1088/0004-6256/147/5/103}

\bibitem[{Hennebelle \& Inutsuka(2019)}]{Hennebelle2019}
Hennebelle, P., \& Inutsuka, S.-I. 2019.
\newblock \url{https://arxiv.org/pdf/1902.00798.pdf
  http://arxiv.org/abs/1902.00798}

\bibitem[{{Hoang} \& {Lazarian}(2016)}]{HL2016}
{Hoang}, T., \& {Lazarian}, A. 2016, \apj, 831, 159,
  \dodoi{10.3847/0004-637X/831/2/159}

\bibitem[{Hopkins(2016)}]{Hopkins2016}
Hopkins, P.~F. 2016, Monthly Notices of the Royal Astronomical Society, 462,
  576, \dodoi{10.1093/mnras/stw1578}

\bibitem[{{Hopkins} {et~al.}(2022){Hopkins}, {Butsky}, {Panopoulou}, {Ji},
  {Quataert}, {Faucher-Gigu{\`e}re}, \& {Kere{\v{s}}}}]{Hopkins2022}
{Hopkins}, P.~F., {Butsky}, I.~S., {Panopoulou}, G.~V., {et~al.} 2022, \mnras,
  516, 3470, \dodoi{10.1093/mnras/stac1791}

\bibitem[{Hopkins {et~al.}(2020)Hopkins, Chan, Garrison-Kimmel, Ji, Su,
  Hummels, Kere{\v{s}}, Quataert, Faucher, \& Ere}]{Hopkins2020}
Hopkins, P.~F., Chan, T.~K., Garrison-Kimmel, S., {et~al.} 2020, MNRAS, 492,
  3465, \dodoi{10.1093/mnras/stz3321}

\bibitem[{Iffrig \& Hennebelle(2017)}]{Iffrig2017}
Iffrig, O., \& Hennebelle, P. 2017, Astronomy {\&} Astrophysics, 604, A70,
  \dodoi{10.1051/0004-6361/201630290}

\bibitem[{Inoue \& Yoshida(2019)}]{Inoue2019}
Inoue, S., \& Yoshida, N. 2019, Monthly Notices of the Royal Astronomical
  Society, 485, 3024, \dodoi{10.1093/mnras/stz584}

\bibitem[{{Inoue} {et~al.}(2018){Inoue}, {Hennebelle}, {Fukui}, {Matsumoto},
  {Iwasaki}, \& {Inutsuka}}]{Inoue2018}
{Inoue}, T., {Hennebelle}, P., {Fukui}, Y., {et~al.} 2018, \pasj, 70, S53,
  \dodoi{10.1093/pasj/psx089}

\bibitem[{Jung {et~al.}(2023)Jung, McClure-Griffiths, Pakmor, Ma, Hill,
  Van~Eck, \& Anderson}]{Jung2023}
Jung, S.~L., McClure-Griffiths, N.~M., Pakmor, R., {et~al.} 2023, Monthly
  Notices of the Royal Astronomical Society, 526, 836,
  \dodoi{10.1093/mnras/stad2811}

\bibitem[{Katz(2022)}]{Katz2022a}
Katz, H. 2022, Monthly Notices of the Royal Astronomical Society, 512, 348,
  \dodoi{10.1093/mnras/stac423}

\bibitem[{Katz {et~al.}(2021)Katz, Martin-Alvarez, Rosdahl, Kimm, Blaizot,
  Haehnelt, Michel-Dansac, Garel, O{\~{n}}orbe, Devriendt, Slyz, Attia, \&
  Teyssier}]{KMA2021}
Katz, H., Martin-Alvarez, S., Rosdahl, J., {et~al.} 2021, Monthly Notices of
  the Royal Astronomical Society, 507, 1254, \dodoi{10.1093/mnras/stab2148}

\bibitem[{Kimm \& Cen(2014)}]{Kimm2014}
Kimm, T., \& Cen, R. 2014, Astrophysical Journal, 788, 121,
  \dodoi{10.1088/0004-637X/788/2/121}

\bibitem[{Kimm {et~al.}(2017)Kimm, Katz, Haehnelt, Rosdahl, Devriendt, \&
  Slyz}]{Kimm2017}
Kimm, T., Katz, H., Haehnelt, M., {et~al.} 2017, Monthly Notices of the Royal
  Astronomical Society, 466, stx052, \dodoi{10.1093/mnras/stx052}

\bibitem[{{King} {et~al.}(2018){King}, {Fissel}, {Chen}, \& {Li}}]{King2018}
{King}, P.~K., {Fissel}, L.~M., {Chen}, C.-Y., \& {Li}, Z.-Y. 2018, \mnras,
  474, 5122, \dodoi{10.1093/mnras/stx3096}

\bibitem[{K{\"{o}}rtgen {et~al.}(2017)K{\"{o}}rtgen, Federrath, \&
  Banerjee}]{Kortgen2017}
K{\"{o}}rtgen, B., Federrath, C., \& Banerjee, R. 2017, Monthly Notices of the
  Royal Astronomical Society, 472, 2496, \dodoi{10.1093/mnras/stx2208}

\bibitem[{Krause {et~al.}(2018)Krause, Irwin, Wiegert, Miskolczi,
  Damas-Segovia, Beck, Li, Heald, M{\"{u}}ller, Stein, Rand, Heesen, Walterbos,
  Dettmar, Vargas, English, \& Murphy}]{Krause2018}
Krause, M., Irwin, J., Wiegert, T., {et~al.} 2018, Astronomy and Astrophysics,
  611, A72, \dodoi{10.1051/0004-6361/201731991}

\bibitem[{Krause {et~al.}(2020)Krause, Irwin, Schmidt, Stein, Miskolczi,
  Carolina Mora-Partiarroyo, Wiegert, Beck, Stil, Heald, Li, Damas-Segovia,
  Vargas, Rand, West, Walterbos, Dettmar, English, \& Woodfinden}]{Krause2020}
Krause, M., Irwin, J., Schmidt, P., {et~al.} 2020, Astronomy and Astrophysics,
  639, \dodoi{10.1051/0004-6361/202037780}

\bibitem[{Kriel {et~al.}(2023)Kriel, Beattie, Federrath, Krumholz, \&
  Hew}]{Kriel2023}
Kriel, N., Beattie, J.~R., Federrath, C., Krumholz, M.~R., \& Hew, J. K.~J.
  2023.
\newblock \url{https://arxiv.org/abs/2310.17036v1
  http://arxiv.org/abs/2310.17036}

\bibitem[{Kroupa(2001)}]{Kroupa2001}
Kroupa, P. 2001, Monthly Notices of the Royal Astronomical Society, 322, 231,
  \dodoi{10.1046/j.1365-8711.2001.04022.x}

\bibitem[{Lacki \& Beck(2013)}]{Lacki2013}
Lacki, B.~C., \& Beck, R. 2013, Monthly Notices of the Royal Astronomical
  Society, 430, 3171, \dodoi{10.1093/mnras/stt122}

\bibitem[{Laursen {et~al.}(2009)Laursen, Sommer-Larsen, \&
  Andersen}]{Laursen2009}
Laursen, P., Sommer-Larsen, J., \& Andersen, A.~C. 2009, Astrophysical Journal,
  704, 1640, \dodoi{10.1088/0004-637X/704/2/1640}

\bibitem[{{Lee} \& {Draine}(1985)}]{LD1985}
{Lee}, H.~M., \& {Draine}, B.~T. 1985, \apj, 290, 211, \dodoi{10.1086/162974}

\bibitem[{{Lopez-Rodriguez}(2023)}]{Lopez-Rodriguez2023}
{Lopez-Rodriguez}, E. 2023, \apj, 953, 113, \dodoi{10.3847/1538-4357/ace110}

\bibitem[{Lopez-Rodriguez {et~al.}(2021)Lopez-Rodriguez, Guerra, Asgari-Targhi,
  \& Schmelz}]{Lopez-Rodriguez2021}
Lopez-Rodriguez, E., Guerra, J.~A., Asgari-Targhi, M., \& Schmelz, J.~T. 2021,
  The Astrophysical Journal, 914, 24, \dodoi{10.3847/1538-4357/abf934}

\bibitem[{{Lopez-Rodriguez} {et~al.}(2020){Lopez-Rodriguez}, {Alonso-Herrero},
  {Garc{\'\i}a-Burillo}, {Gordon}, {Ichikawa}, {Imanishi}, {Kameno},
  {Levenson}, {Nikutta}, \& {Packham}}]{Lopez-Rodriguez2020}
{Lopez-Rodriguez}, E., {Alonso-Herrero}, A., {Garc{\'\i}a-Burillo}, S.,
  {et~al.} 2020, \apj, 893, 33, \dodoi{10.3847/1538-4357/ab8013}

\bibitem[{{Lopez-Rodriguez} {et~al.}(2022){Lopez-Rodriguez}, {Mao}, {Beck},
  {Borlaff}, {Ntormousi}, {Tassis}, {Dale}, {Roman-Duval}, {Subramanian},
  {Martin-Alvarez}, {Marcum}, {Clark}, {Reach}, {Harper}, \&
  {Zweibel}}]{SALSAIV}
{Lopez-Rodriguez}, E., {Mao}, S.~A., {Beck}, R., {et~al.} 2022, \apj, 936, 92,
  \dodoi{10.3847/1538-4357/ac7f9d}

\bibitem[{Mao {et~al.}(2017)Mao, Carilli, Gaensler, Wucknitz, Keeton, Basu,
  Beck, Kronberg, \& Zweibel}]{Mao2017}
Mao, S.~A., Carilli, C., Gaensler, B.~M., {et~al.} 2017, Nature Astronomy, 1,
  621, \dodoi{10.1038/s41550-017-0218-x}

\bibitem[{Martin-Alvarez {et~al.}(2022)Martin-Alvarez, Devriendt, Slyz,
  Sijacki, Richardson, \& Katz}]{Martin-Alvarez2022}
Martin-Alvarez, S., Devriendt, J., Slyz, A., {et~al.} 2022, Monthly Notices of
  the Royal Astronomical Society, 513, 3326, \dodoi{10.1093/mnras/stac1099}

\bibitem[{Martin-Alvarez {et~al.}(2018)Martin-Alvarez, Devriendt, Slyz, \&
  Teyssier}]{Martin-Alvarez2018}
Martin-Alvarez, S., Devriendt, J., Slyz, A., \& Teyssier, R. 2018, Monthly
  Notices of the Royal Astronomical Society, 479, 3343,
  \dodoi{10.1093/mnras/sty1623}

\bibitem[{Martin-Alvarez {et~al.}(2021)Martin-Alvarez, Katz, Sijacki,
  Devriendt, \& Slyz}]{Martin-Alvarez2021}
Martin-Alvarez, S., Katz, H., Sijacki, D., Devriendt, J., \& Slyz, A. 2021,
  MNRAS, 504, 2517, \dodoi{10.1093/mnras/stab968}

\bibitem[{Martin-Alvarez {et~al.}(2023)Martin-Alvarez, Sijacki, Haehnelt,
  Farcy, Dubois, Belokurov, Rosdahl, \& Lopez-Rodriguez}]{Martin-Alvarez2023}
Martin-Alvarez, S., Sijacki, D., Haehnelt, M.~G., {et~al.} 2023, Monthly
  Notices of the Royal Astronomical Society, 525, 3806,
  \dodoi{10.1093/mnras/stad2559}

\bibitem[{Martin-Alvarez {et~al.}(2020)Martin-Alvarez, Slyz, Devriendt, \&
  G{\'{o}}mez-Guijarro}]{Martin-Alvarez2020}
Martin-Alvarez, S., Slyz, A., Devriendt, J., \& G{\'{o}}mez-Guijarro, C. 2020,
  Monthly Notices of the Royal Astronomical Society, 495, 4475,
  \dodoi{10.1093/mnras/staa1438}

\bibitem[{{McKee} \& {Ostriker}(2007)}]{McKeeOstriker2007}
{McKee}, C.~F., \& {Ostriker}, E.~C. 2007, \araa, 45, 565,
  \dodoi{10.1146/annurev.astro.45.051806.110602}

\bibitem[{{Moullet} {et~al.}(2023){Moullet}, {Kataria}, {Lis}, {Unwin},
  {Hasegawa}, {Mills}, {Battersby}, {Roc}, \& {Meixner}}]{Moullet2023}
{Moullet}, A., {Kataria}, T., {Lis}, D., {et~al.} 2023, arXiv e-prints,
  arXiv:2310.20572, \dodoi{10.48550/arXiv.2310.20572}

\bibitem[{Padoan \& Nordlund(2011)}]{Padoan2011}
Padoan, P., \& Nordlund, A. 2011, Astrophysical Journal, 730, 40,
  \dodoi{10.1088/0004-637X/730/1/40}

\bibitem[{Pakmor {et~al.}(2018)Pakmor, Guillet, Pfrommer, G{\'{o}}mez, Grand,
  Marinacci, Simpson, \& Springel}]{Pakmor2018}
Pakmor, R., Guillet, T., Pfrommer, C., {et~al.} 2018, Monthly Notices of the
  Royal Astronomical Society, 481, 4410, \dodoi{10.1093/mnras/sty2601}

\bibitem[{Pakmor {et~al.}(2017)Pakmor, G{\'{o}}mez, Grand, Marinacci, Simpson,
  Springel, Campbell, Frenk, Guillet, Pfrommer, \& White}]{Pakmor2017}
Pakmor, R., G{\'{o}}mez, F.~A., Grand, R. J.~J., {et~al.} 2017, Monthly Notices
  of the Royal Astronomical Society, 469, 3185, \dodoi{10.1093/mnras/stx1074}

\bibitem[{Pakmor {et~al.}(2023)Pakmor, Bieri, van~de Voort, Werhahn, Fattahi,
  Guillet, Pfrommer, Springel, \& Talbot}]{Pakmor2023}
Pakmor, R., Bieri, R., van~de Voort, F., {et~al.} 2023, MNRAS, 000, 1.
\newblock \url{https://arxiv.org/abs/2309.13104v1
  http://arxiv.org/abs/2309.13104}

\bibitem[{Parizot {et~al.}(2006)Parizot, Marcowith, Ballet, \&
  Gallant}]{Parizot2006}
Parizot, E., Marcowith, A., Ballet, J., \& Gallant, Y.~A. 2006, Astronomy and
  Astrophysics, 453, 387, \dodoi{10.1051/0004-6361:20064985}

\bibitem[{Pellegrini {et~al.}(2020)Pellegrini, Reissl, Rahner, Klessen, Glover,
  Pakmor, Herrera-Camus, \& Grand}]{Pellegrini2020}
Pellegrini, E.~W., Reissl, S., Rahner, D., {et~al.} 2020, Monthly Notices of
  the Royal Astronomical Society, 498, 3193, \dodoi{10.1093/mnras/staa2555}

\bibitem[{Pfrommer {et~al.}(2017)Pfrommer, Pakmor, Schaal, Simpson, \&
  Springel}]{Pfrommer2017b}
Pfrommer, C., Pakmor, R., Schaal, K., Simpson, C.~M., \& Springel, V. 2017,
  Monthly Notices of the Royal Astronomical Society, 465, 4500,
  \dodoi{10.1093/mnras/stw2941}

\bibitem[{{Pfrommer} {et~al.}(2022){Pfrommer}, {Werhahn}, {Pakmor},
  {Girichidis}, \& {Simpson}}]{Pfrommer2022}
{Pfrommer}, C., {Werhahn}, M., {Pakmor}, R., {Girichidis}, P., \& {Simpson},
  C.~M. 2022, \mnras, 515, 4229, \dodoi{10.1093/mnras/stac1808}

\bibitem[{Pillepich {et~al.}(2018)Pillepich, Springel, Nelson, Genel, Naiman,
  Pakmor, Hernquist, Torrey, Vogelsberger, Weinberger, \&
  Marinacci}]{Pillepich2018a}
Pillepich, A., Springel, V., Nelson, D., {et~al.} 2018, Monthly Notices of the
  Royal Astronomical Society, 473, 4077, \dodoi{10.1093/mnras/stx2656}

\bibitem[{{Planck Collaboration}(2015)}]{PlanckCollaboration2015}
{Planck Collaboration}. 2015, Astronomy {\&} Astrophysics, 594, A19,
  \dodoi{10.1051/0004-6361/201525821}

\bibitem[{{Planck Collaboration} {et~al.}(2015){Planck Collaboration}, {Ade},
  {Aghanim}, {Alina}, {Alves}, {Aniano}, {Armitage-Caplan}, {Arnaud},
  {Arzoumanian}, {Ashdown}, {Atrio-Barandela}, {Aumont}, {Baccigalupi},
  {Banday}, {Barreiro}, {Battaner}, {Benabed}, {Benoit-L{\'e}vy}, {Bernard},
  {Bersanelli}, {Bielewicz}, {Bond}, {Borrill}, {Bouchet}, {Boulanger},
  {Bracco}, {Burigana}, {Cardoso}, {Catalano}, {Chamballu}, {Chiang},
  {Christensen}, {Colombi}, {Colombo}, {Combet}, {Couchot}, {Coulais}, {Crill},
  {Curto}, {Cuttaia}, {Danese}, {Davies}, {Davis}, {de Bernardis}, {de Rosa},
  {de Zotti}, {Delabrouille}, {Dickinson}, {Diego}, {Donzelli}, {Dor{\'e}},
  {Douspis}, {Dupac}, {Efstathiou}, {En{\ss}lin}, {Eriksen}, {Falgarone},
  {Fanciullo}, {Ferri{\`e}re}, {Finelli}, {Forni}, {Frailis}, {Fraisse},
  {Franceschi}, {Galeotta}, {Ganga}, {Ghosh}, {Giard}, {Giraud-H{\'e}raud},
  {Gonz{\'a}lez-Nuevo}, {G{\'o}rski}, {Gregorio}, {Gruppuso}, {Guillet},
  {Hansen}, {Harrison}, {Helou}, {Hern{\'a}ndez-Monteagudo}, {Hildebrandt},
  {Hivon}, {Hobson}, {Holmes}, {Hornstrup}, {Huffenberger}, {Jaffe}, {Jaffe},
  {Jones}, {Juvela}, {Keih{\"a}nen}, {Keskitalo}, {Kisner}, {Kneissl},
  {Knoche}, {Kunz}, {Kurki-Suonio}, {Lagache}, {Lamarre}, {Lasenby},
  {Lawrence}, {Leonardi}, {Levrier}, {Liguori}, {Lilje}, {Linden-V{\o}rnle},
  {L{\'o}pez-Caniego}, {Lubin}, {Mac{\'\i}as-P{\'e}rez}, {Maino}, {Mandolesi},
  {Maris}, {Marshall}, {Martin}, {Mart{\'\i}nez-Gonz{\'a}lez}, {Masi},
  {Matarrese}, {Mazzotta}, {Melchiorri}, {Mendes}, {Mennella}, {Migliaccio},
  {Miville-Desch{\^e}nes}, {Moneti}, {Montier}, {Morgante}, {Mortlock},
  {Munshi}, {Murphy}, {Naselsky}, {Nati}, {Natoli}, {Netterfield}, {Noviello},
  {Novikov}, {Novikov}, {Oxborrow}, {Pagano}, {Pajot}, {Paoletti}, {Pasian},
  {Pelkonen}, {Perdereau}, {Perotto}, {Perrotta}, {Piacentini}, {Piat},
  {Pietrobon}, {Plaszczynski}, {Pointecouteau}, {Polenta}, {Popa}, {Pratt},
  {Prunet}, {Puget}, {Rachen}, {Reinecke}, {Remazeilles}, {Renault},
  {Ricciardi}, {Riller}, {Ristorcelli}, {Rocha}, {Rosset}, {Roudier},
  {Rusholme}, {Sandri}, {Scott}, {Soler}, {Spencer}, {Stolyarov}, {Stompor},
  {Sudiwala}, {Sutton}, {Suur-Uski}, {Sygnet}, {Tauber}, {Terenzi},
  {Toffolatti}, {Tomasi}, {Tristram}, {Tucci}, {Umana}, {Valenziano},
  {Valiviita}, {Van Tent}, {Vielva}, {Villa}, {Wade}, {Wandelt}, \&
  {Zonca}}]{PlanckXX2015}
{Planck Collaboration}, {Ade}, P.~A.~R., {Aghanim}, N., {et~al.} 2015, \aap,
  576, A105, \dodoi{10.1051/0004-6361/201424086}

\bibitem[{{Planck Collaboration} {et~al.}(2020){Planck Collaboration},
  {Aghanim}, {Akrami}, {Alves}, {Ashdown}, {Aumont}, {Baccigalupi},
  {Ballardini}, {Banday}, {Barreiro}, {Bartolo}, {Basak}, {Benabed}, {Bernard},
  {Bersanelli}, {Bielewicz}, {Bock}, {Bond}, {Borrill}, {Bouchet}, {Boulanger},
  {Bracco}, {Bucher}, {Burigana}, {Calabrese}, {Cardoso}, {Carron}, {Chary},
  {Chiang}, {Colombo}, {Combet}, {Crill}, {Cuttaia}, {de Bernardis}, {de
  Zotti}, {Delabrouille}, {Delouis}, {Di Valentino}, {Dickinson}, {Diego},
  {Dor{\'e}}, {Douspis}, {Ducout}, {Dupac}, {Efstathiou}, {Elsner},
  {En{\ss}lin}, {Eriksen}, {Falgarone}, {Fantaye}, {Fernandez-Cobos},
  {Ferri{\`e}re}, {Finelli}, {Forastieri}, {Frailis}, {Fraisse}, {Franceschi},
  {Frolov}, {Galeotta}, {Galli}, {Ganga}, {G{\'e}nova-Santos}, {Gerbino},
  {Ghosh}, {Gonz{\'a}lez-Nuevo}, {G{\'o}rski}, {Gratton}, {Green}, {Gruppuso},
  {Gudmundsson}, {Guillet}, {Handley}, {Hansen}, {Helou}, {Herranz}, {Hivon},
  {Huang}, {Jaffe}, {Jones}, {Keih{\"a}nen}, {Keskitalo}, {Kiiveri}, {Kim},
  {Krachmalnicoff}, {Kunz}, {Kurki-Suonio}, {Lagache}, {Lamarre}, {Lasenby},
  {Lattanzi}, {Lawrence}, {Le Jeune}, {Levrier}, {Liguori}, {Lilje},
  {Lindholm}, {L{\'o}pez-Caniego}, {Lubin}, {Ma}, {Mac{\'\i}as-P{\'e}rez},
  {Maggio}, {Maino}, {Mandolesi}, {Mangilli}, {Marcos-Caballero}, {Maris},
  {Martin}, {Mart{\'\i}nez-Gonz{\'a}lez}, {Matarrese}, {Mauri}, {McEwen},
  {Melchiorri}, {Mennella}, {Migliaccio}, {Miville-Desch{\^e}nes}, {Molinari},
  {Moneti}, {Montier}, {Morgante}, {Moss}, {Natoli}, {Pagano}, {Paoletti},
  {Patanchon}, {Perrotta}, {Pettorino}, {Piacentini}, {Polastri}, {Polenta},
  {Puget}, {Rachen}, {Reinecke}, {Remazeilles}, {Renzi}, {Ristorcelli},
  {Rocha}, {Rosset}, {Roudier}, {Rubi{\~n}o-Mart{\'\i}n}, {Ruiz-Granados},
  {Salvati}, {Sandri}, {Savelainen}, {Scott}, {Sirignano}, {Sunyaev},
  {Suur-Uski}, {Tauber}, {Tavagnacco}, {Tenti}, {Toffolatti}, {Tomasi},
  {Trombetti}, {Valiviita}, {Vansyngel}, {Van Tent}, {Vielva}, {Villa},
  {Vittorio}, {Wandelt}, {Wehus}, {Zacchei}, \& {Zonca}}]{PlanckXII2020}
{Planck Collaboration}, {Aghanim}, N., {Akrami}, Y., {et~al.} 2020, \aap, 641,
  A12, \dodoi{10.1051/0004-6361/201833885}

\bibitem[{Ponnada {et~al.}(2022)Ponnada, Panopoulou, Butsky, Hopkins, Loebman,
  Hummels, Ji, Wetzel, Faucher-Gigu{\`{e}}re, \& Hayward}]{Ponnada2022}
Ponnada, S.~B., Panopoulou, G.~V., Butsky, I.~S., {et~al.} 2022, Monthly
  Notices of the Royal Astronomical Society, 516, 4417,
  \dodoi{10.1093/mnras/stac2448}

\bibitem[{Ponnada {et~al.}(2023)Ponnada, Panopoulou, Butsky, Hopkins, Skalidis,
  Hummels, Quataert, Kere{\v{s}}, Faucher-Gigu{\`{e}}re, \& Su}]{Ponnada2023a}
---. 2023, MNRAS, 000, 1.
\newblock \url{https://arxiv.org/abs/2309.04526v1}

\bibitem[{{Ponnada} {et~al.}(2023){Ponnada}, {Butsky}, {Skalidis}, {Hopkins},
  {Panopoulou}, {Hummels}, {Kere{\v{s}}}, {Quataert}, {Faucher-Gigu{\`e}re}, \&
  {Su}}]{Ponnada2023b}
{Ponnada}, S.~B., {Butsky}, I.~S., {Skalidis}, R., {et~al.} 2023, arXiv
  e-prints, arXiv:2309.16752, \dodoi{10.48550/arXiv.2309.16752}

\bibitem[{Powell {et~al.}(1999)Powell, Roe, Linde, Gombosi, \&
  De~Zeeuw}]{Powell1999}
Powell, K.~G., Roe, P.~L., Linde, T.~J., Gombosi, T.~I., \& De~Zeeuw, D.~L.
  1999, Journal of Computational Physics, 154, 284,
  \dodoi{10.1006/JCPH.1999.6299}

\bibitem[{Powell {et~al.}(2011)Powell, Slyz, \& Devriendt}]{Powell2011}
Powell, L.~C., Slyz, A., \& Devriendt, J. 2011, Monthly Notices of the Royal
  Astronomical Society, 414, 3671, \dodoi{10.1111/j.1365-2966.2011.18668.x}

\bibitem[{Power {et~al.}(2003)Power, Navarro, Jenkins, Frenk, White, Springel,
  Stadel, \& Quinn}]{Power2003}
Power, C., Navarro, J.~F., Jenkins, A., {et~al.} 2003, Monthly Notices of the
  Royal Astronomical Society, 338, 14, \dodoi{10.1046/j.1365-8711.2003.05925.x}

\bibitem[{Purcell(1979)}]{Purcell1979}
Purcell, E.~M. 1979, The Astrophysical Journal, 231, 404,
  \dodoi{10.1086/157204}

\bibitem[{Rasera \& Teyssier(2006)}]{Rasera2006}
Rasera, Y., \& Teyssier, R. 2006, Astronomy {\&} Astrophysics, 445, 1,
  \dodoi{10.1051/0004-6361:20053116}

\bibitem[{Reissl {et~al.}(2019)Reissl, Brauer, Klessen, \&
  Pellegrini}]{Reissl2019}
Reissl, S., Brauer, R., Klessen, R.~S., \& Pellegrini, E.~W. 2019, The
  Astrophysical Journal, 885, 15, \dodoi{10.3847/1538-4357/ab3664}

\bibitem[{Reissl {et~al.}(2016)Reissl, Wolf, \& Brauer}]{Reissl2016}
Reissl, S., Wolf, S., \& Brauer, R. 2016, Astronomy and Astrophysics, 593, A87,
  \dodoi{10.1051/0004-6361/201424930}

\bibitem[{Roche {et~al.}(2018)Roche, Lopez-Rodriguez, Telesco, Schodel, \&
  Packham}]{Roche2018}
Roche, P.~F., Lopez-Rodriguez, E., Telesco, C.~M., Schodel, R., \& Packham, C.
  2018, \dodoi{10.1016/j.pss.2018.07.007}

\bibitem[{Rodr{\'{i}}guez~Montero {et~al.}(2022)Rodr{\'{i}}guez~Montero,
  Martin-Alvarez, Sijacki, Slyz, Devriendt, \& Dubois}]{Rodriguez-Montero2022}
Rodr{\'{i}}guez~Montero, F., Martin-Alvarez, S., Sijacki, D., {et~al.} 2022,
  Monthly Notices of the Royal Astronomical Society, 511, 1247,
  \dodoi{10.1093/mnras/stab3716}

\bibitem[{Rodr{\'{i}}guez~Montero {et~al.}(2023)Rodr{\'{i}}guez~Montero,
  Martin-Alvarez, Slyz, Devriendt, Dubois, \& Sijacki}]{Rodriguez-Montero2023}
Rodr{\'{i}}guez~Montero, F., Martin-Alvarez, S., Slyz, A., {et~al.} 2023,
  MNRAS, 000, 1.
\newblock \url{https://arxiv.org/abs/2307.13733v1
  http://arxiv.org/abs/2307.13733}

\bibitem[{Rosdahl {et~al.}(2015)Rosdahl, Schaye, Teyssier, \&
  Agertz}]{Rosdahl2015b}
Rosdahl, J., Schaye, J., Teyssier, R., \& Agertz, O. 2015, Monthly Notices of
  the Royal Astronomical Society, 451, 34, \dodoi{10.1093/mnras/stv937}

\bibitem[{Rosdahl \& Teyssier(2015)}]{Rosdahl2015a}
Rosdahl, J., \& Teyssier, R. 2015, Monthly Notices of the Royal Astronomical
  Society, 449, 4380, \dodoi{10.1093/mnras/stv567}

\bibitem[{Rosen \& Bregman(1995)}]{Rosen1995}
Rosen, A., \& Bregman, J.~N. 1995, The Astrophysical Journal, 440, 634,
  \dodoi{10.1086/175303}

\bibitem[{{Sanati} {et~al.}(2024){Sanati}, {Martin-Alvarez}, {Schober},
  {Revaz}, {Slyz}, \& {Devriendt}}]{Sanati2024}
{Sanati}, M., {Martin-Alvarez}, S., {Schober}, J., {et~al.} 2024, arXiv
  e-prints, arXiv:2403.05672, \dodoi{10.48550/arXiv.2403.05672}

\bibitem[{Schmidt(1959)}]{Schmidt1959}
Schmidt, M. 1959, The Astrophysical Journal, 129, 243, \dodoi{10.1086/146614}

\bibitem[{{Seifried} {et~al.}(2019){Seifried}, {Walch}, {Reissl}, \&
  {Ib{\'a}{\~n}ez-Mej{\'\i}a}}]{Seifried2019}
{Seifried}, D., {Walch}, S., {Reissl}, S., \& {Ib{\'a}{\~n}ez-Mej{\'\i}a},
  J.~C. 2019, \mnras, 482, 2697, \dodoi{10.1093/mnras/sty2831}

\bibitem[{Shukurov \& Subramanian(2021)}]{Shukurov2021}
Shukurov, A., \& Subramanian, K. 2021, {Astrophysical Magnetic Fields}
  (Cambridge University Press), \dodoi{10.1017/9781139046657}

\bibitem[{Steinwandel {et~al.}(2019)Steinwandel, Beck, Arth, Dolag, Moster, \&
  Nielaba}]{Steinwandel2019}
Steinwandel, U.~P., Beck, M.~C., Arth, A., {et~al.} 2019, Monthly Notices of
  the Royal Astronomical Society, 483, 1008, \dodoi{10.1093/mnras/sty3083}

\bibitem[{Su {et~al.}(2017)Su, Hopkins, Hayward, Faucher-Gigu{\`{e}}re,
  Kere{\v{s}}, Ma, \& Robles}]{Su2017}
Su, K.~Y., Hopkins, P.~F., Hayward, C.~C., {et~al.} 2017, Monthly Notices of
  the Royal Astronomical Society, 471, 144, \dodoi{10.1093/MNRAS/STX1463}

\bibitem[{Sun {et~al.}(2008)Sun, Reich, Waelkens, \& En{\ss}lin}]{Sun2008}
Sun, X.~H., Reich, W., Waelkens, A., \& En{\ss}lin, T.~A. 2008, Astronomy and
  Astrophysics, 477, 573, \dodoi{10.1051/0004-6361:20078671}

\bibitem[{{Surgent} {et~al.}(2023){Surgent}, {Lopez-Rodriguez}, \&
  {Clark}}]{Surgent2023}
{Surgent}, W.~J., {Lopez-Rodriguez}, E., \& {Clark}, S.~E. 2023, \apj, 954, 53,
  \dodoi{10.3847/1538-4357/ace4c0}

\bibitem[{Tabatabaei {et~al.}(2017)Tabatabaei, Schinnerer, Krause, Dumas,
  Meidt, Damas-Segovia, Beck, Murphy, Mulcahy, Groves, Bolatto, Dale, Galametz,
  Sandstrom, Boquien, Calzetti, Kennicutt, Hunt, Looze, \&
  Pellegrini}]{Tabatabaei2017}
Tabatabaei, F.~S., Schinnerer, E., Krause, M., {et~al.} 2017, The Astrophysical
  Journal, 836, 185, \dodoi{10.3847/1538-4357/836/2/185}

\bibitem[{{Tahani} {et~al.}(2022){Tahani}, {Glover}, {Lupypciw}, {West},
  {Kothes}, {Plume}, {Inutsuka}, {Lee}, {Grenier}, {Knee}, {Brown}, {Doi},
  {Robishaw}, \& {Haverkorn}}]{Tahani2022a}
{Tahani}, M., {Glover}, J., {Lupypciw}, W., {et~al.} 2022, \aap, 660, L7,
  \dodoi{10.1051/0004-6361/202243322}

\bibitem[{Tahani {et~al.}(2022)Tahani, Lupypciw, Glover, Plume, West, Kothes,
  Inutsuka, Lee, Robishaw, Knee, Brown, Doi, Grenier, \&
  Haverkorn}]{Tahani2022bb}
Tahani, M., Lupypciw, W., Glover, J., {et~al.} 2022, Astronomy and
  Astrophysics, 660, A97, \dodoi{10.1051/0004-6361/202141170}

\bibitem[{Teyssier(2002)}]{Teyssier2002}
Teyssier, R. 2002, Astronomy {\&} Astrophysics, 385, 337,
  \dodoi{10.1051/0004-6361:20011817}

\bibitem[{Teyssier {et~al.}(2006)Teyssier, Fromang, \& Dormy}]{Teyssier2006}
Teyssier, R., Fromang, S., \& Dormy, E. 2006, Journal of Computational Physics,
  218, 44, \dodoi{10.1016/j.jcp.2006.01.042}

\bibitem[{Thilker {et~al.}(2023)Thilker, Lee, Deger, Barnes, Bigiel, Boquien,
  Cao, Chevance, Dale, Egorov, Glover, Grasha, Henshaw, Klessen, Koch,
  Kruijssen, Leroy, Lessing, Meidt, Pinna, Querejeta, Rosolowsky, Sandstrom,
  Schinnerer, Smith, Watkins, Williams, Anand, Belfiore, Blanc, Chandar,
  Congiu, Emsellem, Groves, Kreckel, Larson, Liu, Pessa, \&
  Whitmore}]{Thilker2023}
Thilker, D.~A., Lee, J.~C., Deger, S., {et~al.} 2023, The Astrophysical Journal
  Letters, 944, L13, \dodoi{10.3847/2041-8213/acaeac}

\bibitem[{Tillson {et~al.}(2015)Tillson, Devriendt, Slyz, Miller, \&
  Pichon}]{Tillson2015}
Tillson, H., Devriendt, J., Slyz, A., Miller, L., \& Pichon, C. 2015, Monthly
  Notices of the Royal Astronomical Society, 449, 4363,
  \dodoi{10.1093/mnras/stv557}

\bibitem[{T{\'{o}}th(2000)}]{Toth2000}
T{\'{o}}th, G. 2000, Journal of Computational Physics, 161, 605,
  \dodoi{10.1006/jcph.2000.6519}

\bibitem[{Trebitsch {et~al.}(2017)Trebitsch, Blaizot, Rosdahl, Devriendt, \&
  Slyz}]{Trebitsch2017}
Trebitsch, M., Blaizot, J., Rosdahl, J., Devriendt, J., \& Slyz, A. 2017,
  Monthly Notices of the Royal Astronomical Society, 470, 224,
  \dodoi{10.1093/mnras/stx1060}

\bibitem[{Tweed {et~al.}(2009)Tweed, Devriendt, Blaizot, Colombi, \&
  Slyz}]{Tweed2009}
Tweed, D., Devriendt, J., Blaizot, J., Colombi, S., \& Slyz, A. 2009, Astronomy
  {\&} Astrophysics, 506, 647, \dodoi{10.1051/0004-6361/200911787}

\bibitem[{Van De~Voort {et~al.}(2021)Van De~Voort, Bieri, Pakmor, G{\'{o}}mez,
  Grand, \& Marinacci}]{vandeVoort2021}
Van De~Voort, F., Bieri, R., Pakmor, R., {et~al.} 2021, Monthly Notices of the
  Royal Astronomical Society, 501, 4888, \dodoi{10.1093/mnras/staa3938}

\bibitem[{Vazza {et~al.}(2017)Vazza, Br{\"{u}}ggen, Gheller, Hackstein, Wittor,
  \& Hinz}]{Vazza2017}
Vazza, F., Br{\"{u}}ggen, M., Gheller, C., {et~al.} 2017, Classical and Quantum
  Gravity, 34, 234001, \dodoi{10.1088/1361-6382/aa8e60}

\bibitem[{Webber(1998)}]{Webber1998}
Webber, W.~R. 1998, The Astrophysical Journal, 506, 329, \dodoi{10.1086/306222}

\bibitem[{{Werhahn} {et~al.}(2021{\natexlab{a}}){Werhahn}, {Pfrommer}, \&
  {Girichidis}}]{Werhahn2021c}
{Werhahn}, M., {Pfrommer}, C., \& {Girichidis}, P. 2021{\natexlab{a}}, \mnras,
  508, 4072, \dodoi{10.1093/mnras/stab2535}

\bibitem[{{Werhahn} {et~al.}(2021{\natexlab{b}}){Werhahn}, {Pfrommer},
  {Girichidis}, {Puchwein}, \& {Pakmor}}]{Werhahn2021a}
{Werhahn}, M., {Pfrommer}, C., {Girichidis}, P., {Puchwein}, E., \& {Pakmor},
  R. 2021{\natexlab{b}}, \mnras, 505, 3273, \dodoi{10.1093/mnras/stab1324}

\bibitem[{{Werhahn} {et~al.}(2021{\natexlab{c}}){Werhahn}, {Pfrommer},
  {Girichidis}, \& {Winner}}]{Werhahn2021b}
{Werhahn}, M., {Pfrommer}, C., {Girichidis}, P., \& {Winner}, G.
  2021{\natexlab{c}}, \mnras, 505, 3295, \dodoi{10.1093/mnras/stab1325}

\bibitem[{Whittingham {et~al.}(2021)Whittingham, Sparre, Pfrommer, \&
  Pakmor}]{Whittingham2021}
Whittingham, J., Sparre, M., Pfrommer, C., \& Pakmor, R. 2021, Monthly Notices
  of the Royal Astronomical Society, 506, 229, \dodoi{10.1093/mnras/stab1425}

\bibitem[{Whitworth {et~al.}(2023)Whitworth, Smith, Klessen, Mac~Low, Glover,
  Tress, Pakmor, \& Soler}]{Whitworth2023}
Whitworth, D.~J., Smith, R.~J., Klessen, R.~S., {et~al.} 2023, Monthly Notices
  of the Royal Astronomical Society, 520, 89, \dodoi{10.1093/mnras/stad105}

\bibitem[{Widrow(2002)}]{Widrow2002}
Widrow, L.~M. 2002, Reviews of Modern Physics, 74, 775,
  \dodoi{10.1103/RevModPhys.74.775}

\bibitem[{{Yuan} {et~al.}(2024){Yuan}, {Martin-Alvarez}, {Haehnelt}, {Garel},
  \& {Sijacki}}]{Yuan2024}
{Yuan}, Y., {Martin-Alvarez}, S., {Haehnelt}, M.~G., {Garel}, T., \& {Sijacki},
  D. 2024, arXiv e-prints, arXiv:2401.02572, \dodoi{10.48550/arXiv.2401.02572}

\bibitem[{Zubko {et~al.}(2004)Zubko, Dwek, \& Arendt}]{Zubko2004}
Zubko, V., Dwek, E., \& Arendt, R.~G. 2004, The Astrophysical Journal
  Supplement Series, 152, 211, \dodoi{10.1086/382351}

\end{thebibliography}



\end{document}